\documentclass[a4paper,10pt]{article}
\pdfoutput=1
\usepackage[colorlinks=true]{hyperref}
\usepackage[dvipsnames]{xcolor}
\usepackage{graphicx,rotating,hyperref,slashed,amsmath,xcolor,amssymb,amsfonts,colortbl,cite}
\usepackage{microtype}
\usepackage{comment}
\usepackage{tcolorbox}
\makeatletter
\hypersetup{colorlinks,bookmarksopen,bookmarksnumbered,  
linkcolor=blus,pdfstartview=FitH,urlcolor=rossos,citecolor=verde}
\allowdisplaybreaks		 

\hypersetup{%
    urlcolor=black,
    citecolor=black,
    linkcolor=black
}
\usepackage{mciteplus}

\AtBeginDocument{
  \hypersetup{
     urlcolor=NavyBlue,
    citecolor=NavyBlue,
    linkcolor=black,
  }%
}

\usepackage{colortbl}
\definecolor{green3}{cmyk}{0.8, 0., 0.7., 0.}
\definecolor{green2}{cmyk}{0, 1, 0.5, 0}
\definecolor{lightgreen}{cmyk}{0.2, 0, 0.2, 0.2}
\definecolor{lightgray}{cmyk}{0.1,0.2,0,0.1}
\definecolor{lightgray2}{cmyk}{0.4,0.4,0,0.8}
\definecolor{black}{cmyk}{1.0,1.0,1.0,1.0}

\def\bea{\begin{eqnarray}}
\def\eea{\end{eqnarray}}

\def\be{\begin{equation}}
\def\ee{\end{equation}}

\allowdisplaybreaks[1]

\usepackage{colortbl}
\definecolor{lightgreen}{cmyk}{0.2, 0, 0.2, 0.2}
\definecolor{lightgray}{cmyk}{0.1,0.2,0,0.1}
\definecolor{lightgray2}{cmyk}{0.1,0.1,0,0.1}

\makeatletter
\newlength{\apb@width}
\newcommand{\autoparbox}[2][c]{\settowidth{\apb@width}{#2}\parbox[#1]{\apb@width}{#2}}

\makeatother

\numberwithin{equation}{section}

\def\beq{\begin{equation}}
\def\eeq{\end{equation}}

\def\bea{\begin{eqnarray}}
\def\eea{\end{eqnarray}}
\def\eg{{\it e.g.~}}

\def\ie{{\it i.e.~}}
\def\d{{\rm d}}

\def\d{{\rm d}}

\def\nn{\nonumber}

\def\Mp{M_{\rm pl}}

\def\l{{\ell}}

\def\fr{\frac}

\def\0{{\boldsymbol 0}}

\def\x{{\boldsymbol{x}}}

\def\fr{\frac}

\def\etc{\eta_{\rm c}}
\def\dn{\Delta N}

\usepackage{setspace}

\begin{document}

\begin{titlepage}

\setcounter{page}{1} \baselineskip=15.5pt \thispagestyle{empty}

\bigskip\

\vspace{1cm}
\begin{center}

{
{\fontsize{19}{28}\selectfont  \sffamily \bfseries {
{Consistency conditions and primordial black holes
\\  \vskip 0.1cm 
 in single  field inflation }
}}
}
\\
\hskip1cm

\end{center}

\vspace{0.2cm}

\begin{center}
{\fontsize{13}{30}\selectfont Ogan \"Ozsoy$^{\heartsuit}$
\,, Gianmassimo Tasinato$^{\spadesuit}$}
\end{center}

\begin{center}

\vskip 8pt
\textsl{
$\spadesuit$ Department of Physics, Swansea University, Swansea, SA2 8PP, United Kingdom\\
$\heartsuit$ CEICO, Institute of Physics of the Czech Academy of Sciences, Na Slovance 1999/2, 182 21,
Prague.}
\vskip 7pt

\end{center}

\vspace{1.2cm}
\hrule \vspace{0.3cm}
\noindent {\sffamily \bfseries Abstract}\\
We discuss new consistency relations for single field  models of inflation capable of generating
primordial black holes (PBH), and their observational  implications for CMB $\mu$-space distortions. 
 These inflationary models  include a short period of non-attractor evolution:  the scale-dependent profile   of  curvature perturbation  is characterised by a pronounced dip, followed by a rapid growth leading  to a peak responsible for PBH formation. 
 We investigate  the squeezed and the collapsed limits of three
and four point functions of curvature perturbation around the dip, showing that they satisfy consistency relations connecting their  values  to  the total amplification of the curvature spectrum, and to the  duration of the non-attractor era. Moreover, the corresponding non-Gaussian parameters  are scale-dependent  in proximity of the dip, with features that again depend on the amplification  of the spectrum. For typical PBH scenarios requiring an order ${\cal O}(10^7)$ enhancement of the spectrum from large towards small scales, we generally find values $f_{\rm NL}^{\rm sq}\,=\,{\cal O}(10)$ and  $\tau_{\rm NL}^{\rm col}\,=\,{\cal O}(10^3)$  in a range of  scales that can be probed by CMB $\mu$-space distortions. Using these consistency relations, we carefully analyse  how  the  scale-dependence of  non-Gaussian parameters leads to characteristic
features in $\langle \mu T \rangle$ and $\langle \mu \mu \rangle$ correlators,  providing distinctive probes of inflationary PBH scenarios   that can be tested  using well-understood CMB physics.

%
  \vskip 10pt
\hrule

\vspace{0.6cm}
 \end{titlepage}

 \tableofcontents
 
\newpage


\section{Introduction}

Primordial black holes (PBH) might constitute a fraction of the dark matter of our
universe being formed by the gravitational collapse of overdense regions
in the early stages of our universe evolution \cite{Hawking:1971ei,Carr:1974nx,Carr:1975qj}.
 Such overdense regions can be produced by an amplification  of the spectrum of  primordial curvature
 fluctuations from  inflation \cite{Ivanov:1994pa,GarciaBellido:1996qt}. We refer
 the reader to the recent reviews \cite{Carr:2016drx,Sasaki:2018dmp,Carr:2020xqk,Carr:2020gox} for constraints on PBH populations and details on their formation mechanisms. 
 
 \smallskip
 
 Within the framework of single-field inflation, the enhancement of the spectrum of fluctuations can occur during a brief phase of
 non-attractor evolution, for example associated with an inflection point in the inflaton potential (for
 explicit constructions, see \cite{Garcia-Bellido:2017mdw,Germani:2017bcs,Ezquiaga:2017fvi,Motohashi:2017kbs,Ballesteros:2017fsr,Hertzberg:2017dkh,Cicoli:2018asa,Ozsoy:2018flq}). In this case, the would-be decaying mode of curvature fluctuations
is actually active  at superhorizon scales, where it  plays an important role in determining the amplitude 
of the curvature spectrum. In particular, it causes an enhancement of its amplitude by several
orders of  magnitude from large towards small scales -- see Fig \ref{fig:complete-profile2-intro} for a representative example. Despite the  large variety of single-field scenarios, there are universal features that are common to
all models with a short-phase of non--slow-roll  evolution. First, the spectrum as a function of momentum does
not grow faster than $k^4$ in its way towards the peak \cite{Byrnes:2018txb}, see for example Fig \ref{fig:complete-profile2-intro}. Second, the growth of the spectrum is preceded by a pronounced dip, whose position is controlled
by the total enhancement of the spectrum \cite{Tasinato:2020vdk}. In \cite{Ozsoy:2021qrg}, building on
the original idea of \cite{Pajer:2012vz}, we showed that large squeezed non-Gaussianity in proximity
of the dip of the spectrum induces sizeable cross-correlations between CMB $\mu$-distortions
and temperature fluctuations, which can be used as a test of PBH scenarios based on single-field inflation. 
Interestingly, this proposal is  based only on well understood  perturbation theory at large
cosmological scales, and does
not need to consider complex non-linear phenomena occurring at PBH formation. 

\smallskip

In this work we make further steps in exploring this subject, by determining new properties for the statistics of curvature fluctuations around the dip of the spectrum, and their consequences for CMB $\mu$-distortion anisotropies. We first show that appropriate limits of $n$-point correlators of  curvature fluctuations  satisfy  new consistency relations, connecting their maximal values to the total amplification of the spectrum. 
For example, calling $\Pi_T$ the total amplification of the spectrum from large to small scales -- and having
in mind a typical values a  $\Pi_T\,\simeq\,10^7$ for generating PBH -- we focus on the parameters
$f_{\rm NL}^{\rm sqz}$   and $\tau_{\rm NL}^{\rm coll}$   controlling respectively the squeezed bispectrum
and collapsed trispectrum. We find that around the dip these quantities satisfy universal consistency relations, and have maximal values that scale as
$f_{\rm NL}^{\rm sqz}\,\simeq\,\Pi_T^{1/4}\,\simeq\,{\cal O}(10)$,
and $\tau_{\rm NL}^{\rm coll}\,\simeq\,\Pi_T^{1/2}\,\simeq\,{\cal O}(10^3)$ with respect to $\Pi_T$, up to overall 
 order one coefficients  determined by the duration of the non-attractor epoch.  
Moreover, such squeezed and collapsed limits  are scale-dependent (see Fig \ref{fig:plot3}), and their
features are again controlled by the total amplification of the spectrum. We prove our consistency relations
for non-Gaussian parameters in two ways. In section \ref{sec_heur1} we make use of a heuristic approach based
on \cite{Tasinato:2020vdk}, which allows one to get a simple, intuitive understanding of the physics behind the
single-field system we consider. Then, in section \ref{S2}, we re-derive the consistency conditions
using the rigorous gradient expansion approach first introduced \cite{Leach:2001zf}, so to place  the results of section  \ref{sec_heur1} in firmer
footings. In section \ref{PBSandTS}, we study the implications
of our findings for CMB $\mu$-distortions. In this respect,
we go beyond the analysis carried in \cite{Ozsoy:2021qrg}
 by developing the following points:
  First, we show how the information provided by the consistency 
 relations allows us to carry on a more detailed  analysis of $\langle \mu T \rangle$
 correlators,  whose quantitative and qualitative features depend on the properties
 of the scale-dependent squeezed bispectrum.
  Then, we study for the first time the implications for the  $\langle \mu \mu \rangle$  self-correlator 
 of a  scale-dependent
 collapsed trispectrum around the dip. 
  Finally, at the light of the results above,  we  discuss improved estimates
for the detectability of  non-Gaussian consistency relations  with   PIXIE or PRISM-like experiment, and physical implications 
for  PBH populations.  We conclude our work with a Discussion in Section \ref{SecDis}, followed by technical appendixes.





\section{
Consistency relations at the dip: a heuristic  approach}
\label{sec_heur1}
We begin by developing a heuristic, intuitive approach for  characterising the statistics of
  curvature perturbations. We consider single-field inflationary  scenarios
 that include a short phase of non-slow-roll evolution, which   can lead to PBH
 production. We characterize  the  scale-dependence  of  the curvature  perturbation  spectrum,
   that
   in turn implies new consistency relations for  squeezed  limits of   $n$-point functions.
  The results  of these  heuristic considerations   will be supported by a more
 rigorous analysis in section \ref{S2}. Phenomenological implications are then developed in section \ref{PBSandTS}, where
 we apply  our findings  to CMB $\mu$-distortions as a  probe of early universe scenarios
leading to   PBH formation.

 \medskip

  \begin{figure}[h!]
\centering
 \includegraphics[width = 0.55 \textwidth]{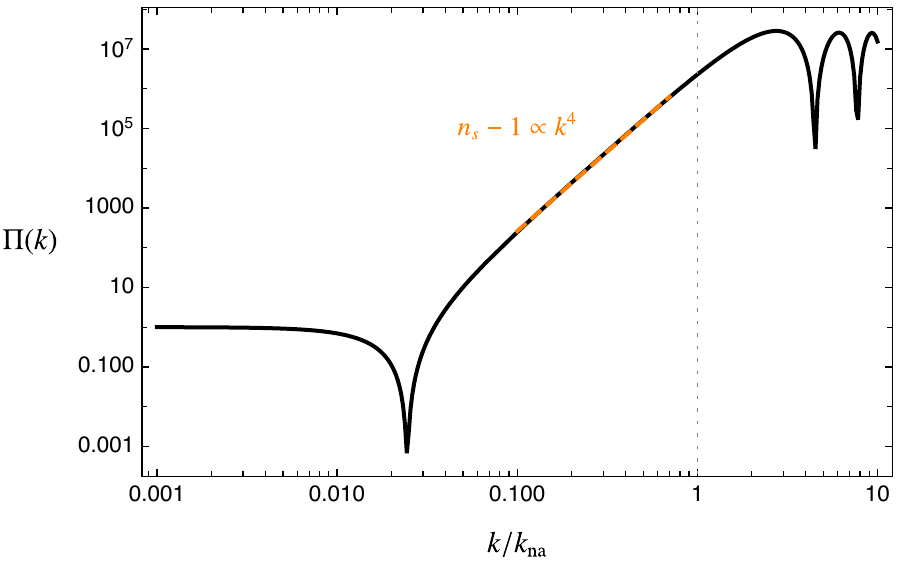}
 \caption{\it{\small   A representative  plot of  the  spectrum  of fluctuations in inflationary scenarios including short phase of non-attractor expansion.  Horizontal axis:    $k/k_{\rm na}$, with
  $k_{\rm na}$  the scale of horizon-crossing  
  at the onset of the non-slow-roll epoch.  Vertical axis:  ratio $\Pi(k)$ of the  spectrum evaluated at scale $k$, versus its large-scale value at $k=0$.  The orange dashed line has
    a profile proportional to $k^4$.}}
 \label{fig:complete-profile2-intro}
\end{figure}

A short phase of  non-attractor
 evolution  during inflation is a 
  common ingredient of inflationary scenarios leading to primordial
  black hole production  \cite{Garcia-Bellido:2017mdw,Ezquiaga:2017fvi,Motohashi:2017kbs,Ballesteros:2017fsr,Hertzberg:2017dkh,Cicoli:2018asa,Ozsoy:2018flq,Mahbub:2019uhl,Ballesteros:2020qam,Liu:2020oqe,Kefala:2020xsx,Ng:2021hll,Inomata:2021uqj,Inomata:2021tpx}.
   During the non-attractor epoch, the slow-roll conditions are violated, and the would-be decaying mode influences
   the dynamics of curvature fluctuations,  enhancing by several orders
   of magnitude the amplitude of the corresponding spectrum.
      Fig \ref{fig:complete-profile2-intro} shows  a representative plot for the enhancement
     of  the spectrum of curvature
  fluctuations caused by a  non-attractor phase. We  notice interesting features that 
  are universal in  single-field models of inflation:
  \begin{itemize}
  \item As first shown in \cite{Byrnes:2018txb}, the spectrum
   towards the peak 
   grows as function of  momentum  $k$ with a power not larger~\footnote{Some exceptions to this conclusion can be made through a prolonged non-attractor era with $\eta_c = -1$ \cite{Carrilho:2019oqg} (See also \cite{Ozsoy:2019lyy}) which can result with a growth rate of $k^5 (\ln k)^2 > k^4$ or through multiple non-attractor phases that exhibit an almost instantaneous transition \cite{Tasinato:2020vdk,Davies:2021loj}.} than $k^4$.
  \item The 
  growth of the spectrum towards the peak  is preceded  by a deep dip, whose 
  properties  depend on the  total amplification of the spectrum, and on the duration   
  of the non-attractor era. 
   \end{itemize} 
  In this section, using  methods based on \cite{Tasinato:2020vdk}, 
  we   discuss   universal properties of curvature fluctuations at the dip of the spectrum, in the limit of 
  short duration of the non-attractor era.  The spectral index around the dip, as well as the squeezed limits of the bispectrum and
  trispectrum in its proximity, satisfy consistency relations that   connect  their  values  with the total  size of amplification
  of the spectrum during the non-attractor regime.     In the next section, using a gradient formalism, we prove our results more rigorously,  and extend them to include a sizeable duration of the non-slow-roll era during inflation.

\smallskip
   
   In the limit of  short duration of  non-attractor regime, the equations controlling the  curvature
  fluctuations and the corresponding spectrum at the dip position simplify considerably.  
We assume a conformally flat Friedmann-Lemaitre-Roberson-Walker (FLRW)  background metric during inflation, with scale factor $a(\tau)$
and $-\infty<\tau\,<\,0$ the conformal time. 
The system consists of three phases:
\begin{enumerate}
\item An initial prolonged phase of  quasi de Sitter expansion, for $\tau<\tau_1$ (we neglect any
slow-roll corrections in this epoch, and in epoch 3 below).
\item  A short phase 
of non-attractor expansion for $\tau_1\le\tau\le \tau_2$, whose  duration is parameterized by $\Delta \tau_A\,=\,\tau_2-\tau_1$. 
\item A final phase of inflationary  quasi de Sitter expansion,  lasting $\Delta \tau_B\,=\,-\tau_2$ for $\tau_2\le\tau\le0$. We  assume
$\Delta \tau_A \ll \Delta \tau_B$, and this relation quantifies the limit of short-duration of the non-attractor process. 
\end{enumerate}
The succession of these  three phases    requires   appropriate matchings  at times
$\tau_{1,2}$: this fact has phenomenological implications that we will  discuss in due time.  

\smallskip

The quadratic action for a Fourier mode of comoving curvature perturbation ${\cal R}_k$ is given by
\be \label{secorac1}
S_k^{(2)}\,=\,\frac12 \int \d \tau
 \,z^2(\tau) \left[  {\cal R}_k'^2(\tau)-k^2 {\cal R}_k^2(\tau)\right],
\ee
with $z(\tau)\,=\,a \dot{\phi}/H$ a function of time only,  dubbed {\it pump field}. During the quasi-de Sitter
 phases of inflationary evolution, $\tau<\tau_1$ and $\tau>\tau_2$, the pump field is proportional to the
 scale factor $a(\tau)$, with (nearly) constant coefficient depending on the inflationary model. During the phase of non-attractor, instead, the time profile of the pump field  changes considerably, potentially inducing amplifications of the spectrum   of curvature fluctuations. 
 
 We define
\be
\beta_h\,\equiv\,\frac{\Delta \tau_A}{\Delta \tau_B} \left[ 2+\left( \frac{\d\,\ln z^2(\tau)}{ \d \,\ln \tau} \right)_{\tau=\tau_1}\right],
\ee
 as a (possibly) large parameter associated with the variations  in the pump field 
 during the non-slow-roll phase $\tau_1\le\tau\le\tau_2$. In the limit of pure de Sitter expansion in the intervals  $\tau<\tau_1$ and $\tau>\tau_2$, and considering small values for the ratio $\Delta \tau_A/\Delta \tau_B$, the work 
  \cite{Tasinato:2020vdk} determined  the following analytical expression for the mode function ${\cal R}_k(\tau)$ during
  the final phase of evolution $\tau>\tau_2$:

 \be
 \label{finRk1}
{\cal R}_k(\tau)\,=\,\frac{{\cal R}_0}{\sqrt{2 \,k^3}}\left[ {\cal C}_1(k)\, e^{- i k \tau} \left(1+ i k \tau\right)
+{\cal C}_2(k)\, e^{i k \tau} \left(1- i k \tau\right) \right]\,,\hskip1cm \tau\ge \tau_2\,.
\ee 

The previous expression contains 
an overall constant normalization   ${\cal R}_0$ common to all modes $k$, whose value depends
on the inflationary model and is not relevant for our arguments. The scale-dependent functions 
${\cal C}_{1,2}(k)$ read as
\bea
\label{solC1}
{\cal C}_1(k)&=&1+\frac{\beta_h}{8 k^2  \Delta \tau_A\,\Delta \tau_B}
\left( 1-e^{2 i k \Delta \tau_A }-2
i k \Delta \tau_A \left(1+ 2 i k  \Delta \tau_B\right)
\right),
\\
\label{solC2}
{\cal C}_2(k)&=&-
\frac{\beta_h\,e^{-2 i k  \tau_2}}{8 k^2    \Delta \tau_A\,\Delta \tau_B}
\left( 1-e^{2 i k \Delta \tau_A }\left(1- 2 i k  \Delta \tau_B\right)-2
i k \left( \Delta \tau_A+  \Delta \tau_B  \right)
\right).
\eea
When $\beta_h=0$, the  configuration  reduces to the usual solution for fluctuations during a
 de Sitter phase of expansion.
  Notice that the solution \eqref{finRk1}, \eqref{solC1}, \eqref{solC2}  satisfies the 
 Bunch-Davies
condition at early times. We refer the reader to \cite{Tasinato:2020vdk} for a complete discussion regarding the formula \eqref{finRk1}.
 
\smallskip

Equation \eqref{finRk1} allows us to compute the spectrum of curvature fluctuations at the end of inflation $\tau=0$, which corresponds to the end of the second slow-roll epoch following the intermediate non-attractor phase. In our context, we are interested to study the enhancement of the curvature spectrum
from large $k\to 0$ towards small scales. Denoting with a prime $\langle \dots \rangle'$  the $n$-point
correlators without the corresponding momentum-conserving $\delta$-functions, we consider the ratio
\be
\Pi(k)\,\equiv\,\frac{\langle {\cal R}_k^2(\tau=0) \rangle'}{\langle {\cal R}_{k=0}^2(\tau=0) \rangle'}
\ee 
of correlation functions evaluated at the end of inflation. The resulting function $\Pi(k)$ quantifies
the amplification of the spectrum from large to small scales, and  depends
on the  scale $k$ of horizon exit for  modes produced at early times during inflation.  In our case it reads:
\be
\Pi(k)\,=\,|{\cal C}_1(k)+{\cal C}_2(k)|^2.
\ee
The maximal value of $\Pi(k)$ at small scales, which we define as $\Pi_{T}$ can be  
analytically determined from the expressions \eqref{solC1}, \eqref{solC2} as a simple function of $\beta_h$:
\be\label{pitoth}
\Pi_{T}\,\equiv\,
\Pi(k\to\infty)\,=\,\left(1+\frac{\beta_h}{2} \right)^2.
\ee
In the limit of large $\beta_h$, the value of $\Pi_{T}$ can be  large, 
 enhancing the spectrum to the values needed   for producing PBHs.  For example, a $\beta_h$
 of orders of  a few thousands can produce an enhancement $\Pi_{T}\,\simeq\,10^7$, the typical enhancement
 of the curvature spectrum
 required for PBH formation mechanisms. 

\medskip

\begin{figure}[h!]
\centering
 \includegraphics[width = 0.495 \textwidth]{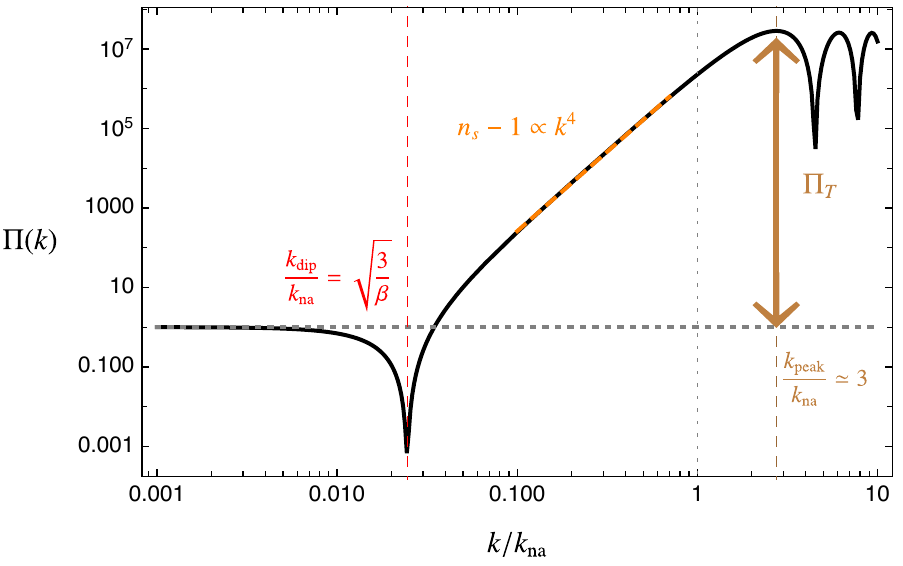}
  \includegraphics[width = 0.495 \textwidth]{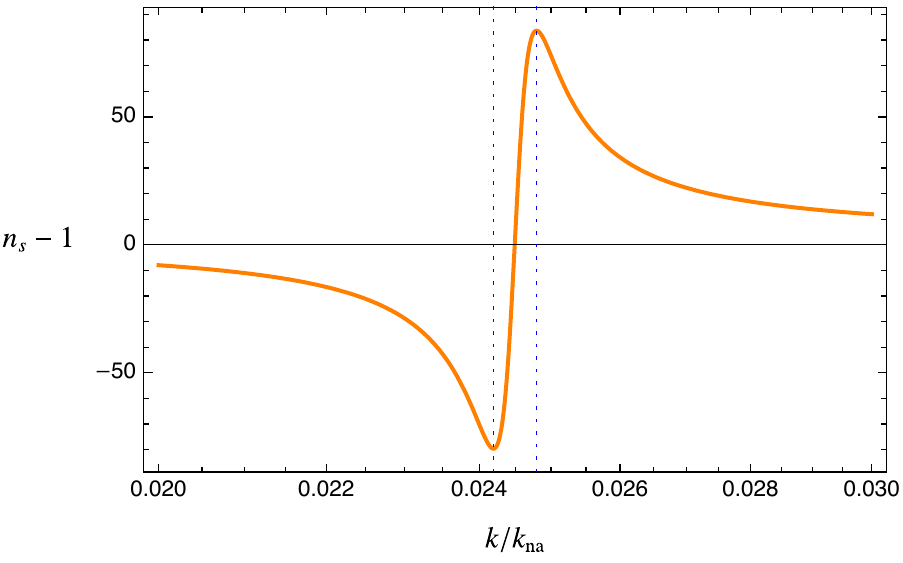}
 \caption{\it{ \small Left: Scale dependence of the Power spectrum and its important global features. Right: Scale dependence of the spectral index $n_s - 1$ around the dip feature. In the right panel, the location of the minimum/maximum of the spectral index $n_s -1$ (see \eg eq. \eqref{nsminmax}) are indicated with blue dashed lines. In both plots we use  $\beta_h=5 \times 10^3$.}}
 \label{fig:plot1}
\end{figure}

\noindent{\bf The position of the dip of the  power spectrum.}
 In the present instance,  we are  interested in using the previous formulas for examining in details the properties
of the dip feature
we notice in Fig \ref{fig:complete-profile2-intro},
which  corresponds to  a  universal property of single-field models with a short   non-attractor phase.
 In particular we are interested to its position, and the scale-dependence of the spectrum in its
proximity.  We take the simplifying limit of very small duration of non-attractor phase, $\Delta \tau_A/\Delta \tau_B\,\to\,0$.  We  then identify the momentum scale $k_{\rm na}$: 
\be 
k_{\rm na}\,=\,1/\Delta \tau_B\,=\,-1/\tau_2\ee
 with
the  scale at which modes start leaving the horizon during the non-attractor era (while for larger scales
$k\,\le\,k_{\rm na}$ modes leave the horizon during the initial de Sitter phase). We 
introduce the dimensionless quantity
\be
\kappa={k}/{k_{\rm na}}.
\ee
to describe the quantity $\Pi(\kappa)$ that characterize the scale dependence of the power spectrum as
\be
\label{pieps0}
\Pi(\kappa)\,=\,1-2 \,\beta_h \,\kappa\,j_1(\kappa)\,\cos{\kappa}+\beta_h^2 \,\kappa^2\,j^2_1(\kappa),
\ee
with $j_{1}(\kappa)$  the spherical Bessel function:
\be
j_1(\kappa)\,=\,\frac{1}{\kappa} \left( \frac{\sin \kappa}{\kappa}-\cos \kappa\right).
\ee
Notice that the spectrum $\Pi(\kappa)$ in eq \eqref{pieps0}  depends
on a single extra parameter $\beta_h$, which is the only free parameter available describing the impact
of the very short phase of non-attractor dynamics. As shown in Figure \ref{fig:plot1}, \eqref{pieps0} describes very well the global features we mentioned above including the intermediate dip feature and the expected $k^4$ growth that follows the dip towards small scales. 

\smallskip

In the limit of large values of $\beta_h$ (and $\kappa \leq 1$), it is straightforward to compute the position of the dip, using $1/\beta_h$
as expansion parameter. For this purpose, we introduce a convenient variable $x$ through the rescaling
\be\label{xvskappa}
\kappa \equiv \sqrt{\frac{3}{\beta_h}} x,
\ee
to express $\Pi(\kappa)$ of eq \eqref{pieps0} as (see \eg eq. \eqref{piapp}),
\be\label{spih}
\Pi (x)\,=\,\left(1-x^2 \right)^2+\frac{3 x^4\left( 6-x^2\right)}{5 \beta_h}+{\cal O}\left(\frac{1}{\beta_h^2}\right).
\ee
To leading order in the large $\beta_h$ limit, $x=1$ therefore locates the position of the universal dip feature \cite{Tasinato:2020vdk}. In terms of the original variable $\kappa = k/k_{\rm na}$, the location and depth of the dip feature are then given by
\beq\label{dippos1}
\fr{k_{\rm dip}}{k_{\rm na}}=\sqrt{\frac{3}{\beta_h}}\,=\,\sqrt{\frac32}\,{\Pi_T^{-1/4}}\quad \longrightarrow \quad \Pi\left(\fr{k_{\rm dip}}{k_{\rm na}}\right)= \frac{3}{2}\,{\Pi_T^{-1/2}},
\eeq
where we used \eqref{pitoth} to relate the position of the dip and its depth to the total enhancement in the spectrum. 
These simple interesting relations connect the position of the dip with an inverse power of the total enhancement $\Pi_T$ of the spectrum. In the limit of large  $\Pi_T$, the position of the dip in momentum space is parametrically well  smaller than the scale at which modes start to leave the horizon during the non-attractor era $k_{\rm na}$, \ie $\kappa_{\rm dip}\,=\,k_{\rm dip}/k_{\rm na}\,\ll\,1$. \emph{The position of dip scale in momentum space therefore corresponds to modes that leave the horizon during the initial attractor slow-roll regime.}
For example, for a typical PBH forming scenario with $\Pi_T\sim 10^7$,
we find $\kappa_{\rm dip}\ \,\sim\,10^{-2}$.  

We will reconsider and prove relations similar to \eqref{dippos1} with more rigorous methods in section \ref{S2} to further relate these features to the additional properties of the system such as the duration of the non-attractor era and the value of the slow-roll parameter during this epoch. 

\medskip

\noindent
{\bf The spectral index.} 
We continue our  discussion by studying the behaviour of the spectral tilt
 \be\label{nsdef}
n_s-1\,=\, \frac{\d \ln \Pi(\kappa)}{\d \ln \kappa},
\ee
around the position of the dip. Using the simplified profile of eq \eqref{spih} and \eqref{xvskappa} in \eqref{nsdef}, we 
present the behaviour of the spectral index around the dip feature in the right panel of Fig \ref{fig:plot1}. The plot shows
that the spectral index can acquire values of order ${\cal O}(10)$ around the dip. We can utilize the same formulas to analytically determine the maximum and the minimum acquired by $n_s -1$, focusing on the zeros of the running defined as
\be\label{as}
\alpha_s\,\equiv\,\frac{\d n_s}{\d \ln k}.
\ee
With this aim, we notice from the right panel of Figure \ref{fig:plot1} that extremal points of the spectral tilt are located in the close vicinity of the dip feature. We therefore we make the convenient substitution in terms of an expansion in the small quantity $1/\beta_h$,
\be\label{kappaex}
\kappa\,=\,\sqrt{\frac{3}{\beta_h}} \left(1+\frac{x_1}{\sqrt{\beta_h}} +\frac{x_2}{{\beta_h}}\right),
\ee 
for two yet to be determined quantities $x_{1,2}$. Using \eqref{spih} and \eqref{xvskappa}, we find that $\alpha_s$ \eqref{as} vanishes with the following choices for the parameters $x_{1,2}$:
\bea
x_1=\pm\frac{\sqrt{3}}{2},
\hskip0.5cm \hskip0.5cm
x_2\,=\,-\frac{9}{40}.
\eea
Then the position of the extrema of the spectral tilt are located in
\beq\label{nsminmax}
\fr{k_{\rm min/max}}{k_{\rm na}}=
\sqrt{\frac{3}{\beta_h}} \left(1\mp \frac{\sqrt{3}}{2\sqrt{\beta_h}} -\frac{9}{40\beta_h}
\right),
\eeq
with corresponding value of the minimum and maximum value of the $n_s -1$:
\bea
(n_{s} -1)_{\rm min/max}=2\mp \frac{2 \,\sqrt{\beta_h}}{\sqrt 3}
\,
=\,2\mp\sqrt{\frac83}\,\,{\Pi_T^{1/4}},
\eea
in good agreement with the profile shown in the right panel of Figure \ref{fig:plot1}. These formulas provide  new consistency relations
for the extremal values of the spectral index around the dip in single-field inflationary models
which include a short non-slow-roll phase. Interestingly, these relations all involve combinations
of the quantity $\Pi_T^{1/4}$, which controls the position of the dip and the properties of the slope of 
the spectrum in its proximity. 

\subsection*{Implications for non-Gaussianity}

\begin{figure}[h!]
\centering
 \includegraphics[width = 0.46 \textwidth]{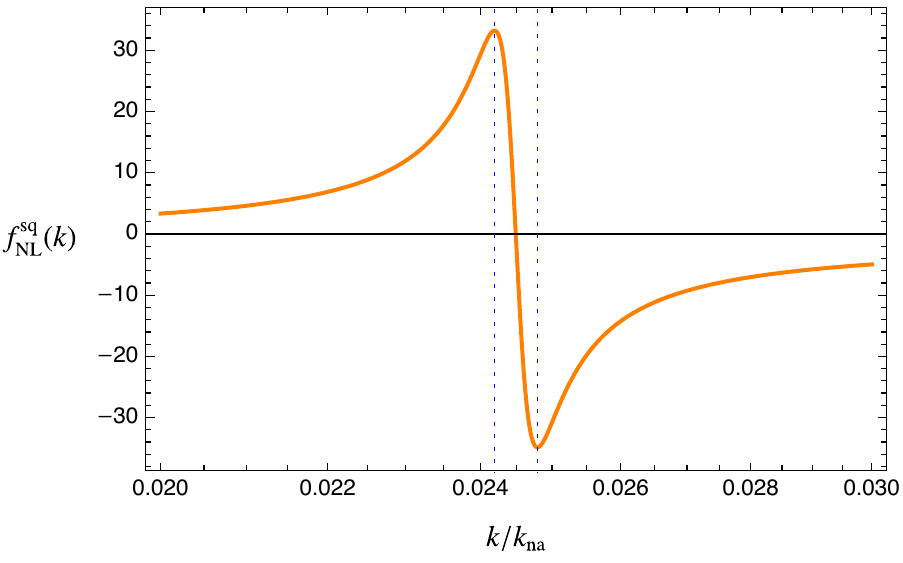}
  \includegraphics[width = 0.47 \textwidth]{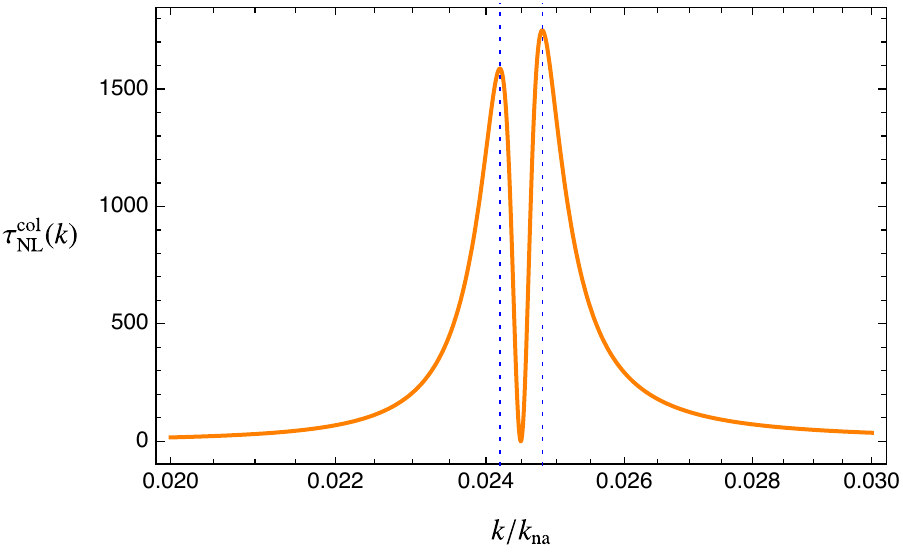}
 \caption{\it Scale dependence of $f_{\rm NL}^{\rm sqz}$ (Left) and $\tau_{\rm NL}^{\rm coll}$ (Right) around the dip scale $k_{\rm dip}$ assuming the consistency relations \eqref{cc} in the squeezed and collapsed limits respectively. In both plots we take $\beta_h=5 \times 10^3$ corresponding to a total enhancement of $\Pi_{T} \simeq 10^7$ as shown in Figure \ref{fig:plot1}.
 }
 \label{fig:plot3}
\end{figure}

We find that the location of the dip feature occurs at momentum scales much smaller than the
scale at which inflationary modes leave the horizon during the non-slow-roll phase. In this regime,
since modes that leave the horizon around the dip scale are associated with the initial single-field slow-roll epoch, 
we can reasonably expect  the validity of  Maldacena consistency relations for  squeezed and collapsed limits 
of $n$-point correlation functions.
Namely, 
defining $P_1\equiv  \langle {\cal R}_{k_1} {\cal R}_{k_1}  \rangle'$, $k_{12}\,=\,|\vec k_1+\vec k_2|$, and 
\bea
f^{\rm sq}_{\rm NL}&=&\frac{5}{12}\,\lim_{k_1\to 0}\,\frac{\langle {\cal R}_{k_1} {\cal R}_{k_2} {\cal R}_{k_3}  \rangle'}{P_1\,P_2}, \hskip0.5cm \hskip0.5cm
\tau_{\rm NL}^{\rm col}\,=\,\frac14\,\lim_{k_{12}\to 0}\frac{\langle {\cal R}_{k_1} {\cal R}_{k_2} {\cal R}_{k_3} {\cal R}_{k_4} \rangle'}{P_1 \,P_3\,P_{12}},
\eea
we expect the following relations
to hold \cite{Maldacena:2002vr,Creminelli:2004yq,Byrnes:2006vq,Suyama:2007bg,Smith:2011if,Assassi:2012zq},
\bea\label{cc}
f_{\rm NL}^{\rm sq}(k)=
\frac{5}{12}\left(1-n_s(k) \right),\quad\quad
\tau_{\rm NL}^{\rm col}(k)=\left( \frac65 \,f_{\rm NL}^{\rm sqz}(k) \right)^2,
\eea
respectively for the squeezed bispectrum and collapsed bispectrum. We represent the scale dependence of these non-linearity parameters in Figure \ref{fig:plot3}. 
Following our discussions above, we find that for large values of the enhancement factor $\Pi_T$ the maximal/minimal value for $f_{\rm NL}^{\rm sq}$ and the extremal values of $\tau_{\rm NL}^{\rm col}$ around 
the dip are given by
\be \label{fnlMAXMIN}
\left(f_{\rm NL}^{\rm sq} \right)_{\rm max/min}\,=\,
\frac{5}{12} \left( \pm \sqrt{\frac83}\,\,{\Pi_T^{1/4}}-1\right),\quad\quad \left(\tau_{\rm NL}^{\rm col} \right)_{\rm ext}\,=\,\frac{3}{5} \, \left( \pm \sqrt{\frac83}\,\,{\Pi_T^{1/4}}-1\right)^2.
\ee
For a total enhancement of $\Pi_T\sim 10^7$, these results yield to $|(f_{\rm NL}^{\rm sq})_{\rm max}|\sim 38$ and $(\tau_{\rm NL}^{\rm col})_{\rm ext}\sim2 \times 10^3$.

We find it interesting that the maximal values of non-Gaussianity parameters are controlled  by the total enhancement of the curvature spectrum, $\Pi_{T}$, the only free parameter that enters in our discussion. As we will learn in the next section, more refined proofs of the new consistency relations lead to the same order of magnitude for the amplitudes of $f_{\rm NL}$, $\tau_{\rm NL}$, up to order-one corrections related to the details of the non-attractor phase. 
Moreover,  we will confirm that  the quantities  $f_{\rm NL}$ and $\tau_{\rm NL}$ are characterized by the specific scale dependence shown in Figure \ref{fig:plot3}, whose features are also  controlled by $\Pi_{T}$.

Before discussing  a more rigorous derivation of these results, we comment on  some of their physical implications,
once we assume that $\Pi_T\,\sim\,   {\cal O}( 10^7)$ as usually required by PBH formation mechanisms:
\begin{itemize}
\item The position of the dip roughly satisfies $k_{\rm dip} \simeq 10^{-2} k_{\rm peak}$, see the left panel of Figure \ref{fig:plot1}.  It  occurs at relatively large scales where modes still leave the horizon during the initial attractor phase. 
\item The consistency relations \eqref{cc} and \eqref{fnlMAXMIN} dictate that in any single-field model of inflation with a non-attractor phase 
we should  expect amplitudes of order $f_{\rm NL}^{\rm sq}\sim {\cal O}(10)$ and $\tau_{\rm NL}^{\rm col}\sim {\cal O}(  10^3)$ for the non-linearity parameters around the dip. 
\item Besides their large magnitude, the parameters $f_{\rm NL}^{\rm sq}$ and $\tau_{\rm NL}^{\rm col}$ acquire a pronounced, specific
scale dependence around the dip feature -- see Figure \ref{fig:plot3}. As we will show  in section \ref{PBSandTS}, the amplitude and the characteristic scale dependence of these parameters play an important role for studying observable quantities --such as the statistics of CMB $\mu$-distortions-- at scales much larger than the peak scale in the power spectrum. This is a distinctive  feature of these scenarios producing PBH, that may allow us to differentiate non-attractor single-field inflation from other models that exhibit large non-Gaussianity at $\mu$-distortion scales.
\item Since the scenarios we consider undergo  phases of non-attractor evolution that need dedicated matching
conditions to connect to distinct  slow-roll phases,  the squeezed limits
of non-Gaussian parameters correspond to {\it physical} quantities, and can not be removed
by  coordinate transformations as in \cite{Pajer:2013ana}. 
Non-Gaussianity in explicit models undergoing non-attractor evolution are studied for example in \cite{Namjoo:2012aa,Martin:2012pe,Chen:2013aj,Chen:2013eea,Finelli:2017fml,Cai:2018dkf} and 
 the subtlety with matching conditions clearly discussed in the recent work
 \cite{Suyama:2021adn}. 
\end{itemize}

\section{The gradient expansion approach to consistency relations}\label{S2}

The aim of this technical section is to use the {\it gradient expansion formalism}\footnote{For a detailed account on the spectral profile of the scalar power spectrum that leads to PBH formation and its applications in the context of induced gravitational waves, we refer the reader to \cite{Ozsoy:2019lyy} where further developments of the gradient expansion formalism is discussed.} -- first introduced in \cite{Leach:2001zf} -- for reproducing and place in firm footing  the heuristic perspective to consistency relations we discussed in the previous section.  

In the gradient expansion  framework, iterative solutions of the comoving curvature perturbation are  generated at any desired order in a small-$k$ expansion, in terms of  analytic functions controlled by  the background evolution. For example, the amplitude of curvature fluctuations at a fiducial  late time $\tau = \tau_*$ (\eg at the reheating surface)  is mapped  to its value at an initial time around horizon exit $\tau = \tau_k$ in terms of a complex-valued $k$-dependent coefficient:
 \beq\label{irf}
\mathcal{R}_k (\tau_*) = \alpha_k \mathcal{R}_k (\tau_k) =  \left( \alpha^R_k + i \alpha^I_{k}\right) \mathcal{R}_k (\tau_k)\,,
\eeq
where in the last equality we split the enhancement factor into its real and imaginary parts: 
\be \label{splitalfa}
\alpha_k \equiv \alpha^R_k + i \alpha^I_k\,.
\ee
Once expanded up to second, $k^2$-order in the  gradient expansion, the real and imaginary parts of this  coefficient are  given by
\begin{align}\label{ar}
\alpha_k^R &= 1 + D(\tau_k)\, v_\mathcal{R}^{R} - F(\tau_k)\, k^2,\\
\alpha_k^I &= D(\tau_k)\, v_{\mathcal{R}}^I \label{ai}\,.
\end{align}
The quantity $v_\mathcal{R}^{R}$ and $v_\mathcal{I}^{R}$ denotes the real and imaginary part of the $k$-dependent fractional velocity of the curvature perturbation evaluated at $\tau = \tau_k$ defined as
\beq\label{fr}
v_\mathcal{R} (\tau_k) = \fr{\mathcal{R}_k'}{3\mathcal{H}_k\mathcal{R}_k} \bigg|_{\tau=\tau_k} \,.
\eeq 
The full $k$ dependence of the expressions \eqref{ar} and \eqref{ai} on super-horizon scales is encoded in
the quantities $v^R_\mathcal{R}$,  $v^I_\mathcal{R}$ (see Appendix \ref{AppA} for details) and in the functions $D(\tau_k),F(\tau_k)$, given by the following nested integrals of the pump field appearing in eq \eqref{secorac1} (see \cite{Leach:2001zf,Ozsoy:2019lyy} for further details):
\begin{align}
\label{Dint} D(\tau)&=3 {\cal H}_k\,\int_\tau^{\tau_*}\,\d \tau'\,\frac{z^2 (\tau_k) }{z^2 (\tau') }\, ,\\ 
\label{Fint}  F(\tau)&=\int_\tau^{\tau_*}\,\frac{d \tau'}{z^2(\tau')}
\,\int^{\tau'}_{\tau_k}\,\d \tau'' z^2(\tau'') \,. 
\end{align}
Expressions \eqref{Dint} and \eqref{Fint} indicate that if the pump field increases with time -- as in standard slow-roll inflation, where $z \propto a(\tau)$ --  the functions $D$, $F$ rapidly decrease to zero after horizon crossing (\ie $\alpha_k \to 1$). In this case, the curvature perturbation in \eqref{irf} settles to a constant shortly after horizon exit ($\mathcal{R}_k (\tau_*) \simeq \mathcal{R}_k (\tau_k)$). On the contrary, in inflationary models containing phases of non-attractor evolution, $z(\tau)$ transiently decreases  and the  functions $D$, $F$ can grow, amplifying  the spectrum of curvature perturbation (\ie $|\alpha_k| \gg 1$ in eq \eqref{irf}) at super-horizon scales:
  see Appendix \ref{AppB}.
 
 \medskip
 
\noindent{\bf The curvature perturbation power spectrum.} We define the late-time power spectrum 
evaluated at $\tau=\tau_*$ as 
\beq\label{psdef}
\langle\mathcal{R}_k(\tau_*)\mathcal{R}_{k'}(\tau_*)\rangle = (2\pi)^3\, P_{\mathcal{R}}(\tau_*,k)\,\delta\left(\vec{k} +\vec{k}'\right) .
\eeq
Using eq. \eqref{irf}, we can then relate the power spectrum at late times to the power spectrum evaluated at  horizon crossing via 
\beq\label{dfps}
P_{\mathcal{R}}(\tau_*,k)\equiv \frac{2\pi^2}{k^{3}} \mathcal{P}_{\mathcal{R}}(\tau_*,k) =  \frac{2\pi^2}{k^{3}} \left[ \,|\alpha_k|^2 \, \mathcal{P}_\mathcal{R}(\tau_k)\,\right] \equiv |\alpha_k|^2 P_\mathcal{R}(\tau_k),
\eeq
where $\mathcal{P}_{\mathcal{R}}(\tau_k) \equiv k^3 P_{\mathcal{R}}(\tau_k,k)/2\pi^2$, and $|\alpha_k|^2 = (\alpha_k^{R})^2 + (\alpha_k^{I})^2$. 
 
 \medskip
 
\noindent
{\bf A non-linear expression for $\mathcal{R}$.}
We assume that $\mathcal{R}_k(\tau_k)$ is a Gaussian random variable: nevertheless, the superhorizon evolution typically introduces
non-linearities. In fact, we can go beyond the linear theory used for  eq. \eqref{irf} to compute 
the bispectrum of the late time curvature perturbation $\mathcal{R}_k(\tau_*)$. For the purpose of deriving an analytic expression for the bispectrum and to study the corresponding consistency relations, we adopt the following non-linear expression for $\mathcal{R}_k(\tau_*)$, first derived in \cite{Takamizu:2010xy,Takamizu:2013wja} 
\begin{align}\label{RNL}
\mathcal{R}_k(\tau_*)&=\alpha_k \mathcal{R}_{k}(\tau_{k})+\fr{F(\tau_k)}{2}\left\{\int \frac{\d^{3} k^{\prime}}{(2 \pi)^{3}}\left[4 k^{\prime 2}-\vec{k}^{\prime}. (\vec{k}-\vec{k}^{\prime})\right] \mathcal{R}_{k'}(\tau_{k'}) \mathcal{R}_{|\vec{k}-\vec{k}'|}(\tau_{|\vec{k}-\vec{k}'|})\right\}\,.
\end{align}
where the last term represents the non-linear contribution parametrized by the convolution of the Gaussian variable $\mathcal{R}_k(\tau_k)$.
\medskip

\noindent{\bf Bispectrum.} We define the late-time bispectrum of curvature perturbations as 
\beq\label{BSdef}
\left\langle\mathcal{R}_{{k}_{1}}(\tau_*) \mathcal{R}_{{k}_{2}}(\tau_*) \mathcal{R}_{{k}_{3}}(\tau_*) \right\rangle=(2 \pi)^{3} B_{\mathcal{R}}\left({k}_{1}, {k}_{2}, {k}_{3}\right) \delta\left(\vec{k}_{1}+\vec{k}_{2}+\vec{k}_{3}\right),
\eeq
and using \eqref{RNL}, at leading order in the non-linear term, the bispectrum is given by \cite{Takamizu:2010xy,Takamizu:2013wja}
\beq\label{3PT}
B_{\mathcal{R}}\left({k}_{1}, {k}_{2}, {k}_{3}\right)=\frac{(2 \pi^{2})^2}{2(k_{1} k_{2} k_{3})^{3}}\left[{\rm Re}[\alpha^{*}_{k_{1}} \alpha_{k_{2}}] F(\tau_{k_{3}})\left\{5\left(k_{1}^{2}+k_{2}^{2}\right)-k_{3}^{2}\right\} k_{3}^{3}\,\mathcal{P}_{\mathcal{R}}(\tau_{k_{1}}) \mathcal{P}_{\mathcal{R}}(\tau_{k_{2}})+\text { perms }\right],
\eeq
where the permutations are cyclic among the three external momenta $\{\vec{k}_1,\vec{k}_2,\vec{k}_3\}$. Correspondingly, we define the scale-dependent non-linearity parameter $f_{\rm NL}$ as
\beq\label{deffnl}
 f_{\rm NL}\left(k_{1}, k_{2}, k_{3}\right) =  \frac{5}{6} \fr{B_{\mathcal{R}}\left(k_{1}, k_{2}, k_{3}\right)}{\left[P_{\mathcal{R}}\left(\tau_*,k_1\right) P_{\mathcal{R}}\left(\tau_*,k_2 \right)+\text { perms }\right]}\,\,,
\eeq
Using \eqref{dfps} and \eqref{3PT} in \eqref{deffnl}, we then obtain the scale-dependent  $f_{\rm NL}$ as
\beq\label{fnlf}
 f_{\rm NL}\left(k_{1}, k_{2}, k_{3}\right) = \fr{5}{12}\fr{\left({\rm Re}[\alpha^{*}_{k_{1}} \alpha_{k_{2}}] F(\tau_{k_{3}})\left\{5\left(k_{1}^{2}+k_{2}^{2}\right)-k_{3}^{2}\right\} k_{3}^{3}\,\mathcal{P}_{\mathcal{R}}(\tau_{k_{1}}) \mathcal{P}_{\mathcal{R}}(\tau_{k_{2}})+\text { perms }\right)}{\left[\left|\alpha_{k_1}\alpha_{k_2}\right|^2 \mathcal{P}_{\mathcal{R}}\left(\tau_{k_1}\right) \mathcal{P}_{\mathcal{R}}\left(\tau_{k_2} \right)k_3^3+\text { perms }\right]}.
\eeq

\medskip

\noindent{\bf Trispectrum.} We define the connected part of the late time trispectrum of curvature perturbation as 
\beq\label{TSdef}
\left\langle\mathcal{R}_{{k}_{1}} \mathcal{R}_{{k}_{2}} \mathcal{R}_{{k}_{3}}  \mathcal{R}_{{k}_{4}} \right\rangle=\langle \mathcal{R}^{\rm G}_{{k}_{1}}(\tau_*) \mathcal{R}^{\rm G}_{{k}_{2}}(\tau_*) \mathcal{R}^{\rm G}_{{k}_{3}}(\tau_*) \mathcal{R}^{\rm G}_{{k}_{4}}(\tau_*)\rangle+ (2 \pi)^{3}\, \delta\left(\vec{K}_{\rm tot}\right) T_{\mathcal{R}}\left({k}_{1}, {k}_{2}, {k}_{3}, {k}_{4}\right) , 
\eeq
where the first term is the Gaussian part of the trispectrum (see appendix \ref{AppD}) and $\vec{K}_{\rm tot} \equiv \sum_{i=1}^{4}\vec{k}_{i}$. 

Utilizing again the non-linear expression \eqref{RNL},  
the leading order scale-dependent trispectrum is found to be 
\begin{align}\label{4PT}
\nn T_{\mathcal{R}}\left({k}_{1}, {k}_{2}, {k}_{3}, k_{4}\right) &= \fr{{\rm Re}[\alpha^{*}_{k_{1}} \alpha_{k_{2}}] F(\tau_{k_{3}})F(\tau_{k_{4}})}{4}\left\{4\left(k_{1}^{2} + k^2_{13}\right)+2 \vec{k}_{1}.\vec{k}_{13}\right\} \left\{4\left(k_{2}^{2} + k^2_{13}\right)-2 \vec{k}_{2}.\vec{k}_{13}\right\} \\
&\quad\quad\quad\quad\quad \quad\quad\quad\quad\quad\quad\quad\quad\quad\,\,\times {P}_{\mathcal{R}}(\tau_{k_{1}}) {P}_{\mathcal{R}}(\tau_{k_{2}})  {P}_{\mathcal{R}}(\tau_{k_{13}})+\text {11 perms},
\end{align}
where we introduce $\vec{k}_{ij} = \vec{k}_i +\vec{k}_j$, $k_{ij} = |\vec{k}_i +\vec{k}_j|$ for $i \neq j$ and perms denote distinct permutations among the external momenta $\{\vec{k}_1,\vec{k}_2,\vec{k}_3,\vec{k}_4\}$. We define the corresponding scale dependent non-linearity parameter $\tau_{\rm NL}$ as 
\beq\label{taunl}
\tau_{\rm NL}(k_1,k_2,k_3,k_4) = \fr{T_{\mathcal{R}}\left({k}_{1}, {k}_{2}, {k}_{3}, {k}_{4}\right)}{P_{\cal R}(\tau_*,k_1)P_{\cal R}(\tau_*,k_2)P_{\cal R}(\tau_*,k_{13})+ {\rm 11\, perms}}.
\eeq

Eqs \eqref{fnlf}, \eqref{4PT} and \eqref{taunl} suggest that the size and the scale-dependence of the $f_{\rm NL}$ and $\tau_{\rm NL}$ parameters depend on the function $ F(\tau_{k})$ and the quantity $\alpha_k$, whose behavior depends on the pump field profile.  Given these preliminary results, we now study the scale dependence of the power spectrum and bispectrum on a representative set-up capable of producing a large PBH population during inflation, to further support the findings of section \ref{sec_heur1}.

\subsection{Spectral profile of the scalar power spectrum and its properties}\label{SPS}

To study the spectral shape and enhancement in the power spectrum, we consider a typical two phase inflationary scenario that instantly connects an initial slow-roll era with $\eta_{\rm sr} =0$, to a slow-roll violating, non-attractor phase with $\etc \leq -6$,  within the time range $\tau_0 \leq \tau \leq\tau_f$. Here  $\eta$ denotes the second slow-roll parameter, $H\,\eta\,\equiv\,d \ln \epsilon/d t$. The first  slow-roll parameter is given by $H\,\epsilon\,=\,-d \ln H/d t$. The pump field $z
(\tau)$ appearing in the quadratic action \eqref{secorac1} is assumed to have a profile (we take $\tau<0$):
\beq\label{zsol1}
z(\tau)=
\left\{    
 \begin{array}{rl}
&z_0\left({\tau}/{\tau_0}\right)^{-1} \hskip1.6cm \,\,{\tau}/{\tau_0}\, \geq\, 1 \,,\\
&z_0 \left({\tau}/{\tau_0}\right)^{-(\eta_{\rm c}+2)/2} \hskip0.6cm\,\, {\tau_f}/{\tau_0}\leq {\tau}/{\tau_0} \leq 1 \,,
\end{array}\right. \,
\eeq 
describing collectively the initial slow-roll and the slow-roll violating phases. Although the set-up
is qualitatively similar to the one discussed in the previous section, in this case we consider a {\it
finite duration} for the non-attractor era denoted by $\Delta N = \ln{(\tau_0/\tau_f)}$ in e-fold numbers. 
 We relate the quantity $z_0$ with a constant slow-roll parameter  $\epsilon_{\rm sr}$
via $z_0 = -a(\tau_0) \sqrt{2\epsilon_{\rm sr}}\Mp$.  For simplicity we parametrize the scale factor as
in de Sitter space: $a = -1/(H \tau)$ with a constant Hubble rate $H$ during inflation, which defines comoving horizon $\tau_0 = (a_0 H)^{-1} = \mathcal{H}_0^{-1}$ at the time of the transition to the non-attractor era.

We  proceed with determining spectral profile of the power spectrum by evaluating the dimensionless power spectrum \eqref{dfps} at the end of the non-attractor era:
\beq\label{psom}
\mathcal{P}_\mathcal{R}(\tau_f,k)  \equiv |\alpha_k|^2  \mathcal{P}_{\mathcal{R}}(\tau_k) = \left[ (\alpha_k^{R})^2 + (\alpha_k^{I})^2 \right]\mathcal{P}_{\mathcal{R}}(\tau_k)\,.
\eeq  

\begin{figure}[t!]
\begin{center}
\includegraphics[scale=0.93]{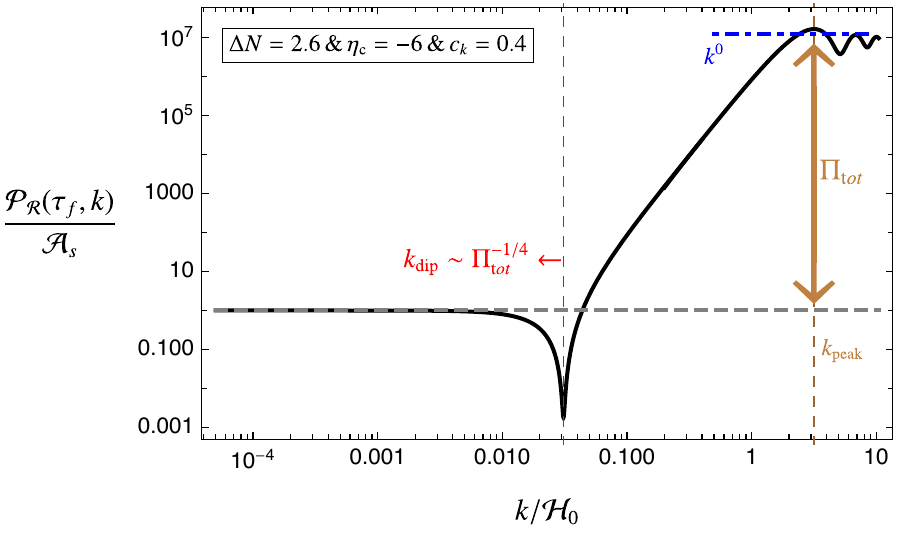}\includegraphics[scale=0.82]{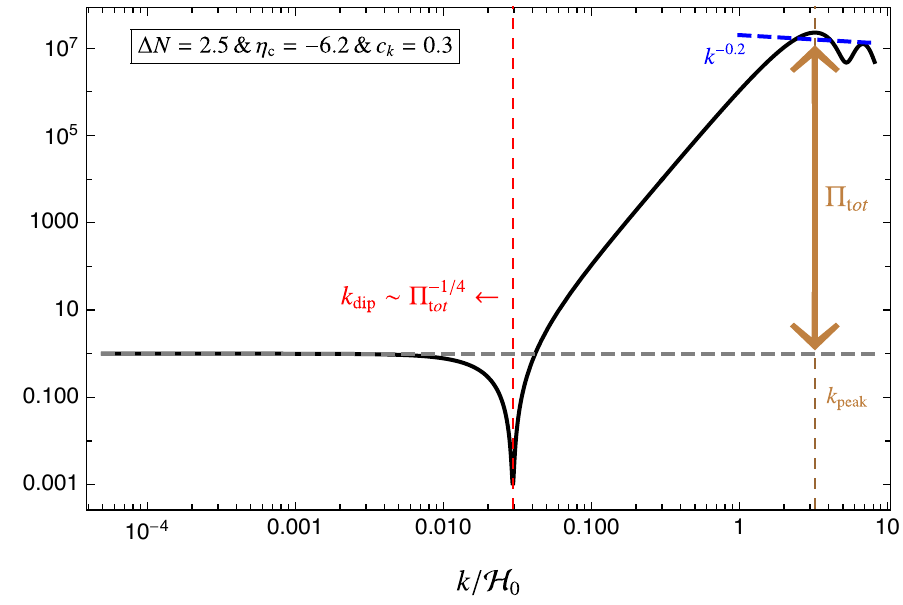}
\end{center}
\caption{\it \small Scale dependence of the power spectrum for an inflationary scenario that includes a transient ultra slow-roll or constant roll phase. The location of the dip feature (red dashed line) and its relation to the global enhancement $\Pi_{\rm tot}$ in the power spectrum is shown. The plots are obtained using the gradient expansion formalism, and normalized with respect to power spectrum at the largest scales, \ie $\mathcal{A}_s = H^2/(8\pi^2\epsilon_{\rm sr}\Mp^2) \simeq 2.1 \times 10^{-9}$.\label{fig:ps}}
\end{figure}

\noindent Using the pump field profile \eqref{zsol1}, we can derive analytic formulas for $\alpha^R_k$ \eqref{ar} and $\alpha^I_k$ \eqref{ai} (see Appendix \ref{AppA} and \ref{AppB}) to characterize the shape of the power spectrum. The resulting profile of the late time power spectrum is shown in Figure \ref{fig:ps} for representative scenarios leading to PBH formation given the pump field profile of \eqref{zsol1}. For understanding  the physical implications of our findings within the gradient expansion formalism, it is convenient to 
 introduce a fixed quantity 
  \beq \label{defOck}
  c_k\,\equiv\,-k\tau_k \, \leq 1\,,
  \eeq
which  determines the size of a mode $k$ with respect to  the horizon $(aH)^{-1}$ at time $\tau = \tau_k$, corresponding to the horizon crossing epoch. {We stress that by virtue of the relation \eqref{defOck} and the super-horizon gradient formalism, all modes we consider in Figure \ref{fig:ps} (and in general in this work) are outside the horizon at the initial time $\tau_k$}. We then distinguish modes whose momenta lie in the following ranges:

\begin{itemize}
\item[i)]  modes that become super-horizon during the initial slow-roll era, \ie modes satisfying $\tau_k/\tau_0 > 1$ or equivalently $k/\mathcal{H}_0 < c_k \leq 1,$ and
\item[ii)] modes that leave the horizon during the non-attractor $\eta_c \leq -6$ phase, $c_k < k/\mathcal{H}_0 $. 
\end{itemize}
Focusing on these regimes separately,  we discuss  below the spectral behavior of $\mathcal{P}_{\mathcal{R}}$ and its global features.

\medskip

\noindent{\bf Total enhancement in the power spectrum.}  Figure \ref{fig:ps} indicates  that independently from the choice of model parameters, the power spectrum reaches its peak at $k_{\rm peak} \simeq 3\mathcal{H}_0$ (See also \cite{Biagetti:2018pjj} for an analysis of the ultra--slow-roll case). To obtain the parametric dependence of the total enhancement on the model parameters, we  evaluate the power spectrum at this scale to determine its height with respect to the largest scales. For this purpose, we first focus on $\mathcal{P}_{\mathcal{R}}(\tau_k)$ as in \eqref{pstaukna} in the non-attractor phase ($c_k < k/\mathcal{H}_0$), which reads
\beq\label{ps2}
\fr{\mathcal{P}_{\mathcal{R}}(\tau_k)}{\mathcal{A}_s}  = \fr{c_k^{2\nu} \pi^2}{4} \left(\fr{k}{\mathcal{H}_0}\right)^{-2\nu+2}\bigg[{f_3^2+2\left(\fr{k}{\mathcal{H}_0}\right)f_3f_4+\left(\fr{k}{\mathcal{H}_0}\right)^2(f_3^2+f_4^2)}\bigg]_{\tau= \tau_k},
\eeq 
where $\nu = (3+\eta_{\rm c})/2$ and $\mathcal{A}_s = {H^2}/{8\pi^2\epsilon_{\rm sr}\Mp^2}$ is the normalization of the power spectrum at very large scales, $k \to 0$. 

In appendix \ref{AppA} we show that the functions $f_3$ and $f_4$ that appear in \eqref{ps2} are given by ($J$, $Y$ being  Bessel functions of first and second kind)
\bea
\label{f3}f_3\left(-k\tau,\fr{k}{\mathcal{H}_0},\nu\right) &=& J_{\nu-1}\left(\fr{k}{\mathcal{H}_0}\right)Y_{\nu}(-k\tau)-Y_{\nu-1}\left(\fr{k}{\mathcal{H}_0}\right)J_{\nu}(-k\tau),\\
\label{f4}f_4 \left(-k\tau,\fr{k}{\mathcal{H}_0},\nu\right) &=& J_{\nu}\left(\fr{k}{\mathcal{H}_0}\right)Y_{\nu}(-k\tau)-Y_{\nu}\left(\fr{k}{\mathcal{H}_0}\right)J_{\nu}(-k\tau).
\eea

Using   \eqref{f3} and \eqref{f4}, it is straightforward to  realise  that the expression in the square brackets of \eqref{ps2} is dominated by the last term around the peak scale. Also, around the scale of the peak, the modulus square of the enhancement factor are well approximated by the following expression 
\beq\label{as2}
|\alpha_k|^2 \simeq (\alpha_k^{R})^2 \simeq \left(\,1 + D(\tau_k) v^R_\mathcal{R}\, \right)^2 \simeq D(\tau_k)^2,
\eeq
where in the last step we take $v^R_\mathcal{R} \to 1$ which can be verified explicitly applying eq  \eqref{fvintr} at scales around $k_{\rm peak}$. Combining \eqref{as2} and \eqref{ps2} evaluated at the peak scale, the total enhancement of the power spectrum from large scales to small scales can be computed by making use of  the expression in \eqref{psom}. This gives, 
\beq\label{pitot}
\Pi_{\rm tot} \equiv \fr{\mathcal{P}_{\mathcal{R}}(\tau_f, k_{\rm peak})}{\mathcal{A}_s} \simeq C(c_k,\nu)\, e^{-2(3+\eta_{\rm c})\Delta N},
\eeq
where $C(c_k, \nu) $ is
\beq
C(c_k, \nu) = \fr{\pi^2\, 3^{2\nu+6}}{16\, \nu^2\, c_k^{2\nu}} \left[ f_3^2(c_k, 3,\nu)+f_4^2(c_k, 3, \nu)\right].
\eeq
Eq  \eqref{pitot} indicates  that the total enhancement in the scalar power spectrum is exponentially sensitive to slow-roll parameter $\eta_{\rm c}\leq -6$ and in particular to the duration $\Delta N$ of the non-attractor phase. Using this expression, one can confirm that for typical parameter choices that leads to a $\Pi_{\rm tot} \simeq10^7$ enhancement required for PBH formation, one obtains $C \simeq \mathcal{O}(1)$.

\medskip
\noindent{\bf The dip  and its properties.} For modes that exit the horizon during the initial slow-roll era, there is a pronounced dip in the power spectrum occurring far away from the peak scale, $k_{\rm dip} \ll \mathcal{H}_0 < k_{\rm peak}$. It is worth mentioning that  the dip  appears due to competing contributions in the power spectrum that are weighted by opposite signs. To determine its location, we focus our attention on the scale dependence of the power spectrum \eqref{psom} for scales satisfying $k/\mathcal{H}_0 < c_k \leq 1$. As we show in Appendix \ref{AppB1}, in this regime the enhancement factor $\alpha_k$ of eq \eqref{splitalfa} is dominated by its real part whose scale dependence can be accurately captured by 
eq. \eqref{appak}:
\beq\label{ars}
\alpha^R_k \simeq \alpha^R_{(0)} \left[ 1 - \beta \left(\fr{k}{\mathcal{H}_0}\right)^{2}\right] + \mathcal{O}\left(\fr{k^3}{\mathcal{H}^3_0}\right)
\eeq
where $\alpha^R_{(0)}$ is an order quantity and $\beta \gg 1$ is an exponentially large number parametrized in terms of the duration $\Delta N$ of the slow-roll violating phase and the value of $\eta_{\rm c}$ in this era. (See eq. \eqref{misc}.)  In Appendix \ref{AppA} -- see in particular eq \eqref{pstauk}  --  we show that the quantity $\mathcal{P}_{\mathcal{R}}(\tau_k)$ is nearly  scale-independent  
for modes that leave the horizon in the initial slow-roll era. This implies  that the scale dependence of the late time power spectrum \eqref{psom} around the dip  is completely dictated by  $\alpha^{R}_k$. Therefore the zero of \eqref{ars} provides an accurate description for the location of the dip  in the power spectrum,  which reads as
\beq\label{kdip}
\fr{k_{\rm dip}}{\mathcal{H}_0}  = \fr{1}{\sqrt{\beta}} \equiv \left[-\fr{\eta_{\rm c}\,e^{-(3+\eta_{\rm c})\Delta N}}{\alpha^R_{(0)}(c_k)\,(\eta_{\rm c} +1)(\eta_{\rm c} +3) }\right]^{-1/2}.
\eeq

 {Notice that by virtue of the above definition \eqref{kdip}, the exponentially large number $\beta$ plays the same role of the parameter $\beta_h$ introduced in the heuristic approach we discussed in the previous section. A clear advantage of the gradient approach is the fact that it makes apparent why such a large number appears in PBH forming inflationary scenarios by relating $\beta$ to the duration $\Delta N$ and $\eta_{\rm c}$ of the slow-roll violating phase as $\beta \propto e^{-(3+\eta_{\rm c})\Delta N} \gg 1$.}

The presence of such a pronounced dip in the spectrum is {a universal feature, being} virtually present in all single field models based on non-attractor evolution that are aiming to generate a sizeable peak   in the power spectrum 
 for producing PBH, say of order  $\Pi_{\rm tot} \simeq 10^{7}$ (see e.g. \cite{Garcia-Bellido:2017mdw,Ezquiaga:2017fvi,Ballesteros:2017fsr,Hertzberg:2017dkh,Cicoli:2018asa,Ozsoy:2018flq,Mahbub:2019uhl,Ballesteros:2020qam,Liu:2020oqe,Kefala:2020xsx}).  Using \eqref{kdip} we can universally relate the location of the dip feature to the total enhancement in the power spectrum in \eqref{pitot}, as first found in \cite{Tasinato:2020vdk}, which reads as
 \beq\label{kdipvspitot}
 \fr{k_{\rm dip}}{\mathcal{H}_0} \simeq \left[-\fr{\sqrt{C(c_k,\nu)}\,\alpha^R_{(0)}(c_k)\,(\eta_{\rm c} +1)(\eta_{\rm c} +3)\,}{ \eta_{\rm c}}\right]^{1/2} \Pi_{\rm tot}^{-1/4}\approx 10^{-2}.
 \eeq
 A close examination of Figure \ref{fig:ps} confirms these arguments and shows that \eqref{kdipvspitot} is a robust relation, valid in all single field inflationary scenarios that can produce a pronounced peak in the power spectrum. Note also that -- considering the relation between the peak scale and $\mathcal{H}_0$ we mentioned before -- we can  connect the peak scale to the dip scale as $k_{\rm dip} \simeq 10^{-2}\,\,k_{\rm peak}$. Relation  \eqref{kdipvspitot} is in agreement with the results
 of section \ref{sec_heur1}, and includes an overall, order-one  factor depending on parameters controlling the
 duration of the dip and the properties of the system.
  
\subsection{Bispectrum in the squeezed limit: a consistency relation around the dip}\label{SBS}

Let us concentrate on modes that exit during the initial slow-roll era, $k/\mathcal{H}_0 < c_k$, to investigate the scale-dependence of the bispectrum. We anticipate that the bispectrum  exhibits features and be amplified around the dip scale $k_{\rm dip}$ in the power spectrum since non-linearities are usually enhanced at the location of rapid changes in the power spectrum \cite{Chluba:2015bqa}. More importantly, we  show that scale dependence of the squeezed bispectrum closely follows the prediction of  Maldacena consistency condition: \ie $f_{\rm NL}(k) = 5 (1-n_s(k))/12$ \cite{Maldacena:2002vr}, proving the heuristic
results of  section \ref{sec_heur1}.
\medskip

\noindent{\bf Consistency condition.} We begin by the squeezed limit of the scale dependent non-linearity parameter\footnote{A scale-dependent
$f_{\rm NL}$ may arise in a various other inflationary contexts, see e.g. the early works \cite{Chen:2005fe,Byrnes:2009pe,Byrnes:2010ft}.}. For scales that exit the horizon during the initial slow-roll era, noting the scale invariance of $\mathcal{P}_{\mathcal{R}}(\tau_k)$ factors, we take the squeezed limit $\vec{k}_3 \to 0$  ($k_1 \simeq k_2 \equiv q$) of \eqref{fnlf} which yields \cite{Takamizu:2010xy,Ozsoy:2021qrg}
\beq\label{fnlsq}
f_{\rm NL}(q,q,k_3\to 0) \equiv f^{\rm sq}_{\rm NL} = \fr{5}{12}\left(\fr{4\,\, {\rm Re}[\alpha^{*}_{q} \alpha_{k_{3}}]\, F(\tau_{k_{q}}) q^2}{\left|\alpha_{q}\alpha_{k_3}\right|^2 }\right)\simeq \fr{5}{12}\left(\fr{4\, \alpha^{R}_{q} \, F(\tau_{k_{q}}) q^2}{\left|\alpha_{q}\right|^2 }\right),
\eeq
where in the last equality we make the approximation $\alpha_{k_3} \to 1$ in the $k_3 \to 0$ limit, and we use the fact that the enhancement factor is dominated by its real part $\alpha^{R}_k$ in the initial slow-roll phase. To prove that the consistency condition holds, we now  show that the term inside the brackets in \eqref{fnlsq} is equivalent to $(1-n_s)$. For this purpose, approximating $\alpha_k \simeq \alpha^{R}_k$ as before, we utilize \eqref{psom} and the definition $n_s - 1\equiv  {\d \ln \mathcal{P}_{\mathcal{R}}}(\tau_f, k)/\d \ln k$ to write
\beq\label{1mns}
(1-n_s) \equiv -\fr{\d \ln \mathcal{P}_\mathcal{R}(\tau_f, q)}{\d \ln q} =- \fr{2\alpha^R_q\,\left(\fr{q}{\mathcal{H}_0}\times \alpha^{R'}_q\right)}{\left|\alpha_{q}\right|^2} \approx \fr{4 \alpha^R_q F(\tau_{k_{q}}) q^2}{\left|\alpha_{q}\right|^2},
\eeq
 \begin{figure}[t!]
\begin{center}
\includegraphics[scale=1.1]{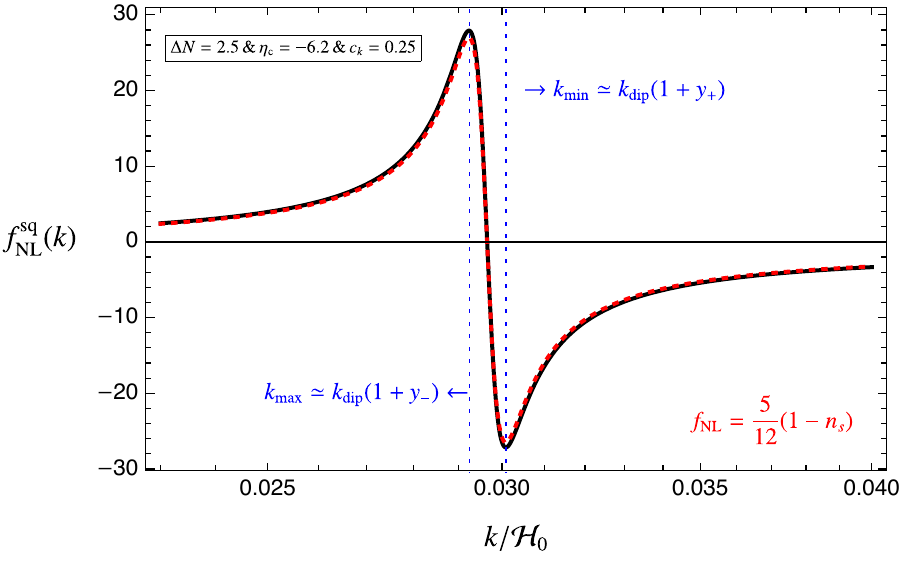}
\end{center}
\caption{\it \small Scale dependence of the squeezed configuration non-linearity parameter $f_{\rm NL}$ around $k_{\rm dip}$ for a transient constant-roll model that generates a $\Pi_{\rm tot}\simeq 10^7$ enhancement in the power spectrum. The accuracy of the consistency relation $f_{\rm NL} = 5(1-n_s)/12$ in capturing the behavior of the $f_{\rm NL}$ in the squeezed limit is shown by red dotted curve. \label{fig:fnl}}
\end{figure}
where the  prime denotes derivative with respect to the normalized wave-number $k/\mathcal{H}_0$. Notice that in the last equality of \eqref{1mns} we take $- (q/\mathcal{H}_0) \alpha^{R'}_q \simeq 2 F(\tau_q) q^2$. Employing further simplifications for $\alpha^{R}_k$ and its derivative (see eqs. \eqref{arapp} and \eqref{arpapp}),  we prove in appendix \ref{AppB1} that this relation holds to a very good approximation. This fact confirms that the consistency condition always holds in the initial slow-roll stage of inflationary scenarios that can generate PBH populations. To be more explicit, in Figure \ref{fig:fnl} we show the behavior of squeezed limit $f_{\rm NL}$ superimposed with $5(1-n_s)/12$ around the dip scale for a representative choice of parameters that can generate a $ \Pi_{\rm tot} \simeq 10^{7}$ enhancement in the power spectrum as required for PBH production. The figure shows  clearly  that the  scale dependence of the non-linearity parameter $f_{\rm NL}$ follows Maldacena consistency condition.   

\medskip

\noindent{\bf Maximal value of $f^{\rm sq}_{\rm NL}$ around the dip.} Another conclusion we can guess from Figure \ref{fig:fnl} is that the maximal amplitude of the squeezed limit non-linearity parameter becomes $|f_{\rm NL}|\simeq {\cal O}(10)$ around $k_{\rm dip}$ which is consistent with the large values obtained by the spectral tilt $|n_s -1|$ around the dip region: see Figure \ref{fig:ps}, and recall the results of section \ref{sec_heur1}. In what follows, we will show that $|f_{\rm NL}|\simeq {\cal O}(10)$ universally holds around the dip scale $k_{\rm dip}$ for any inflationary scenario that support a $\Pi_{\rm tot} \simeq 10^7$ enhancement in the power spectrum. 


 We proceed 
as in section \ref{sec_heur1}: we focus on the maximum and the minimum values obtained by the $n_s - 1$ and specifically the zeros of its running defined in \eqref{as}. 
On the other hand, as Figure \ref{fig:fnl} suggests, the spectral index reaches its maximal values very close to the dip scale $k_{\rm dip}$. Therefore we find it convenient to define a new variable, 
\beq\label{x}
\fr{k}{\mathcal{H}_0} \equiv \fr{x}{\sqrt{\beta}},
\eeq
where $\beta = \beta (\eta_{\rm c}, \Delta N, c_k) \gg 1$ is defined as in \eqref{kdip} (See also eq.  \eqref{misc}). In the parametrization of  eq. \eqref{x}, the dip feature in the power spectrum corresponds to $x = 1$ which in turn overlaps with the zero of the spectral index, since the latter is proportional to the real part of the enhancement factor as can be verified from eq. \eqref{1mns}. Furthermore, we expect the maximal values of $n_s -1$ to be very close to the dip feature. We then define $x = 1 + y$ where $y\ll 1$ and expand the resulting expression using $\alpha_s$ \eqref{as} up to quadratic order in $y$. In this way we obtain a simple quadratic equation  for $y$, and its roots provide us the zeros of the running $\alpha_s$ and hence the location of the maximum and minimum of the spectral index. In particular, in terms of the model parameters, we find  the following solutions,
\beq\label{ypm}
y_\pm \simeq \pm \fr{\bar{D}}{2\, \alpha^R_{(0)}(c_k)\sqrt{\beta}} +  \fr{3 \bar{D}^2}{4\,[\alpha^R_{(0)}(c_k)]^2 \beta},
\eeq
where we define
\beq\label{bard}
\bar{D} = \fr{\alpha^R_{(0)}(c_k) (\eta_{\rm c} + 1)}{(1+c_k^2)\eta_{\rm c}} - \alpha^I_{(0)}(c_k)\, \sqrt{\beta}.
\eeq
We note that although the latter expression appear to be of the order of $\sqrt{\beta} \gg 1$, it is genuinely an order one number due to the small factor $\alpha^I_{(0)}\propto c_k^3 \ll 1$ (See \eg \eqref{misc}). In fact, for any scenario that leads to a $\Pi_{\rm tot} \simeq 10^7$ enhancement in the power spectrum, the second term in \eqref{bard} is comparable to the first term in absolute magnitude and therefore it is a order-one number as one can verify explicitly from \eqref{misc}. Using \eqref{ypm}, the location of the maximum and the minimum  of the spectral index is therefore given by
\beq\label{kpm}
\fr{k_{\pm}}{\mathcal{H}_0} \simeq \fr{k_{\rm dip}}{\mathcal{H}_0} \left(1 \pm y_{\pm}\right) = \fr{1}{\sqrt{\beta}} \left(1 \pm y_{\pm}\right). 
\eeq
The accuracy of these formulas in locating the max/min values of $n_s -1$ is shown in Figure \ref{fig:fnl}. Using these results, we can then determine the maximal values obtained by the non-linearity parameter in the squeezed limit. Utilizing the consistency condition, we plug \eqref{kpm} in \eqref{1mns} and at leading order in the large parameter $\beta$, we found 
\beq
(f^{\rm sq}_{\rm NL})_{\rm max/min} = \fr{5}{12}(1-n_s) \simeq - \fr{10}{3} \fr{\alpha^R_{(0)}(c_k)^2}{\bar{D}^2}\beta\, y_{\mp} \simeq \pm \fr{5}{3} \fr{\alpha^R_{(0)}(c_k)}{\bar{D}}\sqrt{\beta}.
\eeq
Using \eqref{pitot}, we can then re-write the max/min value of the non-linearity parameter in terms of the total enhancement in the power spectrum as
\beq\label{fnlmax}
(f^{\rm sq}_{\rm NL})_{\rm max/min} \simeq \pm \fr{5}{3} \left[-\fr{\alpha^R_{(0)}(c_k)\, \eta_{\rm c}}{ \bar{D}^2\,\sqrt{C(c_k,\nu)}\,(\eta_{\rm c} +1)(\eta_{\rm c} +3)}\right]^{1/2} \Pi_{\rm tot}^{1/4}. 
\eeq
For typical scenarios with $\Pi_{\rm tot} \simeq 10^7$, the expression above evaluates to $(f^{\rm sq}_{\rm NL})_{max/min} \simeq \mathcal{O}(10)$ as can be checked explicitly. This results confirms and supports the findings of section \ref{sec_heur1}. 

\subsection{Trispectrum in the collapsed limit and its consistency relation}\label{STS}
We now focus on the scale dependence of the trispectrum around the dip feature, \ie for $k/{\mathcal{H}_0} < c_k$. As we  show below, the resulting  scale dependent trispectrum closely tracks a specific single-field relation  \cite{Byrnes:2006vq,Suyama:2007bg,Smith:2011if,Assassi:2012zq}
\be\label{SY}
\tau^{\rm col}_{\rm NL} = \fr{36}{25}\left(f^{\rm sq}_{\rm NL}\right)^2.
\ee
 in the collapsed limit $|\vec{k}_{12}| \to 0$. 

\medskip

\begin{figure}[t!]
\begin{center}
\includegraphics[scale=1.1]{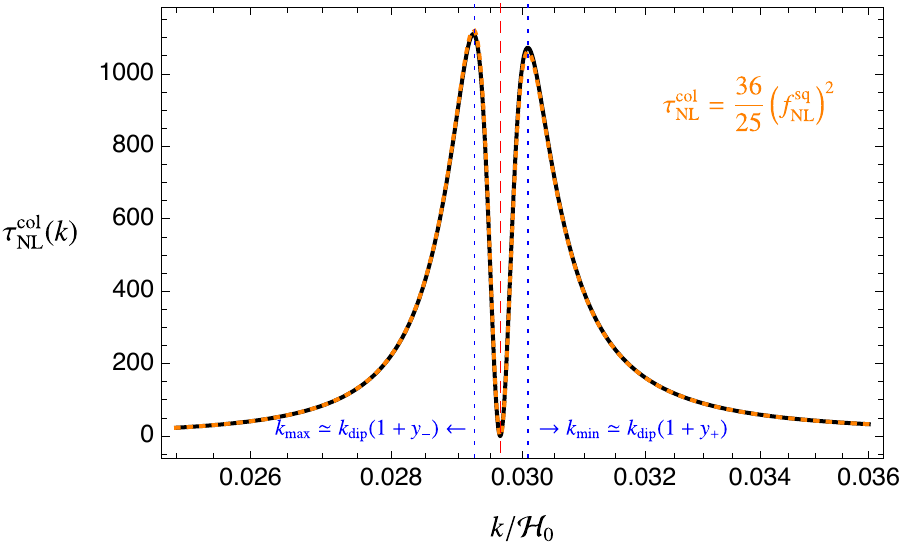}
\end{center}
\caption{\it \small Scale dependence of $\tau_{\rm NL}$ around $k_{\rm dip}$ in the collapsed limit for the transient constant-roll scenario with the same model parameters provided in Figure \ref{fig:fnl}. $\tau^{\rm col}_{\rm NL}$ as predicted by the consistency relation \eqref{SY} is shown by orange dotted curve. $\tau^{\rm col}_{\rm NL}$ becomes maximal at the same wave numbers (blue dotted vertical lines) where $f_{\rm NL}$ in the squeezed configuration does.\label{fig:taunl}}
\end{figure}

\noindent{\bf Consistency relation.} To prove tha relation \eqref{SY}  holds for scales that exit the horizon in the initial slow-roll era, we take the collapsed limit  $\vec{k}_{12} \to 0$ ($\vec{k}_{34} \to 0$) of eq. \eqref{taunl} assuming a symmetric folded kite configuration for the external momenta, \ie $k_1\simeq k_2 \simeq k_3 \simeq k_4 = q$ with $k_{12}, k_{34}\to 0$. In this limit, noting the expressions \eqref{4PT} and \eqref{dfps}, we find that $\tau_{\rm NL}$ \eqref{taunl} reduces to 
\beq\label{taunlcol}
\lim_{k_{12}\to 0}\tau_{\rm NL} \equiv \tau^{\rm col}_{\rm NL}(q, k_{12}) = 4 \fr{|\alpha_q|^2 \left(F(\tau_q)\, q^2\right)^2}{|\alpha_q|^4\,|\alpha_{k_{12}}|^2} \simeq 4\fr{(\alpha^R_q)^2  \left(F(\tau_q)\, q^2\right)^2}{|\alpha_q|^4},
\eeq
where in the last step we approximately take $\alpha_{k_{12}} \simeq 1$ in the $k_{12} \to 0$ limit and assume $\alpha_q \simeq \alpha^R_q$  during the initial slow-roll stage. Using the squeezed limit expression for $f^{\rm sq}_{\rm NL}$ \eqref{fnlsq}, it  simply follows from \eqref{taunlcol} that  relation \eqref{SY} holds in a non-trivial scale dependent manner during the initial slow-roll stage, \ie around the dip feature in the power spectrum. 
To illustrate these points concretely, we show in Figure \ref{fig:taunl} the scale dependence of the $\tau^{\rm col}_{\rm NL}$ around the dip scale $k_{\rm dip}$. We clearly see that $\tau^{\rm col}_{\rm NL}$ satisfies the   relation \eqref{SY} and therefore it becomes maximal at wave numbers where $|f^{\rm sq}_{\rm NL}|$ does: \ie at $k_{\pm}$ as given by eq \eqref{kpm}. Using \eqref{fnlmax} and \eqref{SY}, the maximal value that $\tau^{\rm col}_{\rm NL}$ can acquire 
is found to be
\beq
\left(\tau^{\rm col}_{\rm NL}\right)_{\rm max} = -\fr{4\,\alpha^R_{(0)}(c_k)\, \eta_{\rm c}}{ \bar{D}^2\,\sqrt{C(c_k,\nu)}\,(\eta_{\rm c} +1)(\eta_{\rm c} +3)}\, \Pi_{\rm tot}^{1/2}.
\eeq
Therefore for any slow-roll violating transient phase that can generate a $\Pi_{\rm tot} \simeq 10^{7}$ enhancement in the power spectrum, the non-linearity parameter becomes $\tau^{\rm col}_{\rm NL} \simeq \mathcal{O}(10^{3})$ around the dip feature, in agreement
with the heuristic results of section \ref{sec_heur1}.  

Before we conclude this section, we stress that our findings on the scale dependence of the bispectrum in Section \ref{SBS} agree well both qualitatively and quantitatively with the previous studies focusing on the same issue using numerical techniques (See \eg \cite{Passaglia:2018ixg,Gao:2021vxb,Ragavendra:2020sop}). As we mentioned earlier in \cite{Ozsoy:2021qrg}, the gradient expansion formalism has the advantage of analytic control that allows us to transparently capture the features of n-point scalar correlation functions using a few parameters such as the duration of the transient non-attractor era $\Delta N$ and the slow-roll parameter $\eta_{\rm c}$ in this phase. More importantly, using the gradient formalism we proved analytically in Section \ref{SBS} and \ref{STS} that consistency conditions\footnote{See also \cite{Sreenath:2014nca} and \cite{Ragavendra:2020sop} for an investigation on the Maldacena's consistency condition in slow-roll violating inflationary scenarios.} $f^{\rm sq}_{\rm NL} = 5(1-n_s)/12$ and $\tau^{\rm col}_{\rm NL} = {36}\left(f^{\rm sq}_{\rm NL}\right)^2/{25}$ \eqref{SY} hold for modes that leave the horizon during the initial slow-roll era.

\section{CMB $\mu$-distortions  and non-Gaussianity  around the dip}\label{PBSandTS}

In the previous section we studied consistency conditions for non-Gaussian parameters
around the dip of the spectrum.  If the dip feature occurs at relatively large scales, say $10\,$Mpc$^{-1}\,\le k_{\rm dip}\,\le 10^4\,$Mpc$^{-1}$, the properties
of the resulting curvature spectrum  can be probed through CMB $\mu-$distortions, using well controlled CMB 
physics in the linear regime  \cite{Hu:1994bz,Khatri:2011aj} (see \cite{Chluba:2011hw,Chluba:2019kpb} for a review).

This possibility  was suggested in \cite{Ozsoy:2021qrg}, building on the ideas first developed in \cite{Pajer:2012vz}. 
 In this section we further develop and extend these arguments. We stress that the range of scales we identified for $ k_{\rm dip}$ above is well motivated for PBH populations with astrophysical masses. For example, given  that $k_{\rm dip}/k_{\rm peak}\sim 10^{-2}$, with a dip scale located at $k_{\rm dip}\,\simeq\, 10^3-10^4\,$Mpc$^{-1}$ relates to a peak scale $k_{\rm peak}\,\simeq\, 10^5-10^6\,$Mpc$^{-1}$,  that correspond to the formation of PBHs in the mass 
 range $M_{\rm PBH}\simeq 1-100\,M_\odot$ (see e.g. \cite{Carr:2020xqk,Garcia-Bellido:2017fdg}).

CMB $\mu$-distortions can impose  constraints
on PBH formation mechanisms \cite{Byrnes:2018txb,Unal:2020mts}; we wish to emphasize  that,  in case
of detection, the statistics of $\mu$-distortions at large scales  provide information on the  physics sourcing PBHs at much smaller, non-linear scales.
This is possible thanks to the coupling between large and small scales through the squeezed and collapsed limits of non-Gaussianity.

Our starting point are the consistency relations of sections \ref{sec_heur1}, \ref{S2} which imply that for an ${\cal O}(10^7)$ enhancement in the spectrum, we obtain $f_{\rm NL}^{\rm sqz}\,\sim\, {\cal O}(10)$, $\tau_{\rm NL}^{\rm coll}\,\sim\, {\cal O}(10^3)$ around the dip feature for any model of single-field inflation including a short period of non-attractor evolution. Additionally, we find that the squeezed non-Gaussian parameters has features specific of these scenarios -- see Fig \ref{fig:plot3}.

As first proposed in \cite{Pajer:2012vz} (and further explored in \cite{Ganc:2012ae,Emami:2015xqa,Chluba:2016aln,Dimastrogiovanni:2016aul,Ravenni:2017lgw,Orlando:2021nkv,Remazeilles:2021adt}) non-Gaussianity at $\mu$-distortion scales
have distinctive consequences for  $\langle \mu T \rangle$ correlators \footnote{See also \cite{Ota:2016mqd,Remazeilles:2021adt} for the influence of  primordial non-Gaussianities on $\mu E$ and $\mu B$  \cite{Orlando:2021nkv} cross correlations.} among CMB $\mu$-distortions and  temperature
fluctuations -- sensitive to the squeezed limit of the bispectrum -- and for $\langle \mu \mu \rangle$  self correlations --
sensitive to the collapsed limit of the trispectrum. In this section we go beyond work in \cite{Ozsoy:2021qrg}
 by developing the following points
 \begin{itemize}
 \item First, we show how the information provided by the consistency 
 relations allows us to carry on a more detailed  analysis of $\langle \mu T \rangle$
 correlators,  whose quantitative and qualitative features depend on the properties
 of the squeezed  bispectrum. See Section \ref{PmuT}.
 \item Then, we study for the first time the implications for the  $\langle \mu \mu \rangle$  selfcorrelator 
 of a  scale-dependent
 collapsed trispectrum around the dip. See Section \ref{Pmumu}.
 \item Finally, at the light of the results above,  we  discuss improved estimates
for the detectability of  non-Gaussian consistency relations  with   PIXIE or PRISM-like experiments, and physical implications 
for  PBH populations.  See Section \ref{S3p5}.
 \end{itemize}

Before covering the points above, we present some preliminary formulas \footnote{We refer
the reader to \cite{Chluba:2011hw}, or section 3.1 of \cite{Ozsoy:2021qrg} for a mini-introduction of CMB $\mu$-distortions and 
additional motivations for the formulas that follow.} that we are going to use extensively. We begin by relating the initial curvature perturbation $\mathcal{R}_{k}$ to the harmonic coefficients of the CMB temperature anisotropies $\Theta(\hat{n}) = \sum_{lm} a^{T}_{lm}\,Y_{lm}(\hat{n})$ and CMB distortion anisotropies $\mu(\hat{n}) = \sum_{lm} a^{\mu}_{lm}\,Y_{lm}(\hat{n})$.  The coefficients $a_{l m}^{T,\,\mu}$
 associated with $\mathcal{R}_{k}$ are given by \cite{Pajer:2012vz,Ganc:2012ae},
\begin{align}
\label{aT} a_{l m}^{T} &=\frac{12 \pi}{5}(-i)^{l} \int \frac{\mathrm{d}^{3} k}{(2 \pi)^{3}} \,\mathcal{R}_{k}\, \Delta_{l}(k)\, Y_{l m}^{*}(\hat{k}),\\
\nn a_{l m}^{\mu} &\simeq 18.4 \pi(-i)^{l} \int \frac{\mathrm{d}^{3} k_{1} \mathrm{~d}^{3} k_{2}}{(2 \pi)^{6}} Y_{l m}^{*}\left(\hat{k}_{+}\right) \,\mathcal{R}_{k_{1}}\, \mathcal{R}_{k_{2}} \,W\left(\frac{k_{+}}{k_{s}}\right) j_{l}\left(k_{+} \chi_{*}\right) \\
& \quad\quad\quad\quad\quad\quad\quad\quad\quad\quad\quad\quad\quad\quad\quad\times\left\langle\cos \left(c_s k_{1} \tau\right) \cos \left(c_{s} k_{2} \tau\right)\right\rangle_{p}\left[e^{-\left(k_{1}^{2}+k_{2}^{2}\right) / k_{D}^{2}}\right]_{z_f}^{z_i},\label{mua}
\end{align}
where $p$ denotes time averaging over the period acoustic oscillations; $\vec{k}_+ \equiv \vec{k}_{1} + \vec{k}_2$; $c_s$ is the sound speed of the radiation perturbations; $\chi_{*}=\tau_{0}-\tau_{*} \simeq 14\, \mathrm{Gpc}$ is the comoving distance between the last scattering surface and today;
and $\Delta_{l}(k)$ is the  transfer function during radiation dominated universe. Moreover,
in \eqref{mua},  $W(k)= 3k^{-3} [\sin(k)-k\cos(k)]$ is a top-hat filter function in Fourier space that smears the dissipated energy over a volume of radius $k_s^{-1} \gtrsim k_D(z_f)^{-1}$ where $k_D$ is the diffusion damping scale during  radiation domination. It depends on redshift as $ k_{D}(z)  \simeq [{(1+z)}/{10^5}]^{3 / 2}\,\, 130\, \mathrm{Mpc}^{-1} \,$ and the range of $z$ associated with the $\mu$ distortions is given by
 \be\label{muband1}
 z_{f} \equiv 5 \times 10^{4}< z <2 \times 10^{6} \equiv z_{i}.
 \ee 
We then define angular correlators involving anisotropies labeled by $\{i,j\}$ as
\beq\label{defac}
\left\langle\left(a_{l m}^{i}\right)^{*} a_{l^{\prime} m^{\prime}}^{j}\right\rangle=\delta_{l l^{\prime}} \delta_{m m^{\prime}} C_{l}^{i j}\,, \hskip1cm i\,=\,{\mu, T}.
\eeq

In what comes next, we make use of  the definition  \eqref{defac} to study angular correlations $\langle \mu T \rangle$ and $\langle \mu \mu \rangle$, and to relate them to scale dependent bispectrum and trispectrum present around the dip scale of the PBH forming inflationary scenarios. 

\subsection{Phenomenology of the scale dependent squeezed bispectrum: $C^{\mu T}_l$}\label{PmuT}

We start discussing
 cross correlations between $\mu$ distortions and temperature anisotropies $\Theta$.  As we will see,
  the specific scale-dependent profile of the squeezed bispectrum can considerably enhance such 
  cross correlations in comparison with more standard non-Gaussian models.
   
  \smallskip
  
 Using
 the definition \eqref{BSdef} of the bispectrum together with \eqref{aT} and \eqref{mua}, the cross correlations $C^{\mu T}_l$ 
 for a squeezed bispectrum ($\vec{k}_+ = \vec{k}_{1} + \vec{k}_2\to 0$) can be expressed as 
  \cite{Pajer:2012vz,Ganc:2012ae},
\beq\label{clmutf}
 C_{l}^{\mu T} \simeq \frac{27.6}{20 \pi^{3}} \int \d \ln k_{+}\, \Delta_{l}(k_{+})\,j_{l}\left(k_{+} \chi_*\right)\int \d \ln q\,  \bigg[k_{+}^{3} q^{3} B_{\mathcal{R}}\left(q, q, k_+\to 0\right)\bigg]
  \left[e^{-2q^2 / k^{2}_{D}(z)}\right]_{z_f}^{z_i},
\eeq
where we take the filter function $W \to 1$ in the squeezed limit $k_+ \ll k_s \simeq k_{d}(z_f)$, and we relabel $|\vec{k}_1| \simeq |-\vec{k}_2| \equiv q$. Using the general definition \eqref{deffnl} of the non-linearity parameter, the squeezed limit bispectrum inside the $\d \ln q$ integral in \eqref{clmutf} can be expressed as 
\beq\label{bssqg}
 B_{\mathcal{R}}\left(q, q, k_+\to 0\right) \simeq \fr{12}{5} \, \underbrace{f^{\rm sq}_{\rm NL}(q) E(q)}_{\equiv f^{\rm eff}_{\rm NL}(q)}\,\, P_{\mathcal{R}}(\tau_q) P_{\mathcal{R}}(\tau_{k_+}).
\eeq
In \eqref{bssqg}, we defined an effective $f^{\rm eff}_{\rm NL}(q)$ by combining the squeezed limit $f^{\rm sq}_{\rm NL}$ with a scale-dependent enhancement factor
\be
E(k)\, = \,\{\Pi(k), |\alpha_k|^2\},
\ee
which can assume two values, depending on whether we consider the heuristic formulas of section  \ref{sec_heur1}
for the consistency relations, or the gradient expansion formulas of section
 \ref{S2}.  For the rest of this section, we consider in parallel the two cases, normalizing $E$  assuming that  
$E(k_+ \to 0)\to 1$ as we did in \eqref{bssqg}.

Noticing that  $P_{\mathcal{R}}(\tau_p) = 2\pi^2 \mathcal{P}_\mathcal{R}(\tau_p)/p^3$ and $ \mathcal{P}_\mathcal{R}(\tau_p) \simeq 2.1 \times 10^{-9}$ for mode exit during the initial slow-roll era (see e.g. \eqref{pstauk}), we insert \eqref{bssqg} in \eqref{clmutf} to express the angular cross correlation as
\beq\label{CLmuT}
C_{l}^{\mu T} \simeq 2.7 \times 10^{-17}\,\frac{2 \pi}{l(l+1)}\,\, b_{(\mathrm{pbh})}(l),
\eeq
where we define 
\beq\label{bpbh}
b_{(\mathrm{pbh})}(l) \equiv \frac{6 l(l+1)}{\ln \left(\frac{k_{D}\left(z_{i}\right)}{k_{D}\left(z_{f}\right)}\right)} \int \mathrm{d} \ln k_+\, \,\Delta_{l}(k_+)\,\, j_{l}\left(k_+ \chi_{*}\right)  \int \mathrm{d} \ln q\,\, {f}_{\mathrm{NL}}^{\mathrm{eff}}(q)\left[e^{-2 q^{2} / k_{D}^{2}(z)}\right]_{z_f}^{z_i},
\eeq
as the key quantity  that parameterizes the multipole $l$ dependence and the size of the angular correlator $\langle \mu T\rangle$ of eq \eqref{CLmuT}.  Notice that this quantity depends on the transfer function $\Delta_l (k)$ which we decompose as \cite{Chluba:2016aln} 
\be\label{tf}
\Delta_l (k) = \rho(l) \Delta^{\rm sw}_l(k)\,.
\ee 
 $\Delta^{\rm sw}_l(k)$ is the transfer function in the large-scale Sachs-Wolfe (SW) limit
\be\label{tfsw}
\Delta^{\rm sw}_l(k)\, =\,\frac13 j_{l}(k \chi_*),
\ee
while 
\beq\label{rhol}
\rho(l) \simeq 1.08\left[1-0.022 l-1.72 \times 10^{-4} l^{2}+2 \times 10^{-6} l^{3}-4.56 \times 10^{-9} l^{4}\right],
\eeq
is an analytic fit that includes high-$l$ corrections to the SW approximation \cite{Chluba:2016aln}. Inserting \eqref{tf} in \eqref{bpbh} and using \eqref{tfsw}, we can analytically carry the integral over long momenta $k_+$ in \eqref{bpbh},  and  factorize the result as
\beq\label{bpbhl}
b_{(\rm pbh)} (l) = \rho(l)\, b^{\rm sw}_{(\rm pbh)},
\eeq
with
\beq\label{bpbhsw}
b^{\rm sw}_{(\mathrm{pbh})} = \left[{\ln \left(\frac{k_{D}\left(z_{i}\right)}{k_{D}\left(z_{f}\right)}\right)}\right]^{-1}  \int \mathrm{d} \ln q\,\, {f}_{\mathrm{NL}}^{\mathrm{eff}}(q)\left[e^{-2 q^{2} / k_{D}^{2}(z)}\right]_{z_f}^{z_i}.
\eeq

\smallskip

\begin{figure}[t!]
\begin{center}
\includegraphics[scale=0.85]{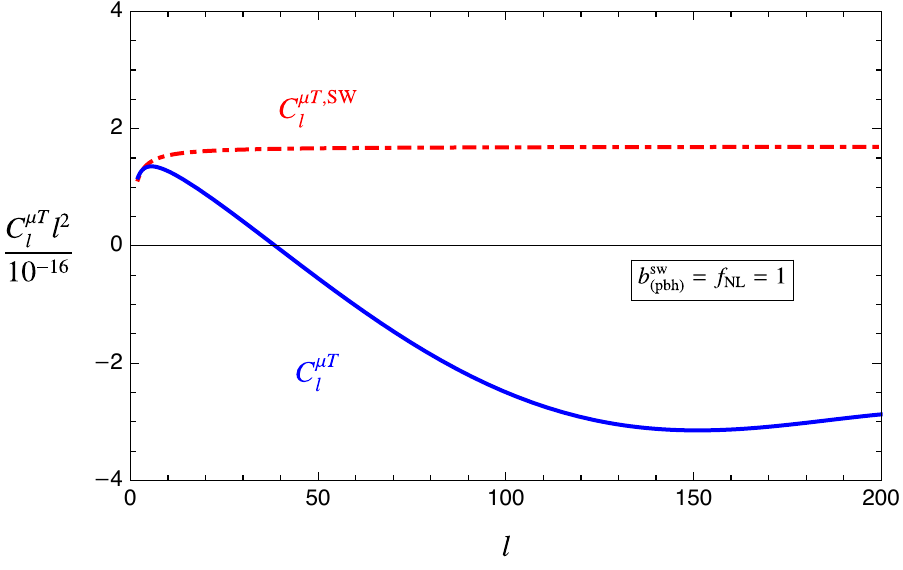} \includegraphics[scale=0.88]{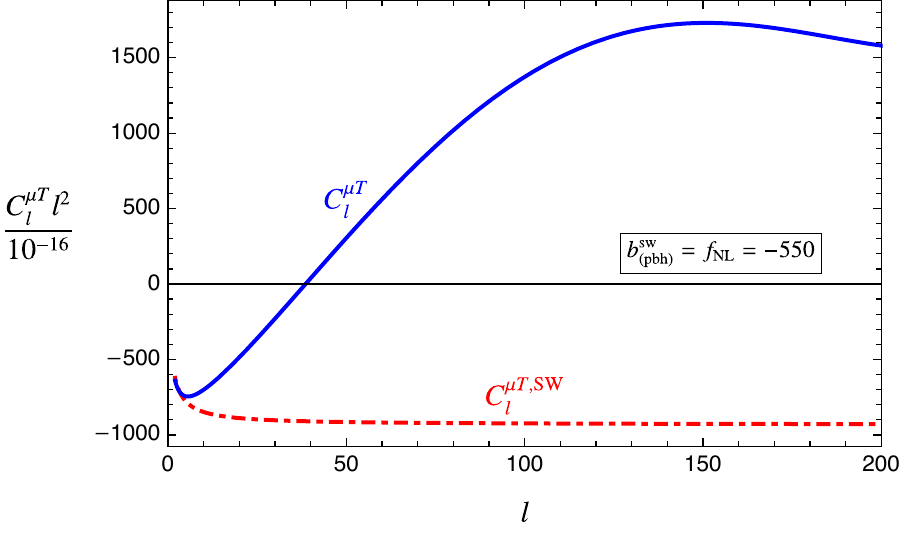} 
\end{center}
\caption{\it \small $C_{l}^{\mu T}$ (see eqs. \eqref{CLmuTF} and \eqref{CLmuTsw}) for a scale independent purely local type bispectrum with $f_{\rm NL} = 1$ (Left) and $f_{\rm NL} = -550$ (Right). \label{fig:clloc}}
\end{figure}

\noindent
The function $b^{\rm sw}_{(\mathrm{pbh})}$ is normalized
in such a way that for 
a purely local spectrum with  constant non-linearity parameter, ${f}_{\mathrm{NL}}^{\mathrm{eff}}(q) \,= \, f_{\rm NL}$, we get    $b^{\rm sw}_{(\mathrm{pbh})} = f_{\rm NL}$. In this case, 
we  reproduce the standard results for $C^{\mu T}_l$ in the SW limit \cite{Pajer:2012vz,Ganc:2012ae} as we show in the left panel of Figure \ref{fig:clloc}.
However, a non-trivial scale dependent of ${f}^{\rm eff}_{\rm NL}$ in \eqref{bpbhsw} can significantly impact the amplitude $C^{\mu T}_l$\eqref{CLmuT}. We will concretely see an example of this fact in the PBH forming inflationary scenarios we are analyzing. 

To summarize the formulas so far, for inflationary models that can produce a sizeable peak in the power spectrum, the amplitude and scale dependence ($l$) of the angular correlator $C^{\mu T}_l$ can be determined by 
\beq\label{CLmuTF}
C_{l}^{\mu T} \simeq 2.7 \times 10^{-17}  \frac{2 \pi}{l(l+1)} \rho(l)\, b^{\rm sw}_{(\mathrm{pbh})} \equiv \rho(l)\, C_{l}^{\mu T, {\rm sw}},
\eeq
where we defined the angular $\langle \mu T\rangle$ cross correlation in the Sachs-Wolfe limit $l\to0$ as
\beq\label{CLmuTsw}
C_{l}^{\mu T,{\rm sw}} = 2.7 \times 10^{-17} \frac{2 \pi}{l(l+1)} \, b^{\rm sw}_{(\mathrm{pbh})}.
\eeq

\medskip

\noindent
{\bf Consistency relations and the function $b^{\rm sw}_{(\mathrm{pbh})}$.}   
The integral of eq \eqref{bpbhsw} which provides the quantity  $b^{\rm sw}_{(\mathrm{pbh})}$ depends
on the scale-dependent non-linearity parameter ${f}_{\mathrm{NL}}^{\mathrm{eff}}(q)$ of eq \eqref{bssqg}. We now make use of the information 
given by the
consistency relation  for the squeezed bispectrum of sections  \ref{sec_heur1} and \ref{S2} 
($f^{\rm sq}_{\rm NL}(q) = 5(1-n_s(q))/12$) to characterise
the  size of $b^{\rm sw}_{(\mathrm{pbh})}$  and its dependence on the dip position $k_{\rm dip}$. We show that for interesting values of  $k_{\rm dip}$ this quantity is large and negative, and its value depends on the total amplification of the spectrum, as well as  on properties of the non-attractor phase. 

\smallskip

In Appendix \ref{AppC}, we derive a power-law expression for $f^{\rm eff}_{\rm NL}(q)$: see eq    \eqref{fnleffgappf}. 
Using it, we can express $b^{\rm sw}_{(\mathrm{pbh})}$ in terms of  the following polynomial 
\beq\label{bpbhswf}
b_{(\mathrm{pbh})}^{\mathrm{sw}}=\frac{5}{3\ln \left[{k_{d}\left(z_{i}\right)}/{k_{d}\left(z_{f}\right)}\right]} \sum_{n=2}^{n_f} {\tilde{c}_{E}^{(n)}}\frac{\Gamma(n / 2)}{2^{(n+2) / 2}}\left(\frac{k_{D}(z)}{ k_{\rm dip}}\right)^{n}\Bigg|_{z_{f}} ^{z_{i}},
\eeq
where the coefficients $\tilde{c}_E$,  with $E = \{\Pi, |\alpha|^2\}$, include  as above both  the heuristic
and gradient expansion approaches. They  are given in \eqref{cfnlgrad} and \eqref{cfnlheur}. In the same way, 
 $n_f = \{5,6\}$ depending on which approach one adopts.
 
This formula depends on $ k_{\rm dip} $, which we take within   the $\mu$-distortion band of eq \eqref{muband1} as
\beq\label{muera}
k_D(z_f) \simeq 46\, {\rm Mpc^{-1}} < k_{\rm dip} < k_D(z_i) \simeq 11 600\, {\rm Mpc^{-1}}.
\eeq
The reason for this choice is two folds. First of all, if we take $k_{\rm dip} \lesssim k_D(z_f)$ then  the peak of the power spectrum would lie within the range of scales associated with  $\mu$-distortions, where we have stringent constraints on the peak amplitude of the power spectrum from COBE and FIRAS which limits $\langle\mu\rangle \lesssim 10^{-5}$ \cite{Mather:1993ij,Fixsen:1996nj}. 
Instead, when choosing  $k_{\rm dip} > k_D(z_f)$, the interesting effects of the scale dependent bispectrum around the dip scale 
are no longer  present,  as can be realized from \eqref{bpbhswf} which behaves as $b_{(\mathrm{pbh})}^{\mathrm{sw}} \to 0$ in the $k_{D}(z_i)/k_{\rm dip} \to 0$ limit. Another concrete consequence of the choice of scales \eqref{muera} is that the amplitude of  $b_{(\mathrm{pbh})}^{\mathrm{sw}}$ is controlled by the upper limit shown in eq \eqref{bpbhswf}, namely by powers of  $k_{D}(z_i)/k_{\rm dip}$. For the derivations we present below, we will repeatedly make us such simplification. 

\smallskip
 In the regime of interest \eqref{muera}, we find that $b_{(\mathrm{pbh})}^{\mathrm{sw}}$  acquires a minimum at a critical value of the ratio $k_{\rm dip}/k_D(z_i)$, whose
 location is 
 
\beq\label{bswmin}
\left(\fr{k_{\rm dip}}{k_D(z_i)}\right)^{n_f - 4}_{\rm min} = -\fr{n_f\,{\Gamma(n_f/2)}}{2^{n_f/2}}\, \fr{\tilde{c}^{(n_f)}_E}{\tilde{c}^{(4)}_E},
\eeq
At its minimum \eqref{bswmin}, the amplitude of $b_{(\mathrm{pbh})}^{\mathrm{sw}}$ \eqref{bpbhswf} is negative and its final value is  set by a competition between the $n_f$ and $n=4$ terms in the sum \eqref{bpbhswf}.  Using \eqref{bswmin} in \eqref{bpbhswf}, we can obtain this negative value at the minimum as
\beq\label{bmin}
b_{(\mathrm{pbh})}^{\mathrm{sw}}\bigg|_{\rm min}\,\simeq\,- \fr{5}{3 \ln \left[{k_{D}\left(z_{i}\right)}/{k_{D}\left(z_{f}\right)}\right]}\fr{(n_f -4)}{8 n_f}\,\, |\tilde{c}^{(4)}_E| \left(\fr{k_D(z_i)}{k_{\rm dip}}\right)_{\rm min}^4,
\eeq
\begin{figure}[t!]
\begin{center}
\includegraphics[scale=0.87]{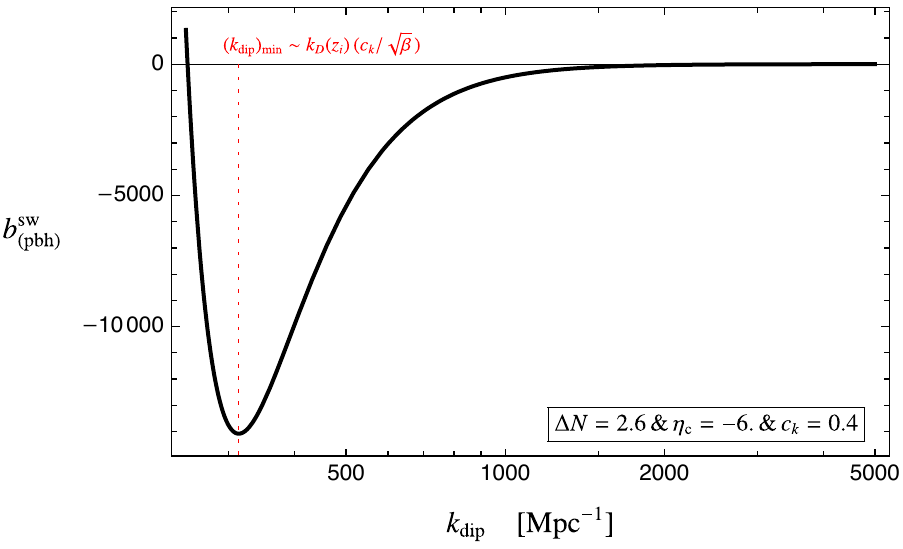} \includegraphics[scale=0.87]{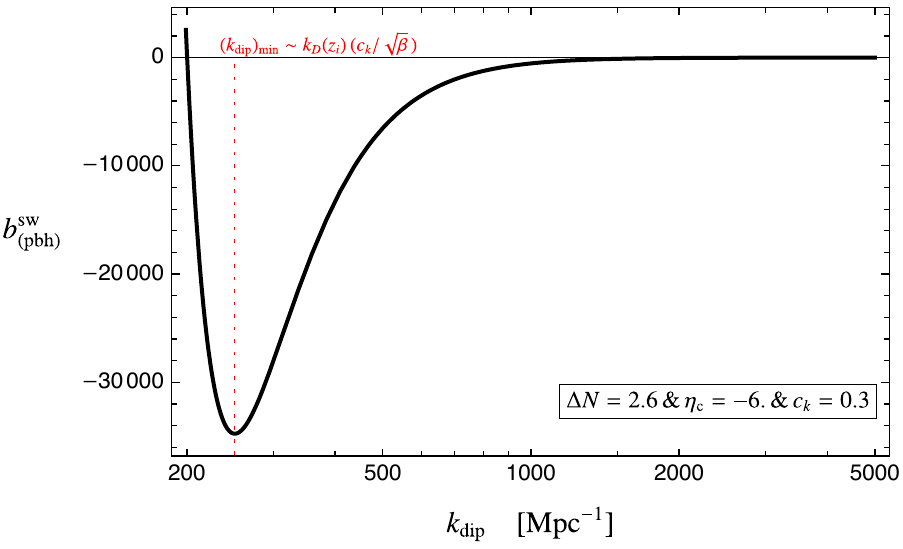} 
\end{center}
\caption{\it \small The quantity $b^{\rm sw}_{(\rm pbh)}$ \eqref{bpbhswf} vs $k_{\rm dip}$ within the gradient expansion formalism for inflationary scenario that contains a transient slow-roll violating phase that is characterized by the parameter choices: $\left\{\Delta N=2.6, \eta_{\mathrm{c}}=-6, c_{k}=0.4\right\}$  (Left) and $\left\{\Delta N=2.6, \eta_{\mathrm{c}}=-6, c_{k}=0.3\right\}$ (Right). \label{fig:bpbh}}
\end{figure}
\begin{figure}[h!]
\begin{center}
\includegraphics[scale=0.93]{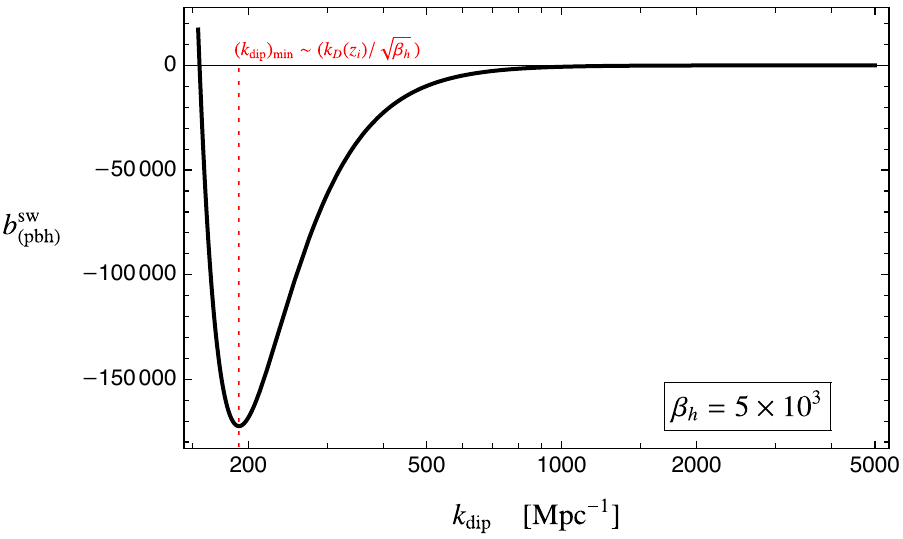} 
\end{center}
\caption{\it \small The quantity $b^{\rm sw}_{(\rm pbh)}$ \eqref{bpbhswf} vs $k_{\rm dip}$ in the heuristic approach of section \ref{sec_heur1} for an inflationary scenario that can generate a $\Pi_{T} \simeq 10^{7}$ growth in the power spectrum.\label{fig:bpbh2}}
\end{figure}
where we use the fact $\tilde{c}^{(4)}_{E} < 0$ for both the approaches we are focusing in. We can be more explicit in discussing
separately the two cases we are considering:
\begin{itemize}
\item
\noindent{\it Heuristic approach:} Using the formulas \eqref{bswmin} and \eqref{bmin}, we can relate the location of the minimum of $b^{\rm sw}_{(\rm pbh)}$, as well as its amplitude at that position to the  free parameter $\beta_h$, 
 by setting $n_f = 6$, and making use of the coefficients we provide in \eqref{cfnlheur}.
 
  At leading order in the large parameter $\beta_h \gg 1$, this procedure gives
\beq\label{pbmaxh}
\left(\fr{k_D(z_i)}{k_{\rm dip}}\right)^{2}_{\rm min} \simeq 0.74 \beta_h\simeq 1.5\, \Pi_T^{1/2}\quad \longrightarrow \quad b_{(\mathrm{pbh})}^{\mathrm{sw}}\bigg|_{\rm min} \simeq -7\times 10^{-3}\,\, \beta_h^2 \simeq -2.8 \times 10^{-2}\,\, \Pi_T,
\eeq
where we use $\beta_h \simeq 2 \Pi_T^{1/2}$ from \eqref{pitoth}. For inflationary scenarios with a total of $\Pi_T \simeq 10^{7}$ enhancement, these results predict $k_D(z_i) \simeq 69 \,k_{\rm dip, min}$ and $b_{(\mathrm{pbh})}^{\mathrm{sw}}\big|_{\rm min} \simeq - \mathcal{O}(10^{5})$. Importantly,  the size of $|b_{(\mathrm{pbh})}^{\mathrm{sw}}|$ is much larger than the maximal values of $|f_{\rm NL}^{\rm sqz}|$ around the dip position, as
we analyzed in section  \ref{sec_heur1} in eq \eqref{fnlMAXMIN}. This is due to the fact that $b_{(\mathrm{pbh})}^{\mathrm{sw}}$ involves an
integration over momenta that -- picking up contributions from the scale-dependent
squeezed bispectrum -- considerably enhances its value with respect to the constant $f_{\rm NL}^{\rm sq}$  case (see comment after eq \eqref{bpbhsw}). This fact improves the chances of detection.   


\item
\noindent{\it Gradient expansion formalism:} In this approach, we use \eqref{cfnlgrad} together with \eqref{bswmin} and \eqref{bmin} to determine the location of the minimum and the resulting $b_{(\mathrm{pbh})}^{\mathrm{sw}}\big|_{\rm min}$ in terms of model parameters as
\beq\label{pbmaxg}
\left(\fr{k_D(z_i)}{k_{\rm dip}}\right)_{\rm min} \simeq \fr{0.34}{c_k} \fr{\eta_{\rm c} (1+c_k^2) }{(\eta_{\rm c} + 1)} \sqrt{\beta} \quad \longrightarrow \quad b_{(\mathrm{pbh})}^{\mathrm{sw}}\bigg|_{\rm min} \simeq -7.5 \times 10^{-3} \left(\fr{0.34}{c_k}\right)^4  \fr{\eta^4_{\rm c} (1+c_k^2)^4}{(\eta_{\rm c} + 1)^4} \beta^2,
\eeq 

where dependence of $\beta \gg 1$ on the properties of the non-attractor era is given by \eqref{misc}. For typical parameter choices that leads to a $\Pi_{\rm tot} \simeq 10^{7}$ enhancement in the power spectrum (see Figure \ref{fig:ps}), we have $\beta \simeq 10^{3}$ and \eqref{pbmaxg} predict $k_D(z_i) \simeq (40-50) \,k_{\rm dip, min}$ and $b_{(\mathrm{pbh})}^{\mathrm{sw}}\big|_{\rm min} \simeq - \mathcal{O}(10^{4})$. 
\end{itemize}

Comparing the result obtained from the heuristic approach and the gradient expansion formalism, we notice  that although they agree  for the location of the minimum $k_{\rm dip,min}$, the amplitude of $b^{\rm sw}_{(\rm pbh)}$ at the minimum differs by an order of magnitude.
{ However,   expression \eqref{pbmaxg} indicates that for smaller choices of the model parameter $c_k$ the overall amplitude $|b^{\rm sw}_{(\rm pbh)}|$ and the location of the minimum of $b^{\rm sw}_{(\rm pbh)}$ tend to agree better with the heuristic approach, as in this case $k_{\rm dip,min}$ shifts to slightly larger scales, causing the overall amplitude $|b^{\rm sw}_{(\rm pbh)}|_{\rm min}$ to  become larger.} 
We confirm these findings in Figures \ref{fig:bpbh} and \ref{fig:bpbh2}, where we present the dependence of $b^{\rm sw}_{(\rm pbh)}$ on the dip scale $k_{\rm dip}$ in both the heuristic approach and gradient expansion formalism for representative parameter choices that can generate substantial growth in the power spectrum, as required for PBH formation. 

\begin{figure}[t!]
\begin{center}
\includegraphics[scale=0.87]{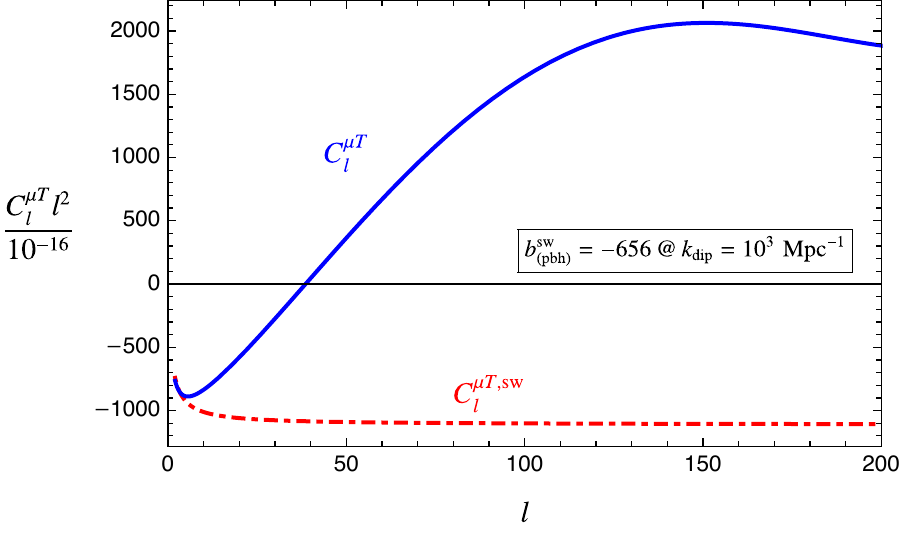} \includegraphics[scale=0.88]{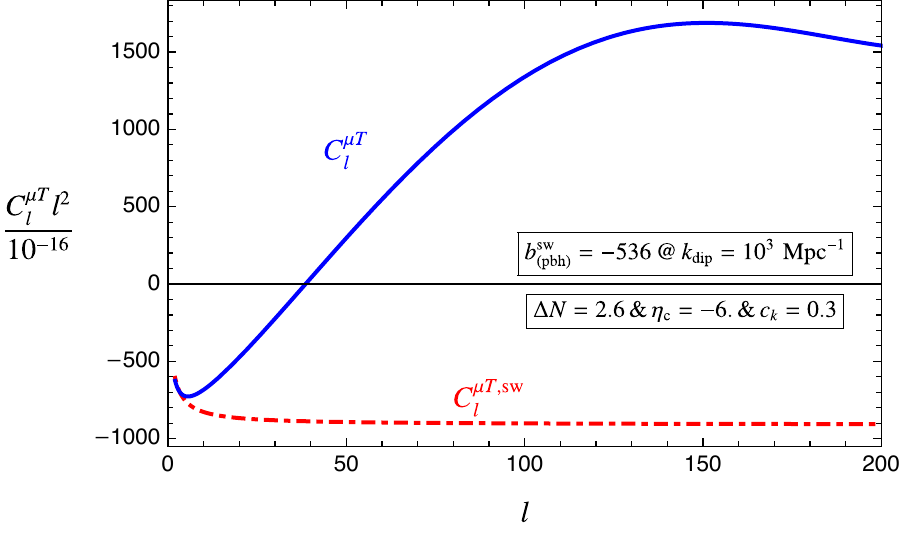} 
\end{center}
\caption{\it \small $C_{l}^{\mu T}$ for inflationary models that can generate a $\Pi_{\rm tot} \simeq 10^{7}$ enhancement in the power spectrum. The left panel represents the results obtained in the heuristic approach with $\beta_h = 5 \times 10^{3}$. In the right panel, we show the corresponding $C_{l}^{\mu T}$ using the formulas in the gradient formalism where the parameter choices that characterize the slow-roll violating phase is indicated in the boxes. \label{fig:cl}}
\end{figure}

{\bf The multipole dependence of  $C^{\mu T}_l$.}
 We collect  these results to consider the scale dependence ($l$) of angular cross correlation $C^{\mu T}_l$. For concreteness, we set $k_{\rm dip} = 10^{3} \, {\rm Mpc^{-1}}$ to first determine the amplitude of $b^{\rm sw}_{(\rm pbh)}$ from \eqref{bpbhswf}. We then represent in Figure \ref{fig:cl}  the multipole dependence of the quantity $C^{\mu T}_l$ using \eqref{rhol}, \eqref{CLmuTF} and \eqref{CLmuTsw}. We notice that, in addition to the enhancement of the amplitude of the angular correlator $C^{\mu T}_l \propto b^{\rm sw}_{(\rm pbh)}$ for $|b^{\rm sw}_{(\rm pbh)}| \gg 1$, its scale dependence from large (small $l$) to small scales (large $l$). This is because $b^{\rm sw}_{(\rm pbh)} < 0$ for the phenomenologically interesting $k_{\rm dip}$ values we are focusing defined in \eqref{muera}. As explained in detail in \cite{Ozsoy:2021qrg}, the origin of this behavior can be traced back to the change of sign of the scale dependent ${f}_{\mathrm{NL}}^{\mathrm{eff}}$ in \eqref{bpbhsw} which occurs at $q = q_{\rm dip}$ followed by its growth in the negative direction. This result implies that for inflationary models that is capable to generate PBH populations, $\mu$ distortions become anti-correlated with temperature anisotropies at large scales.  {We note that a similar profile for $C^{\mu T}_l$ can be generated by a standard local type non-Gaussianity that exhibit a large and negative scale-independent  non-linearity parameter $f_{\rm NL} = -\mathcal{O}(100)$ as we show in the right panel of Figure \ref{fig:clloc}.} 
  
\subsection{Phenomenology of scale dependent trispectrum: $C^{\mu\mu}_l$} \label{Pmumu}

We now investigate non-Gaussian\footnote{Disconnected part of the 4-pt function also leads to a Gaussian contribution for $\langle\mu \mu\rangle$ with $l\neq0$. We analyze  this contribution in Appendix \ref{AppD} and find that it can be neglected compared to the $C^{\mu \mu}_{l, {\rm NG}}$ we investigate in this section.} contribution to $\langle \mu \mu\rangle$ self correlation induced by the scale dependent trispectrum in PBH forming inflationary models. 
 {With this aim, we focus our attention to the collapsed limit $k_+ \to 0$ of the curvature perturbation 4-pt function. In this limit, the non-Gaussian contribution \eqref{clmumuf} to $\langle\mu\mu\rangle$ reads as}
\begin{align}\label{clmumu1}
C_{l, {\rm NG}}^{\mu \mu} \simeq \frac{2.65}{\pi^{5}} \int \d \ln k_{+}\, j_{l}^2\left(k_{+} \chi_*\right)\int \d \ln q\,\d \ln p\,\,k_{+}^{3} q^{3}p^{3} \lim_{k_{+}\to 0}T_{\mathcal{R}}\left(q, p\right) \left[e^{-2q^2 / k^{2}_{D}(z)}\right]_{z_f}^{z_i} \left[e^{-2p^2 / k^{2}_{D}(z)}\right]_{z_f}^{z_i},
\end{align}
where  in the collapsed limit we use  $\vec{k}_1 \to -\vec{k}_2 \Rightarrow k_1 = k_2 \equiv q$ and $\vec{k}_3 \to -\vec{k}_4 \Rightarrow k_3 = k_4 \equiv p$. The expression \eqref{clmumu1} simplifies further if we consider symmetric folded kite configuration $q = p$. To see this fact explicitly, we first focus on the trispectrum in this configuration which reads as  
\beq\label{tscol}
 \lim_{k_{+}\to 0}T_{\mathcal{R}}\left(q, k_+\right) \simeq 4\, \underbrace{\tau^{\rm col}_{\rm NL}(q) E(q)^2}_{\equiv\,[\fr{6}{5}{f}^{\rm eff}_{\rm NL}(q)]^2 }  P_{\mathcal{R}}(\tau_q) P_{\mathcal{R}}(\tau_q)  P_{\mathcal{R}}(\tau_{k_{+}}),
\eeq
where we make use of the consistency relation \eqref{SY} and the definition \eqref{bssqg} of ${f}^{\rm eff}_{\rm NL}$.  The terms under the braces in \eqref{tscol} can be identified as the effective $\tau_{\rm NL}$, namely $\tau^{\rm eff}_{\rm NL}(q) \equiv {36}{f}^{\rm eff}_{\rm NL}(q)^2/25$, which  represents a scale-dependent generalization of the standard local type trispectrum with a constant $\tau_{\rm NL}$ \cite{Okamoto:2002ik,Boubekeur:2005fj}. 
\begin{figure}[t!]
\begin{center}
\includegraphics[scale=0.85]{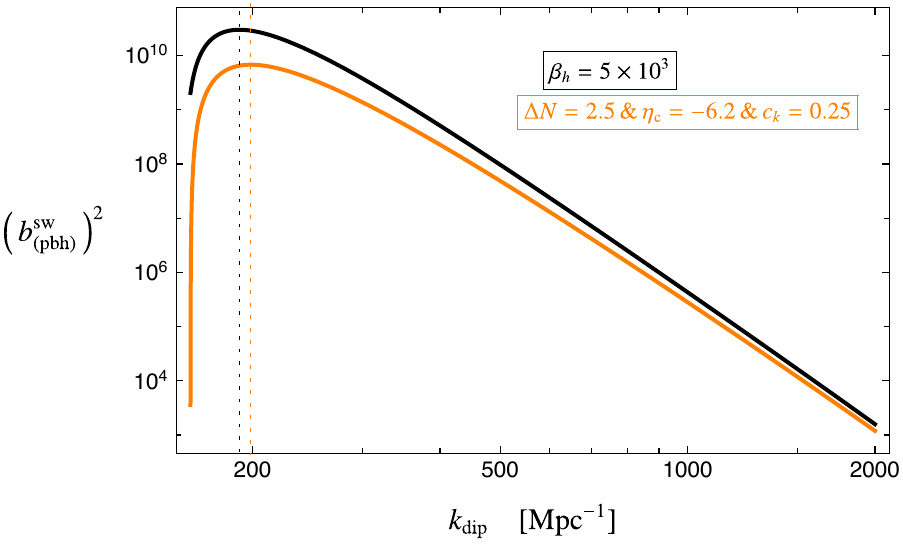} \includegraphics[scale=0.89]{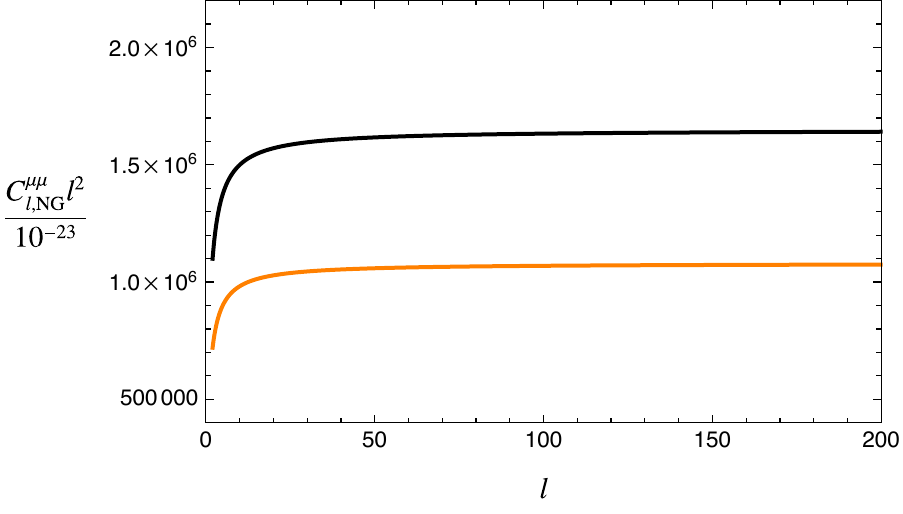} 
\end{center}
\caption{\it \small Left: Logarithmic plot of $b^{\rm sw\,\,^2}_{(\rm pbh)}$ as a function of the location of the dip feature for inflationary models that can generate a peak required  for PBH formation. Right: Scale invariance of $C^{\mu\mu}_l$ for the parameter choices shown in the left panel. The location of the dip feature in the power spectrum is taken as $k_{\rm dip} = 10^3\,{\rm Mpc^{-1}}$ for both curves, corresponding to $b^{\rm sw\,\,^2}_{(\rm pbh)} \simeq \{4.3 \times 10^{5}, 2.8 \times 10^{5}\}$ for the black and orange curves respectively.\label{fig:clmumu} }
\end{figure}
Inserting \eqref{tscol} into \eqref{clmumu1} and focusing on $q=p$ configuration, we can  describe the non-Gaussian $\langle \mu\mu \rangle$ selfcorrelation as 
\beq\label{CLmumuf}
C_{l,{\rm NG}}^{\mu \mu} \simeq 6.1 \times 10^{-24} \, \fr{2\pi}{l(l+1)} \left( b^{\rm sw}_{(\rm pbh)} \right)^2,
\eeq
where $b^{\rm sw}_{(\rm pbh)}$ is defined as in \eqref{bpbhsw}. This result implies that, in single-field  PBH inflationary scenarios, $\langle \mu \mu \rangle$  is scale invariant, \ie $l(l+1) 
C_l^{\mu\mu} = {\rm constant}$ similarly to a purely local form trispectrum \cite{Pajer:2012vz}. However,  its amplitude can be enhanced by a factor of $b^{\rm sw\,\,^2}_{(\rm pbh)} \gg 1$ compared to the latter. In particular, considering a $|b^{\rm sw}_{(\rm pbh)}| \simeq {\cal O}( 10^{2})$  which arises in  scenarios with a dip feature located at  $k_{\rm dip} \approx 10^{3}\, {\rm Mpc^{-1}}$, we can obtain a total enhancement of $\mathcal{O}(10^{5})$. Another point worth stressing is the fact that $C^{\mu\mu}_{l, {\rm NG}} >0$ for all multipoles,  as it should be clear from \eqref{CLmumuf}. We illustrate these points in Figure \ref{fig:clmumu} where we show $b^{\rm sw\,\,^2}_{(\rm pbh)}$ and $C^{\mu\mu}_l$ for representative scenarios that can produce a large peak in the power spectrum. The left panel of the Figure informs us that  $b^{\rm sw\,\,^2}_{(\rm pbh)}$ is maximal  at the same location \eqref{bswmin} where $b^{\rm sw}_{(\rm pbh)}$ has a minimum.

\subsection{Prospects of  detectability of $\mu$ distortion anisotropies}\label{S3p5}

We now develop Fisher forecasts to estimate the  detectability of the signals we investigated in Sections \ref{PmuT} and \ref{Pmumu}. We
 use a 1 $\times$ 1 Fisher information matrix for the parameter $b^{\rm sw}_{(\rm pbh)} \equiv b$ \cite{Zaldarriaga:1996xe}, 
\beq\label{FM}
F = \sum_{l} \sum_{X,Y} \fr{\partial C^{\mu X,S}_l}{\partial b} {\rm Cov}^{-1}(C^{\mu X}_l C^{\mu Y}_l)  \fr{\partial C^{\mu Y,S}_l}{\partial b}, 
\eeq
where the sum runs over $X,Y = \{T, \mu\}$; $S$ refers to the theoretical predictions we derived in eqs. \eqref{CLmuTF} and \eqref{CLmumuf} and ${\rm Cov^{-1}}$ is the inverse of the covariance matrix,  defined as \cite{Zaldarriaga:1996xe}
\beq\label{cov}
{\rm Cov}(C^{\mu X}_l C^{\mu Y}_l) = \fr{1}{2l+1}\left[(C^{\mu X,S}_l + C^{\mu X,N}_l)(C^{\mu Y,S}_l + C^{\mu Y,N}_l)+(C^{\mu\mu,S}_l + C^{\mu\mu,N}_l)(C^{XY,S}_l + C^{XY,N}_l)\right],
\eeq
where $N$ represents the experimental noise. Using  \eqref{FM} and \eqref{cov}, the signal to noise ratio (SNR) is given by $(S/N)^2 = b^2 F$. In the derivation of the  components of the covariance matrix, some simplifications can be made. First of all, note that for scales we are interested in $2 < l < \mathcal{O}(100)$, the experimental noise for $TT$ correlator can be neglected $C^{TT,S}_l \gg C^{TT,N}$, with  
\beq\label{CLTT}
C_l^{TT,S} = \fr{36\pi}{25} \int \d \ln k \, \mathcal{P}_{\mathcal{R}}(\tau_f,k) \Delta_l^2(k).
\eeq
Similarly, $\mu$ and $T$ instrumental noises are uncorrelated and therefore we can set $C^{\mu T, N}_{l} = 0$ in \eqref{cov}. On the other hand, as can be confirmed from \eqref{CLmumuf} and the discussion it follows, $\mu\mu$ correlation is dominated by the instrumental noise, \ie $C^{\mu\mu,S}_l \ll C^{\mu\mu,N}_l$, \ie for phenomenologically interesting values we are focusing where $(b^{\rm sw}_{(\rm pbh)})^2 \lesssim 10^{5}$. For a PIXIE like experiment  \cite{Kogut:2011xw}, this noise can be modeled as  $C^{\mu\mu, N}_l \simeq 4\pi \, \mu_{\rm min}^2\, e^{l^2/84^2}$ \cite{Pajer:2012vz} where $\mu_{\rm min}$ denotes the minimum detectable $\mu$ distortion signal. Finally, for the $X=Y =T$ component of the covariance matrix another simplification arise by noticing that  $(C^{\mu T}_l)^2 \ll C_l^{\mu\mu,N} C^{TT}_l$. This relation can be  confirmed noticing the SW limit of \eqref{CLTT}: $C^{TT,{\rm sw}}_l = 2\pi \mathcal{A}_s / (25 l(l+1))$ with $\mathcal{A}_s = 2.1 \times 10^{-9}$, \eqref{CLmuTF} and the relation for $C^{\mu\mu, N}_l$ above. In light of these arguments, SNR yields as 
\beq\label{StoNdef}
\left(\frac{S}{N}\right)^{2}=\sum_{l = 2}^{l_f}(2 l+1) \left[ \frac{\left(C_{l}^{\mu T,S}\right)^{2}}{C_{l}^{T T,S} C_{l}^{\mu \mu, N}} + 2\fr{C^{\mu\mu,S}_l}{C^{\mu\mu,N}_l}+ 2 \left(\fr{C^{\mu\mu,S}_l}{C^{\mu\mu,N}_l}\right)^2\right],
\eeq
where we carry the sum up to $l_{f} = 200$. To accurately estimate the first term in \eqref{StoNdef}, we require the knowledge of $C^{TT}_l$ in \eqref{CLTT} by taking into account the  transfer function $\Delta_l$. In this respect, \cite{Ganc:2012ae,Chluba:2016aln} found that the contribution from the first term in \eqref{StoNdef} corresponds to    $40 \% $ of the result  obtained  by adopting the SW limit, \ie by taking $\rho(l) \to 1$ in \eqref{CLmuTF},  and adopting $C^{TT,{\rm sw}}_l = 2\pi \mathcal{A}_s / (25 l(l+1))$ \footnote{It should be noted that we ignore the enhancement of the power spectrum in \eqref{CLTT} to arrive this expression. This is justified because the smallest scales we are interested in corresponds to $l_f =200$,  while  for the scenarios we are focusing in this work, the enhancement in the power spectrum occurs for scales corresponding to $l \gg l_f$.}. At the same time, the
 second and  third term in eq \eqref{StoNdef} can be evaluated directly from our results of section \eqref{Pmumu}. We obtain
\beq\label{son}
\fr{S}{N} \simeq 0.99 \times 10^{-3}  \left(\fr{10^{-8}}{\mu_{\rm min}}\right) |b^{\rm sw}_{(\rm pbh)} | + 2.15 \times 10^{-8}  \left(\fr{10^{-8}}{\mu_{\rm min}}\right)^2  |b^{\rm sw}_{(\rm pbh)}|^2,
\eeq

\begin{figure}[t!]
\begin{center}
\includegraphics[scale=0.87]{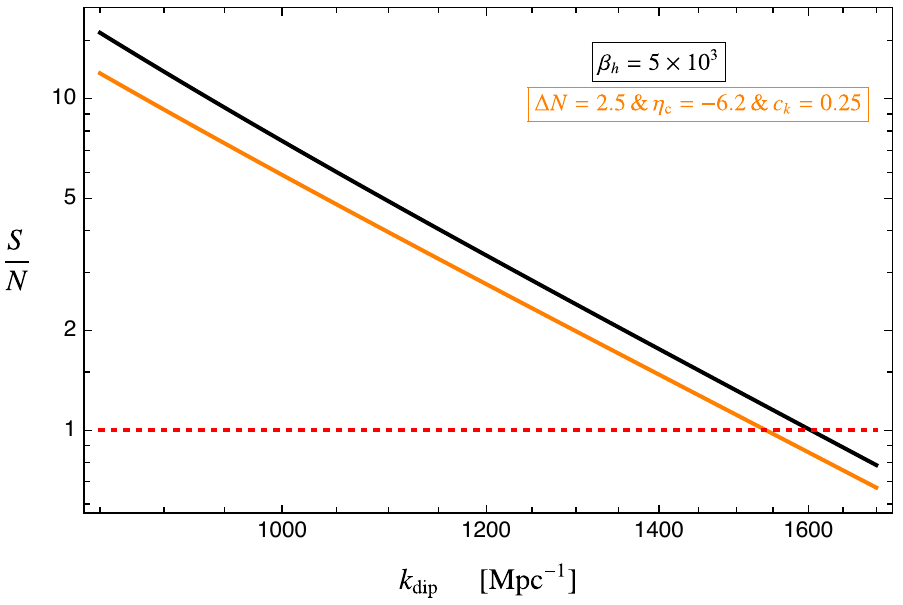} \includegraphics[scale=0.88]{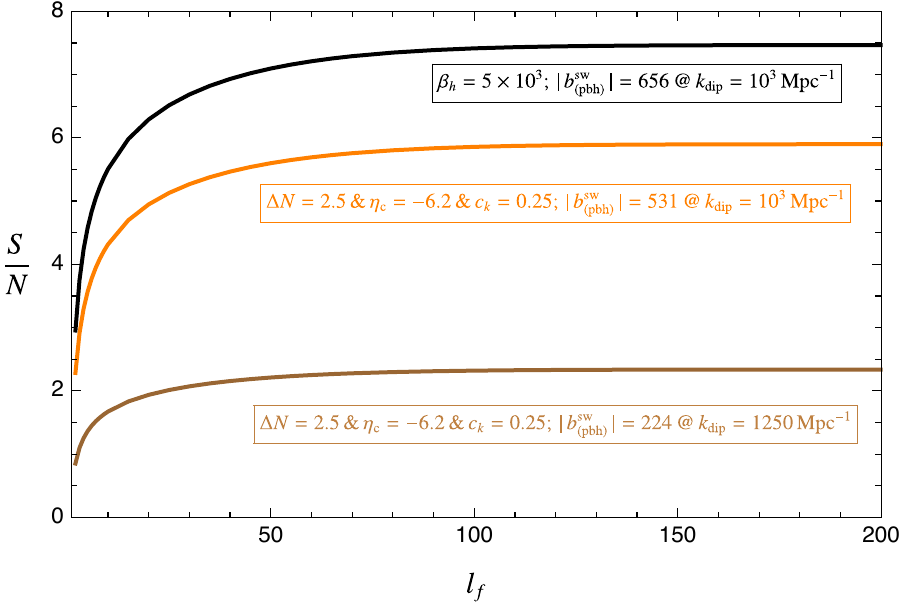} 
\end{center}
\caption{\it \small The cumulative signal to noise ratio \eqref{son} as a function of the dip location in the power spectrum for $k_{\rm dip} \gtrsim 830 \, {\rm Mpc}^{-1}$ (Left). SNR as a function of maximum multipole $l_f$ at fixed $k_{\rm dip}$ and hence $| b^{\rm sw}_{(\rm pbh)} |$ (Right).  \label{fig:snr}}
\end{figure}

\noindent where we normalized the minimum detectable distortion to $\mu_{\rm min}\approx 10^{-8}$, as relevant for a PIXIE-like experiment \cite{Chluba:2013pya}. In \eqref{son}, the first contribution corresponds to the sum of the first two terms in \eqref{StoNdef} which are of the same order of magnitude for   any  values of  $b$ and $\mu_{\rm min}$. In fact, the second term  is weighted by the cross component of the covariance matrix, and it  contributes twice as large as the $\mu T$ (the first) term in \eqref{StoNdef}. At the same time, the last term represents the contribution to SNR from $\mu \mu$ alone: it is generically sub-dominant compared to the first two terms in \eqref{StoNdef}, but it nevertheless provides  a non-negligible contribution. These results  imply that $\langle \mu \mu \rangle$ self correlations improve the prospects for the observability of distortion anisotropies.

Using \eqref{son}, we argue that an inflationary scenario with $ |b^{\rm sw}_{(\rm pbh)}| \gtrsim 989$ is detectable at $1\sigma$ level for a PIXIE like experiment. We note from \eqref{bpbhswf} that such values of $ |b^{\rm sw}_{(\rm pbh)}|$ can be obtained for PBH forming scenarios where $k_{\rm dip}$ lies close its smallest allowed values ($k_{\rm dip} \gtrsim 830\, {\rm Mpc}^{-1}$) dictated by the limits on average distortion $\mu \lesssim 10^{-5}$ (See figure \ref{fig:avmu}). Moreover, for an experimental design comparable to PRISM \cite{Andre:2013nfa} with $\mu_{\rm min} = 10^{-9}$, this situation can be improved since a smaller value  $| b^{\rm sw}_{(\rm pbh)} |\gtrsim 99$ is required for the detectability. Focusing our attention to the latter, in the left panel of Figure \ref{fig:snr} we present the SNR \eqref{son} in terms of the location of the dip scales $k_{\rm dip}$ allowed by the $\mu \lesssim 10^{-5}$ limit. As can be also inferred from \eqref{bpbhswf}, for $k_{\rm dip} \gtrsim 1600 \, {\rm Mpc}^{-1}$ SNR drops below unity. However it satisfies $S/N > 1$ for the allowed region of $830 \, {\rm Mpc}^{-1} \lesssim k_{\rm dip} \lesssim 1600 \, {\rm Mpc}^{-1}$. In the right panel, we show cumulative SNR as a function of $l_f$ for three different scenarios defined in this range of scales together with their corresponding $| b^{\rm sw}_{(\rm pbh)} |$. (Recall that $l_f$ corresponds to the upper
limit of the sum in eq \eqref{StoNdef}.) This implies  that our SNR estimate does not improve significantly and saturates for $l_f \gtrsim 100$. 

\medskip
\noindent
{\bf $\mu$-distortion anisotropies as a probe of PBH populations.} We conclude our analysis with 
some implications of our findings for PBH populations.
In inflationary scenarios where the curvature  power spectrum has a pronounced peak of order $\Pi_{\rm tot} \simeq 10^7$ located at   wave-number $k_{\rm peak} \simeq 100 k_{\rm dip} \gg k_{\rm CMB} = 0. 05 \, {\rm Mpc}^{-1}$, PBHs may have formed during the radiation dominated era upon horizon re-entry of modes whose wave-number is  comparable with the peak scale 
  \cite{Hawking:1971ei,Carr:1974nx}. Assuming that the power spectrum is sufficiently peaked, we can relate the mass of the PBH  today to the location of the dip scale in the power spectrum as \cite{Unal:2020mts,Ozsoy:2018flq},
\beq\label{Mpbh}
M_{\mathrm{pbh}, 0} =\mathcal{A}\, \mathcal{M}\, M_{\rm pbh,f} \simeq \mathcal{A}\, \mathcal{M}\,\left(\fr{\gamma}{0.2}\right)\left(\frac{k_{\rm dip}}{10^{4}\, \mathrm{Mpc}^{-1}}\right)^{-2} 2.4 M_{\odot},
\eeq
where $\mathcal{A}$ and $\mathcal{M}$ indicate the amount mass gain that can arise due to accretion and merger effects for the corresponding PBH seed mass $M_{\rm pbh,f}$ at the time of formation. $\gamma$ is the ratio of the PBH mass to the mass within the causal horizon and can take values between $\gamma = 0.2$ \cite{Carr:1975qj,Carr:2009jm} and $\gamma = 0.8$ \cite{Germani:2018jgr} depending on the assumptions about PBH formation in the radiation dominated era. 

Considering the phenomenologically interesting values of dip scales $830 \, {\rm Mpc}^{-1} \lesssim k_{\rm dip} \lesssim 1600 \, {\rm Mpc}^{-1}$ we identified above, eq \eqref{Mpbh} implies that PBHs with $M_{\mathrm{pbh}, 0} =  10-100 M_{\odot}$ can be probed by $\mu$ distortion anisotropies assuming negligible accretion and merger coefficients:  $\mathcal{A},\mathcal{M} \to 1$. Taking into account these effects within the ranges $10^{5}\geq \mathcal{A}\geq 1$, $10^{5}\geq\mathcal{M}\geq 1$ \cite{Garcia-Bellido:2017aan,Unal:2020mts}, a detection of $\mu$ distortion anisotropies through $\langle\mu T\rangle$ and $\langle \mu\mu\rangle$ angular correlators can be therefore considered as a useful tool to distinguish astrophysical vs primordial the origin of super massive black holes (SMBH) with masses $M_{\mathrm{pbh}, 0} = 10^6 - 10^{9}\, M_{\odot}$ today.

\section{Discussion}\label{SecDis}
In single field-inflationary models that are capable of generating PBH populations, the power spectrum of curvature perturbation has interesting universal features such as the presence of a pronounced dip. Focusing on the heuristic approach introduced in \cite{Tasinato:2020vdk} and gradient expansion formalism \cite{Leach:2001zf} (see sections \ref{sec_heur1} and \ref{S2}), we explicitly demonstrated that the position of the dip in momentum space is uniquely determined by the global enhancement $\Pi \simeq 10^{7}$ of the power spectrum via $k_{\rm dip} \simeq \Pi^{-1/4} k_{\rm peak} \ll k_{\rm peak}$, implying its occurrence on scales much larger than the peak scale associated with PBH formation. More importantly, 
 in sections \ref{SBS} and \ref{STS} we analyzed  consistency relations for $n$-point correlators ($n =3,4$) of curvature perturbation in the vicinity of the  dip feature.  We found that non-Gaussianity parameters satisfy the  conditions $f^{\rm sq}_{\rm NL} = 5(1-n_s)/12$ and $\tau^{\rm col}_{\rm NL} = (6 f^{\rm sq}_{\rm NL}/5)^2$ in a non-trivial  scale dependent manner, allowing us to derive a new set of consistency conditions in terms of the  the global enhancement $\Pi \simeq 10^{7}$ in the power spectrum and relate their scale dependence to its slope. In scenarios where the dip feature lies within the scale range \eqref{muera} where $\mu$-distortions are generated, the characteristic scale dependence of such $n$-point correlators offers us a unique chance to probe the underlying PBH formation mechanism at relatively large scales through the CMB spectral $\mu$-distortion anisotropies.

In fact, 
developing upon the ideas first presented in \cite{Ozsoy:2021qrg},  we explored the implications of the consistency conditions for the bispectrum and trispectrum on the cross correlation between spectral distortions and temperature anisotropies $\langle \mu T \rangle$ and distortion self correlations $\langle \mu \mu\rangle$. In this context, in section \ref{PmuT}, we studied $\langle \mu T \rangle$ angular correlator induced by the squeezed limit bispectrum derived from the consistency relation $f^{\rm sq}_{\rm NL} = 5(1-n_s)/12$. We found that the pronounced characteristic scale dependence of the bispectrum can alter the amplitude and overall multipole dependence of $\langle \mu T \rangle$ significantly with respect to more standard cases (See section \ref{PmuT}). These results confirm the findings obtained earlier in \cite{Ozsoy:2021qrg} and put them in a firmer footing through the use of consistency condition. In Section \ref{Pmumu}, utilizing the consistency relation $\tau^{\rm col}_{\rm NL} = (6 f^{\rm sq}_{\rm NL}/5)^2$, we studied for the first time the influence of the enhanced collapsed limit trispectrum present around the dip scale and found that it induces sizeable distortion self-correlations $\langle \mu \mu \rangle$ that are scale invariant: $l (l+1) C^{\mu\mu}_l = {\rm constant}$. Including the information that we can gain from the $\mu$ self-correlations, in section \ref{S3p5}, we then showed that the prospects of detectability of $\mu$ distortion anisotropies are enhanced compared to considering $\mu T$ correlations alone \cite{Ozsoy:2021qrg}. In particular, we found that for phenomenologically allowed and interesting values of dip location in the range $830 \, {\rm Mpc}^{-1} \lesssim k_{\rm dip} \lesssim 1600 \, {\rm Mpc}^{-1}$, $\mu$-distortion anisotropies -- induced by non-Gaussian consistency relations -- should be observable for a PIXIE or PRISM like experimental design. 

Considering the relation between the dip and peak scale in the power spectrum $k_{\rm peak} \simeq 100 k_{\rm dip}$, spectral distortion anisotropies we derived in this work can be utilized to identify the formation mechanism of BHs with masses $M_{\rm pbh,0} \simeq 10-100\, M_{\odot}$ today, and/or SMBHs with $M_{\rm pbh,0} \simeq 10^{6}-10^{9}\, M_{\odot}$ taking into account strong accretion and merger effects (see Section \ref{S3p5}). Furthermore, the properties  we studied in this work can also be also considered as a useful tool for discriminating inflationary models of PBH formation as in some scenarios based on particle production \cite{Bugaev:2013fya,Garcia-Bellido:2016dkw,Domcke:2017fix,Ozsoy:2020ccy}, the dip feature is not present and therefore unlikely to produce interesting $\mu$-distortion anisotropies at large scales.  

This work can be extended in a few directions. First of all, our analysis on the squeezed limit bispectrum and the collapsed limit trispectrum does not include finite but sub-leading corrections of order $\mathcal{O}(k_+/q)$ in terms of the soft momenta $k_{+}$. It would be interesting to include such corrections to  to investigate the impact of \emph{not so squeezed and collapsed limit} $n$-point correlators of curvature perturbation on the $\mu T$ and $\mu \mu$ angular correlators. Finally, it would be interesting to extend our analysis to derive predictions on the $\mu$ distortion anisotropies for scenarios including multiple scalar fields \cite{Palma:2020ejf,Fumagalli:2020adf,Braglia:2020taf}. Such an analysis would be helpful in comparing the general single field predictions we derived in this work and guide us towards a better understanding for the formation mechanism of BHs with astrophysically relevent masses. 
\subsection*{Acknowledgments}
We would like to thank Enrico Pajer and  Konstantinos Tanidis for useful discussion pertaining this work. 
The work of O\"O is supported by the European Structural and Investment Funds and the Czech Ministry of Education, Youth and Sports (Project CoGraDS-CZ.02.1.01/0.0/0.0/15003/0000437). GT is partially funded by the STFC grant ST/T000813/1. 
\begin{appendix}
\section{The power spectrum $\mathcal{P}_\mathcal{R}(\tau_k)$ and fractional velocity $v_{\mathcal{R}}$}\label{AppA}

For the initial slow-roll phase, the standard solution for the curvature perturbation that reduces to the standard Bunch Davies vacuum can be written as \cite{Ozsoy:2019lyy},
\beq\label{cpsr}
\mathcal{R}^{\rm sr}_k = \fr{i H}{\Mp} \fr{e^{-ik\tau}}{\sqrt{4\epsilon_{\rm sr} k^3}}~(1+ik\tau),\quad\quad\quad \tau_k/\tau_0 > 1,
\eeq
with $\epsilon_{\rm sr} \ll 1$ is the slow-roll parameter which we assume to be constant adopting $\eta = \dot{\epsilon}/\epsilon H \to 0$. Using the solution \eqref{cpsr}, the real and imaginary part of $v_{\mathcal{R}}$ \eqref{fr} can be derived as
\beq\label{vsr}
v^{R}_{\mathcal{R}}(c_k) = -\fr{c_k^2}{3(1+c_k^2)},\quad
v^{I}_{\mathcal{R}}(c_k)  = -\fr{c_k^3}{3(1+c_k^2)},\quad\quad \tau_k/\tau_0 > 1,
\eeq
where $c_k$ is defined as in \eqref{defOck}. Notice that the imaginary part of $v_{\mathcal{R}}$ in \eqref{vsr} includes an extra factor of $c_k$ compared to the real part. We note that unless $c_k =1$, this translates into an extra suppression for the imaginary part of the fractional velocity and hence the imaginary part of $\alpha_k$ \eqref{ai} as we will show explicitly below. On the other hand, using \eqref{cpsr}, the power spectrum evaluated at around horizon crossing is given by
\beq\label{pstauk}
\mathcal{P}_{\mathcal{R}}(\tau_k) = \fr{k^3}{2\pi^2}|\mathcal{R}_k(\tau_k)|^2 = \fr{H^2}{8 \pi^2\epsilon_{\rm sr}\Mp^2} \left(1 + c_k^2\right),\quad\quad\quad \tau_k/\tau_0 > 1.
\eeq 

Next, we need to determine $\mathcal{R}(\tau_k)$, $v_{\mathcal{R}}$ and $\mathcal{P}_{\mathcal{R}}(\tau_k)$ in the non-attractor era, \ie for $\tau_k/\tau_0 < 1$. This was done in \cite{Ozsoy:2021qrg} using a matching procedure for $\mathcal{R}_k$ and its derivative at the transition time $\tau = \tau_0$. The resulting fractional velocity in the non-attractor era, \ie for $\tau_k/\tau_0 < 1$ is given by \cite{Ozsoy:2021qrg} 
\bea
\label{fvintr}v^{R}_{\mathcal{R}}(\tau) &=&-\fr{y}{3}\left[\fr{f_1f_3-y_0\left(f_1f_4+f_2f_3\right)+y_0^2\left(f_1f_3+f_2f_4\right)}{f_3^2-2y_0f_3f_4+y_0^2\left(f_3^2+f_4^2\right)}\right],\\\nn\\
\label{fvinti}v^{I}_{\mathcal{R}}(\tau) &=&-\fr{y}{3}\left[\fr{y_0^2\left(f_1f_4-f_2f_3\right)}{f_3^2-2y_0f_3f_4+y_0^2\left(f_3^2+f_4^2\right)}\right],
\eea
where we defined $y\equiv-k\tau$ and the functions $f_{n} = f_n(y,y_0,\nu)$, for $n=1,2,3,4$ in terms of the Bessel function of the first and second kind as
\bea\label{ffrel}
\nn f_1(y,y_0,\nu) &=& J_{\nu-1}(y_0)Y_{\nu-1}(y)-Y_{\nu-1}(y_0)J_{\nu-1}(y),\\
f_2(y,y_0,\nu) &=& J_{\nu}(y_0)Y_{\nu-1}(y)-Y_{\nu}(y_0)J_{\nu-1}(y),
\eea
satisfying the following relations $f_4 = f_1(y,y_0,\nu+1), f_3 = -f_2(y_0,y,\nu)$ with $\nu = (3+\eta_{\rm c})/2$. 
The continuity of the real and the imaginary part of the fractional velocity in passing from slow-roll to non-attractor era can be confirmed explicitly from the eqs in \eqref{fvintr} and \eqref{fvinti} at $\tau_k=\tau_0$. Finally, the power spectrum evaluated at $\tau = \tau_k$ for modes that leave the horizon during the non-attractor phase ($\tau_k/\tau_0 < 1$) is given by  \cite{Ozsoy:2021qrg}
\bea\label{pstaukna}
\mathcal{P}_{\mathcal{R}}(\tau_k)  = \fr{H^2}{8\pi^2\epsilon_{\rm sr}\Mp^2}\fr{c_k ^{2\nu}\pi^2}{4}\left(\fr{k}{\mathcal{H}_0}\right)^{2-2\nu}\bigg[f_3^2-2y_0f_3f_4+y_0^2(f_3^2+f_4^2)\bigg]_{\tau= \tau_k}.
\eea

Similarly, the continuity of the power spectrum can be confirmed explicitly by evaluating \eqref{pstaukna} at the transition point $\tau_k=\tau_0$ at which it reduces to \eqref{pstauk}.

\section{The functions $D(\tau_k)$,  $F_k(\tau_k)$ and the enhancement factor $\alpha_k$}\label{AppB} 

For the two phase background model parametrized by the pump field profile in eq. \eqref{zsol1}, scale dependent functions $D(\tau_k)$, \,$F(\tau_k)$ (see eqs. \eqref{Dint} and \eqref{Fint}) are calculated in \cite{Ozsoy:2019lyy} for modes that exit during the initial slow-roll era ($k/\mathcal{H}_0 < c_k$) and in \cite{Ozsoy:2021qrg} for mode exit during the non-attractor era ($e^{\Delta N}>k/\mathcal{H}_0 > c_k$). In particular, for mode exit during the initial slow-roll stage, these functions are found to be
\begin{align}
\label{dafa1}
 D(\tau_k) &= 1 - \fr{3}{(\eta_{\rm c}+3) c_k^3} \left[e^{-(\eta_{\rm c}+3)\Delta N}+\fr{\eta_{\rm c}}{3}\right]\left(\fr{k}{\mathcal{H}_0}\right)^{3},\\\nn
F(\tau_k)\,k^2 &=\fr{c_k^2}{6} -\left[\fr{ \eta_{\rm c}\,e^{-(\eta_{\rm c}+3)\dn}}{(\eta_{\rm c}+3)(\eta_{\rm c}+1)}+\fr{\eta_{\rm c}}{2(\eta_{\rm c}+3)}\right]\left(\fr{k}{\mathcal{H}_0}\right)^{2}+\left[\fr{e^{-(\eta_{\rm c}+3)\dn}}{(\eta_{\rm c}+3) c_k}+\fr{\eta_{\rm c}}{3(\eta_{\rm c}+3)c_k}\right]\left(\fr{k}{\mathcal{H}_0}\right)^{3},
\end{align}
where $\tau_f/\tau_0 \equiv x_f = e^{-\dn}$ with $\Delta N$ denoting the duration of the non-attractor era. On the other hand, for modes that exit during the non-attractor era, $e^{\Delta N}>k/\mathcal{H}_0 > c_k$ they are given by

\begin{align}\label{dafa2}
\nn D(\tau_k) &=-\fr{3}{\eta_{\rm c}+3}\ -\fr{3\, e^{-(\eta_{\rm c}+3)\dn}}{(\eta_{\rm c}+3)c_k^{\eta_{\rm c}+3}}\,\left(\fr{k}{\mathcal{H}_0}\right)^{(\eta_{\rm c}+3)},\\
{F(\tau_k)\,}k^2 &= \fr{e^{-(\eta_{\rm c}+3)\dn}}{(\eta_{\rm c}+1)(\eta_{\rm c}+3)c_k^{\eta_{\rm c}+1}}\,\left(\fr{k}{\mathcal{H}_0}\right)^{(\eta_{\rm c}+3)} + \fr{c_k^2}{2(\eta_{\rm c}+3)} -\frac{e^{-2 \Delta N}}{2\left(\eta_{c}+1\right)} \left(\fr{k}{\mathcal{H}_0}\right)^{2}.
\end{align}
Together with the fractional velocities $v^R_{\mathcal{R}}$, $v^I_{\mathcal{R}}$ we found in \eqref{vsr}, \eqref{fvintr} and \eqref{fvinti}, one can make use of the formulas \eqref{dafa1} and \eqref{dafa2} to describe full spectral behavior of the enhancement factor $\alpha_k$ from large to small scales in a continuous way by utilizing the formulas \eqref{ar} and \eqref{ai}.
\subsection{$\alpha_k$ for mode exit during the initial slow-roll era.}\label{AppB1} 
In this subsection, using the formulas of the previous section, we provide an expression for $\alpha_k$ for $k/\mathcal{H}_0 < c_k \leq 1$ that we utilize in the main text extensively.

We begin by noticing that for inflationary scenarios that contains a transient $\Delta N \sim \mathcal{O}(1)$ slow-roll violating $\eta_{\rm c} \leq -6$ phase, the terms in the square brackets of \eqref{dafa1} are dominated by the exponential factors so we can further simplify the quantities $D(\tau_k)$ and $F(\tau_k) k^2$. Then noting the definitions \eqref{ar},\eqref{ai} and \eqref{vsr}, the real and imaginary part of the enhancement factor can be parametrized as  
\begin{align}
\label{appak}
\alpha^R_k &\simeq \alpha^R_{(0)} \left[ 1 - \beta \left(\fr{k}{\mathcal{H}_0}\right)^{2} + \fr{c_k\,(\eta_{\rm c} + 1)}{(1+c_k^2)\,\eta_{\rm c}}\beta \left(\fr{k}{\mathcal{H}_0}\right)^{3}  \right],\\\nn
\alpha^I_k &\simeq \alpha^I_{(0)} - \alpha^R_{(0)}  \fr{\,(\eta_{\rm c} + 1)}{(1+c_k^2)\,\eta_{\rm c}}\beta  \left(\fr{k}{\mathcal{H}_0}\right)^3,
\end{align}
where we defined an exponentially large number $\beta = \beta(\eta_{\rm c}\ \Delta N, c_k)$, $ \alpha^R_{(0)}$ and $ \alpha^I_{(0)}$ as 
\begin{align}\label{misc}
\,\,\alpha^R_{(0)} = 1-\fr{c_k^2}{6} - \fr{c_k^2}{3(1+c_k^2)},\,\,\,\,\alpha^I_{(0)} = -\fr{c_k^3}{3(1+c_k^2)},\,\,\,\, \beta = -\fr{\eta_{\rm c}\,e^{-(3+\eta_{\rm c})\Delta N}}{\alpha^R_{(0)}\,(\eta_{\rm c} +1)(\eta_{\rm c} +3)}.
\end{align}
Note that $\alpha^R_{(0)}$ and $\alpha^I_{(0)}$ parametrize the initial values of the real and imaginary part of $\alpha_k$ in the large scale limit $k\to 0$, respectively. The parameter $\beta \gg 1$ we introduced is instead characterize the amplification that higher order terms in the gradient expansion of $\alpha_k$ obtains for a non-attractor era $\eta_{\rm c} \leq -6$ with a duration of $\Delta N$ e-folds.  As it is clear from \eqref{misc}, for $\Delta N \to 0$, $\beta \to \mathcal{O}(1)$, higher order terms in the gradient expansion becomes insignificant for providing  an enhancement of the power spectrum at small scales $k/\mathcal{H}_0 \to 1$.
 \begin{figure}[t!]
\begin{center}
\includegraphics[scale=0.88]{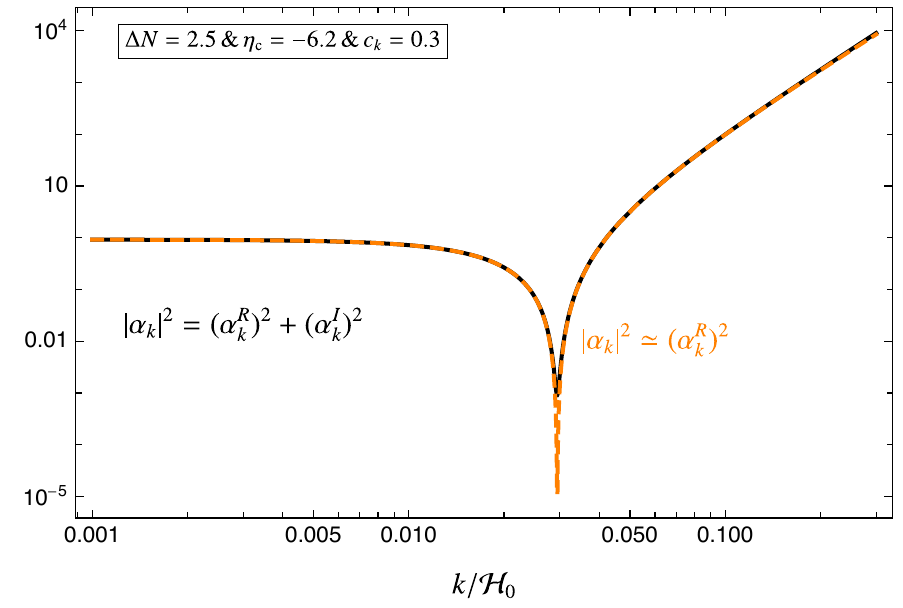}\includegraphics[scale=0.88]{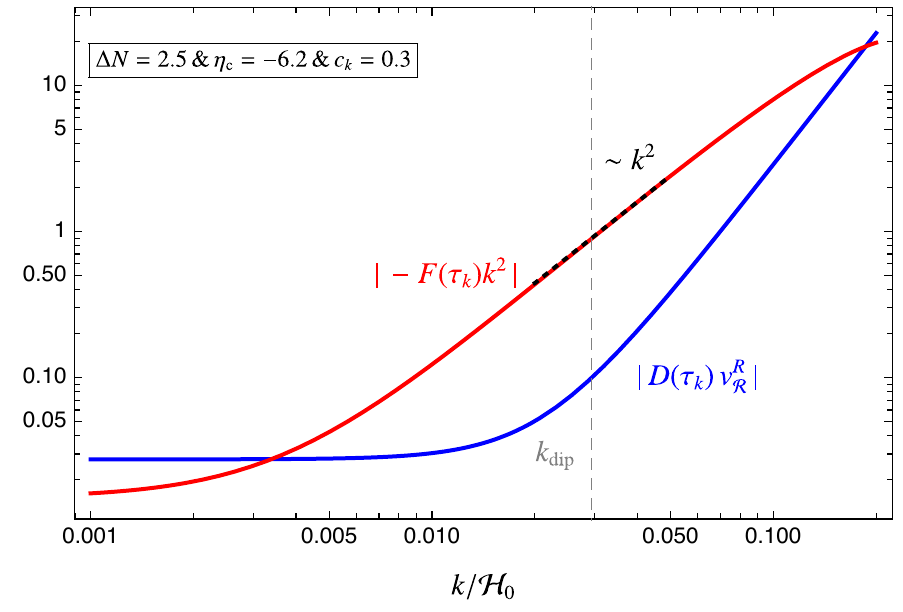}
\end{center}
\caption{\it \small The exact behavior of $|\alpha_k|^2$ (black solid) and the approximation $|\alpha_k|^2 \simeq (\alpha_k^R)^2$ \eqref{akrapp} (orange dashed) from large to small scales associated with the initial slow-roll era, $k/\mathcal{H}_0 < c_k$ (Left). Comparison of the individual terms in $\alpha^{R}_k = 1 + D(\tau_k) v^R_{\mathcal{R}} - F(\tau_k) k^2$ (Right).\label{fig:RvsI}}
\end{figure}

Finally, for the purpose of calculating scale dependent $f^{\rm sq}_{\rm NL}$ and $\tau^{\rm col}_{\rm NL}$ using the consistency conditions we derived in Section \ref{SBS} and \ref{STS}, we derive a simplified expression for the enhancement factor $|\alpha_k|^2$ in the gradient expansion formalism. As we discuss explicitly below, for modes associated with the initial era we can  neglect the imaginary part of the enhancement factor $\alpha^{I}_k \ll \alpha^{R}_k$ and hence $|\alpha^R_k|^2 \simeq (\alpha^R_k)^2$. Furthermore, using \eqref{appak},   truncating the resulting $(\alpha^R_k)^2$ to fifth order in $k$ provides a very accurate description to the exact result: 
\begin{align}\label{akrapp}
\nn |\alpha_k|^2 & \simeq (\alpha^R_{(0)})^2 \left[ 1 - 2\beta \left(\fr{k}{\mathcal{H}_0}\right)^{2} + \fr{2 c_k\,(\eta_{\rm c} + 1)}{(1+c_k^2)\,\eta_{\rm c}}\beta \left(\fr{k}{\mathcal{H}_0}\right)^{3}  + \beta ^2\left(\fr{k}{\mathcal{H}_0}\right)^{4} -\fr{2 c_k\,(\eta_{\rm c} + 1)}{(1+c_k^2)\,\eta_{\rm c}}\beta^2\left(\fr{k}{\mathcal{H}_0}\right)^{5} \right],\\
&\simeq (\alpha^R_k )^2. 
\end{align}
Using the relation between scales \eqref{kdip}, we can rewrite \eqref{akrapp} in a compact way as
\beq\label{akrappf}
 |\alpha_k|^2 \simeq (\alpha^R_k )^2 \simeq \sum^5_{n=0} c^{(n)}_{|\alpha|^2} \left(\fr{k}{k_{\rm dip}}\right)^n,
\eeq
where in terms of the parametrization \eqref{misc} above, the coefficients of the sum are given by
\beq\label{casq}
c^{(0)}_{|\alpha|^2} = (\alpha^R_{(0)})^2,\,\,c^{(1)}_{|\alpha|^2}=0,\,\,\,\,c^{(2)}_{|\alpha|^2} = -2 c^{(4)}_{|\alpha|^2} =  -2\, (\alpha^R_{(0)})^2,\,\,\,\,c^{(3)}_{|\alpha|^2} = - c^{(5)}_{|\alpha|^2}=\fr{2 \,  (\alpha^R_{(0)})^2\,c_k\,(\eta_{\rm c} + 1)}{(1+c_k^2)\,\eta_{\rm c}\sqrt{\beta}}.
\eeq 

\smallskip
\noindent{\bf Comparison between $\alpha^R_k $ and $\alpha^I_k$.} Another important point we will use repeatedly in the main text is the fact that imaginary part of $\alpha_k$ is subdominant compared to its real counterpart for modes associated with the initial slow-roll era. To see this, we can recall \eqref{ai} and notice that $\alpha^I_k = D(\tau_k) v^I_{\mathcal{R}} \ll D(\tau_k) v^R_{\mathcal{R}} \subset \alpha^R_k$ which holds as long as we assume $c_k \ll 1$ implying $v^R_{\mathcal{R}}\gg v^I_{\mathcal{R}}$ as can be inferred from \eqref{vsr}. We would like to remind that the choice $- k\tau_k = c_k \ll 1$ is a natural one considering that in the gradient expansion formalism $c_k \ll 1$ simply implies that all the $k$ modes we focus are outside the horizon at the initial time $\tau_k$. To explicitly check the statements above, we compare the exact $|\alpha_k|^2$ quantity with the approximate $|\alpha_k|^2 \simeq (\alpha_k^R)^2$ relation \eqref{akrapp} in Figure \ref{fig:RvsI}. We observe that two expressions match really well for all scales except at the position of the dip $k_{\rm dip}$ where the approximate relation $|\alpha_k|^2 \simeq (\alpha_k^R)^2$ generically leads to a more pronounced dip feature compared to the full expression $|\alpha_k|^2 = (\alpha_k^R)^2 +  (\alpha_k^I)^2$.

In this work, except for the part where we study the global shape of the power spectrum (see Section \ref{SPS}), we will utilize the approximation  $\alpha_k \simeq \alpha_k^R$ especially when we derive the scale dependence of bispectrum and trispectrum around the dip feature $k_{\rm dip}$ (See \eg Section \ref{SBS} and \ref{STS}). We note however that whenever $\alpha_k$ appears in the denominator of an expression, we kept its full expression $\alpha_k = \alpha^R_k + i \alpha^I_k$ to avoid unphysical divergences that might appear around especially the dip scale. 

\smallskip
\noindent{\bf $\alpha^R_k $ around the dip scale.} In addition to the approximation we mentioned above, one can make further simplifications for the real part of the enhancement factor around the location of the dip feature of the power spectrum. In particular, an analysis of the individual terms that constitutes $\alpha^R_k$ \eqref{ar} for $k/\mathcal{H}_0 < c_k$ reveals that $D(\tau_k) v^R_{\mathcal{R}} \ll F(\tau_k) k^2$ as can be also observed from the right panel of Figure \ref{fig:RvsI}. Therefore around the $k \sim k_{\rm dip}$, we can simply approximate $\alpha^{R}_k$ as
\beq\label{arapp}
\alpha^R_k \simeq 1 - F(\tau_k) k^2. 
\eeq
The relation above automatically implies that $(\alpha^{R}_k)' = - (F(\tau_k) k^2)' $ where prime denotes a derivative with respect to the normalized wave-number $k/\mathcal{H}_0$. Recall that for the discussion that follows equation \eqref{1mns} in section \ref{SBS}, we require $- (k/\mathcal{H}_0) (\alpha^{R}_k)' = - (F(\tau_k) k^2)'  \simeq 2 F(\tau_k) k^2 $ to prove that Maldacena's consistency condition $f_{\rm NL} = 5 (1-n_s)/12$ is satisfied around the dip feature. To show this explicitly, notice from the Figure \ref{fig:RvsI} that $F(\tau_k) k^2$ in \eqref{dafa1} can be truncated to second order in the $k$ expansion around the dip feature. Keeping this in mind, using \eqref{dafa1} up to second order in $k$, we can therefore write
\beq\label{arpapp}
- \fr{k}{\mathcal{H}_0}\, (\alpha^{R}_k)' =  \fr{k}{\mathcal{H}_0}\,  (F(\tau_k) k^2)'  \simeq 2 \left(F(\tau_k) k^2 - \fr{c_k^2}{6}\right) \simeq 2 F(\tau_k) k^2,
\eeq
where in the last equality we ignored the term proportional to $c_k^2 \ll 1$.
Indeed, we utilize the approximation \eqref{arpapp} (together with \eqref{arapp}) in the derivation of the general, scale dependent consistency condition between bispectrum and power spectrum (See section \ref{SBS}). 

\subsection{$\Pi(k)$ for mode exit during the initial slow-roll era}

We now provide a simplified power law expression for the scale dependence of the enhancement factor $\Pi(\kappa)$ of Section \ref{sec_heur1} for modes that exit the horizon in the initial slow-roll stage $\kappa = k/k_{\rm na} < 1$. For this purpose, we find it sufficient to adopt a small $\kappa$ expansion of the expression \eqref{pieps0} up to $k^6$ order which yields
\beq\label{piapp}
\Pi(k) \simeq 1 - \fr{2\beta_h}{3}\left(\fr{k}{k_{\rm na}}\right)^2 + \left[\fr{2\beta_h}{5} + \fr{\beta_h^2}{9}\right]\left(\fr{k}{k_{\rm na}}\right)^4-\fr{\beta_h^2}{45}\left(\fr{k}{k_{\rm na}}\right)^6.
\eeq  
By explicitly comparing with \eqref{pieps0}, we found that \eqref{piapp} reproduces the exact scale dependence of the power spectrum very accurately for modes associated with the initial slow-roll era, $k/k_{\rm na} < 1$. Using the relation \eqref{dippos1} between the scales, we can rewrite \eqref{piapp} in terms of the dip scale $k_{\rm dip}$ as
\beq\label{pidip}
\Pi(k) \simeq \sum^{6}_{n=0} c^{(n)}_{\Pi} \left(\fr{k}{k_{\rm dip}}\right)^n,
\eeq
where in terms of the only free parameter $\beta_h$ of the heuristic approach the coefficients $c_{\Pi}$ are given by
\beq\label{cpi}
c^{(0)}_{\Pi} = 1,\,\,\,\, c^{(1)}_{\Pi}=0,\,\,\,\,\, c^{(2)}_{\Pi} = -2,\,\,\,\, c^{(4)}_{\Pi} = \left[1 + \fr{18}{5\beta_h}\right],\,\,\,\, c^{(6)}_{\Pi} = -\fr{3}{5\beta_h},\,\,\,\, c^{(3)}_{\Pi}=c^{(5)}_{\Pi} = 0.
\eeq
\section{Squeezed limit $f^{\rm eff}_{\rm NL}$ and average $\mu$ distortions}\label{AppC}
Building upon our results in the previous appendix, we now would like to identify the effective non-linearity parameter defined in \eg \eqref{bssqg} for both approaches (heuristic vs gradient expansion) we focus in this work.  Using the consistency condition $f^{\rm sq}_{\rm NL} = 5 (1-n_s)/12$ and noting the definition \eqref{bssqg}, we can describe the squeezed limit ($k_+ \to 0$) $f^{\rm eff}_{\rm NL}$ in terms of small momenta $q$ as 

\beq\label{fnleffgapp}
{f}^{\rm eff}_{\rm NL}(q) \simeq  -\fr{5}{12} q  \fr{\d E(q)}{\d q}, 
\eeq
where we denote the enhancement factors collectively as $E(q) = \{\Pi(q), |\alpha_q|^2\}$ for heuristic and gradient approaches respectively. Using the expressions \eqref{akrappf} and \eqref{pidip} we can then describe scale dependent $f^{\rm eff}_{\rm NL}$ in terms of simple power law expansion. For both approaches we undertake, this power law expression can be expressed collectively as 
\beq\label{fnleffgappf}
{f}^{\rm eff}_{\rm NL}(q) = \fr{5}{3}\sum^{n_f}_{n = 2} \tilde{c}^{(n)}_{\rm E} \left(\fr{q}{q_{\rm dip}}\right)^n, 
\eeq
where $n_f =\{5,6\}$ for the gradient/heuristic approach respectively. In terms of the model parameters that characterize  the both approaches, the coefficients of this expansion are given by
\beq\label{cfnlgrad}
\tilde{c}^{(2)}_{\rm |\alpha|^2}  = - \tilde{c}^{(4)}_{\rm |\alpha|^2} = (\alpha^R_{(0)})^2,\,\,\, \tilde{c}^{(3)}_{\rm |\alpha|^2} = -\fr{5}{3} \tilde{c}^{(5)}_{\rm |\alpha|^2} = -\fr{3}{2}\fr{(\alpha^R_{(0)})^2\,c_k\,(\eta_{\rm c} + 1)}{(1+c_k^2)\,\eta_{\rm c}\sqrt{\beta}},
\eeq
in the gradient expansion formalism and
\beq\label{cfnlheur}
\tilde{c}^{(2)}_{\Pi} = 1,\,\,\,\,\, \tilde{c}^{(3)}_{\Pi} = \tilde{c}^{(5)}_{\Pi} =0,\,\,\,\,\,  \tilde{c}^{(4)}_{\Pi} = - \left[1 + \fr{18}{5\beta_h}\right],\,\,\,\,\,  \tilde{c}^{(6)}_{\Pi} = \fr{9}{10 \beta_h},
\eeq
in the heuristic approach of Section \ref{sec_heur1}. In Section \ref{PmuT} and  \ref{Pmumu}, we utilize the formulas \eqref{fnleffgappf}, \eqref{cfnlgrad} and \eqref{cfnlheur} to calculate the amplitude of $\langle \mu T \rangle$ and $\langle \mu \mu \rangle$ angular correlators as a function of the location of the dip scale $q_{\rm dip}$ in the power spectrum.

\begin{figure}[t!]
\begin{center}
\includegraphics[scale=0.95]{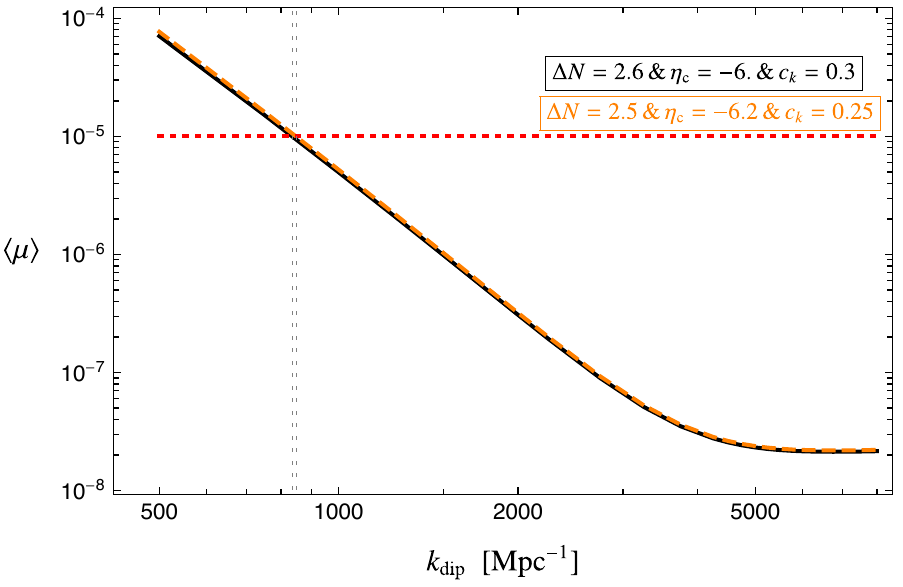}
\end{center}
\caption{\it \small Average $\mu$ distortion with respect to the location of the dip scale in the power spectrum within PBH forming inflationary scenarios. \label{fig:avmu}}
\end{figure}

\noindent{\bf Average $\langle \mu \rangle$.} In the following discussion, we provide $\langle \mu \rangle$ induced due to the scale dependence of the power spectrum in PBH forming single-field inflationary models. For concreteness, we will utilize gradient expansion formalism to describe average $\langle \mu \rangle$ in the sky as a function of the scale $k_{\rm dip}$ beyond which most of the enhancement in the scalar power spectrum occurs. To compute average distortions we start with the following expression\footnote{More refined expressions that relates $\langle \mu \rangle$ to the primordial scalar power spectrum can be found in the earlier works \cite{Chluba:2012we,Chluba:2013dna}.} \cite{Pajer:2012vz},
\beq\label{avmu}
\langle\mu\rangle \simeq 2.3 \int \mathrm{d} \ln k\, \mathcal{P}_{\mathcal{R}}(\tau_f, k)\, \left[e^{-2 k^{2} / k_{D}^{2}(z)}\right]_{z_f}^{z_i}, 
\eeq
where $\mathcal{P}_{\mathcal{R}}(\tau_f, k) = |\alpha_k|^2 \mathcal{P}_{\mathcal{R}}(\tau_k)$ and assuming the scale invariance, $\mathcal{P}_\mathcal{R}(\tau_k) \simeq \mathcal{A}_s \simeq 2.1 \times 10^{-9}$. Using \eqref{akrappf} and \eqref{casq} in \eqref{avmu}, we then estimate the amplitude of $\langle \mu \rangle$ in terms of the location of the dip scale $k_{\rm dip}$. For this purpose, we take the integral in \eqref{avmu} over the scales associated $\mu$ era, in particular from $k_{\rm in}/ {\rm Mpc^{-1}} = 1$ to $k_{\rm fin}/ {\rm Mpc^{-1}} = k_D(z_i) = 11 600$. Results obtained in this way are presented in Figure \ref{fig:avmu} for two representative set of parameter choices that leads to $\Pi_{\rm tot} \simeq 10^7$ growth in the power spectrum. In this plot, the gray dotted lines indicate the threshold value of $k_{\rm dip}$ allowed by the current constraints $\langle \mu \rangle \lesssim 10^{-5}$ such that the right hand side of these vertical lines belongs to the allowed choices of the dip scale. In light of these results, we find it convenient to make the choice $k_{\rm dip} = 10^3\, {\rm Mpc^{-1}}$ to derive predictions for the $\langle \mu T\rangle$ and $\langle \mu \mu \rangle$ correlators (see Sections \ref{PmuT} and \ref{Pmumu}).

We find it worth stressing that our estimates here can be regarded as rough indicator of average distortion in PBH forming single field inflationary models. This is because in realistic models (see \eg \cite{Ozsoy:2018flq,Ozsoy:2020kat}), the spectral shape and amplitude of the power spectrum prior to the dip scale can differ substantially compared to the simple assumption $\mathcal{P}_\mathcal{R}(\tau_k) \simeq \mathcal{A}_s \simeq 2.1 \times 10^{-9}$ we are undertaking here. As a mild indicator of such scenarios, by introducing a red tilt to $\mathcal{P}_{\mathcal{R}}(\tau_k)$ for scales prior to $k_{\rm dip}$ one can confirm that the overall amplitude of $\langle\mu\rangle$ becomes smaller than the estimates we present here. 

\section{Gaussian and non-Gaussian contributions to $C^{\mu\mu}_l$}\label{AppD}
In this appendix, we derive  expressions for the Gaussian and non-Gaussian contribution to the $C^{\mu\mu}_l$ that arise through the connected and disconnected part of the trispectrum respectively. To calculate the distortion angular auto-correlations, we require  
\begin{align}\label{muac}
\nn \left\langle a_{l m}^{\mu^{*}} a_{l^{\prime} m^{\prime}}^{\mu}\right\rangle&= (18.4\pi)^2 i^{l} (-i)^{l'} \int \fr{\d^3 k_1 \d^3 k_2 \d^3 k_3 \d^3 k_4}{(2\pi)^{12}}\,\, Y_{lm}(\hat{k}_+)\,  Y^{*}_{l'm'}(\hat{k}'_+)\,\, W\left(\frac{k_+}{k_{s}}\right) W\left(\frac{k'_+}{k_{s}}\right) \\\nn
&\quad\quad\quad\quad\times  j_{l}\left(k_{+} \chi_*\right) j_{l'}\left(k'_{+} \chi_*\right) \langle\cos{(c_s k_1 \tau)}\cos{(c_s k_2 \tau)}\rangle_p\,\,\langle\cos{(c_s k_3 \tau)}\cos{(c_s k_4 \tau)}\rangle_p\\
&\quad\quad\quad\quad\times  \left[e^{-(k_{1}^{2}\,+\,k^2_2) / k^{2}_{D}(z)}\right]_{f}^{i}\, \left[e^{-(k_{3}^{2}\,+\,k_4^2) / k^{2}_{D}(z)}\right]_{f}^{i} \langle \mathcal{R}_{{k}_{1}}(\tau_f) \mathcal{R}_{{k}_{2}}(\tau_f) \mathcal{R}^{*}_{{k}_{3}}(\tau_f) \mathcal{R}^{*}_{{k}_{4}}(\tau_f)\rangle,
\end{align}
where we used \eqref{mua} with \eqref{defac} and defined $\vec{k}_+ = \vec{k}_1 + \vec{k}_2$, $\vec{k}'_{+} = \vec{k}_3 + \vec{k}_4$. Assuming $\mathcal{R}_k$ is Gaussian at leading order (see \eg \eqref{RNL}), we can generically split the 4-pt correlator in \eqref{muac} into its connected (non-Gaussian) and disconnected (Gaussian) parts as  
\beq\label{4PTF}
\left\langle\mathcal{R}_{{k}_{1}}\mathcal{R}_{{k}_{2}} \mathcal{R}^{*}_{{k}_{3}} \mathcal{R}^{*}_{{k}_{4}}\right\rangle= \langle \mathcal{R}^{\rm G}_{{k}_{1}}(\tau_f) \mathcal{R}^{\rm G}_{{k}_{2}}(\tau_f) \mathcal{R}^{\rm G^{*}}_{{k}_{3}}(\tau_f) \mathcal{R}^{\rm G^{*}}_{{k}_{4}}(\tau_f)\rangle + \langle \mathcal{R}_{{k}_{1}}(\tau_f) \mathcal{R}_{{k}_{2}}(\tau_f) \mathcal{R}^{*}_{{k}_{3}}(\tau_f) \mathcal{R}^{*}_{{k}_{4}}(\tau_f)\rangle_{\rm c}\, .
\eeq
Taking into account all possible pair contractions and using \eqref{psdef}, the disconnected part of the 4-pt correlator \eqref{4PTF} is given by
\begin{align}\label{g4pt}
\langle \mathcal{R}^{\rm G}_{{k}_{1}} \mathcal{R}^{\rm G}_{{k}_{2}} \mathcal{R}^{\rm G^{*}}_{{k}_{3}} \mathcal{R}^{\rm G^{*}}_{{k}_{4}}\rangle &= (2\pi)^6 \bigg\{ \delta(\vec{k}_+)\,\delta(\vec{k}'_+)\, P_{\mathcal{R}}(\tau_f,k_1)\, P_{\mathcal{R}}(\tau_f,k_3)\\\nn
&\quad\quad\quad + P_{\mathcal{R}}(\tau_f,k_1) P_{\mathcal{R}}(\tau_f,k_2) \left[\delta(\vec{k}_1 - \vec{k}_4)\delta(\vec{k}_2 - \vec{k}_3) + \delta(\vec{k}_1 - \vec{k}_3)\delta(\vec{k}_2 - \vec{k}_4) \right]\bigg\}.
\end{align}
On the other hand, the connected part of \eqref{4PTF} can be defined as
\beq\label{C4PT}
 \langle \mathcal{R}_{{k}_{1}}(\tau_f) \mathcal{R}_{{k}_{2}}(\tau_f) \mathcal{R}^{*}_{{k}_{3}}(\tau_f) \mathcal{R}^{*}_{{k}_{4}}(\tau_f)\rangle_{\rm c} = (2\pi)^3 \delta\left(\vec{k}_+ - \vec{k}'_+\right) T_{\mathcal{R}}\left({k}_{1}, {k}_{2}, {k}_{3}, {k}_{4}\right),
\eeq 
where $T_{\mathcal{R}}$ is the trispectrum. Noting \eqref{g4pt} and \eqref{C4PT}, we can then make use of \eqref{4PTF} in \eqref{muac} to separate the angular distortion auto-correlator into Gaussian and non-Gaussian part as $C^{\mu\mu}_l = C^{\mu\mu}_{l,\rm G} + C^{\mu\mu}_{l,\rm NG}$. 

\smallskip
\noindent{\bf Gaussian $C^{\mu\mu}_l$}. Inserting \eqref{g4pt} in \eqref{muac}, notice that the contribution associated with the first term in \eqref{g4pt} vanishes unless the index of the spherical Bessel functions are zero, \ie $l=l'=0$ ($m = m' = 0$). Therefore, after carrying the integrals over $\d^3 k_2$ and $\d^3 k_4$, the contribution of the first term is given by 
\begin{align}\label{mumug1}
\nn \left\langle a_{l m}^{\mu,*} a_{l^{\prime} m^{\prime}}^{\mu}\right\rangle_{\rm G,1} &= 4\pi\, \delta_{l l'} \delta_{m m'} \delta_{l0}\, (2.3)^2 \int \fr{\d^3 k_1 \d^3 k_3}{(2\pi)^6} P_{\mathcal{R}}(\tau_f, k_1) P_{\mathcal{R}}(\tau_f, k_3)\, e^{-2k_1^2/k_D^2(z)} \bigg|_{z_f}^{z_i} \, e^{-2k_1^2/k_D^2(z)} \bigg|_{z_f}^{z_i},\\
&= 4\pi\, \delta_{l l'}\, \delta_{m m'}\, \delta_{l0}\, \langle \mu \rangle^2,
\end{align}
where the average distortion is given by \eqref{avmu}. \textcolor{black}{\eqref{mumug1} describes the average $\langle \mu \mu\rangle$ in the sky (\ie a monopole $l=0$) and should be subtracted from the total $\langle \mu \mu\rangle$. Here our focus is on the non-trivial anisotropies with $l \neq 0$ that originates from the second term in \eqref{g4pt}.} Taking the integrals over $\d^3 k_3$ and $\d^3 k_4$ in \eqref{muac}, this contribution to the Gaussian $C^{\mu\mu}_{l,{\rm G}}$ reads as  
\begin{align}
\nn \left\langle a_{l m}^{\mu,*} a_{l^{\prime} m^{\prime}}^{\mu}\right\rangle_{\rm G,2} &= (18.4\pi)^2 i^{l} (-i)^{l'} \int \fr{\d^3 k_1 \d^3 k_2}{(2\pi)^{6}}\,\, Y_{lm}(\hat{k}_+)\,  Y^{*}_{l'm'}(\hat{k}_+)\,\, W\left(\frac{k_+}{k_{s}}\right)^2 j_{l}\left(k_{+} \chi_*\right) j_{l'}\left(k_{+} \chi_*\right) \\\nn
&\quad\quad\quad\quad\quad\quad\quad\quad\times  \langle\cos{(c_s k_1 \tau)}\cos{(c_s k_2 \tau)}\rangle^2_p \,\left(\left[e^{-(k_{1}^{2}\,+\,k^2_2) / k^{2}_{D}(z)}\right]_{z_f}^{z_i}\right)^2\\
&\quad\quad\quad\quad\quad\quad\quad\quad\times  \, \left[\fr{2\pi^2}{k_1^3} \mathcal{P}_{\mathcal{R}}(\tau_f,k_1)\right] \left[\fr{2\pi^2}{k_2^3} \mathcal{P}_{\mathcal{R}}(\tau_f,k_2)\right].
\end{align}
We then make the transformation $\d^3 k_2 \to \d^3 k_+$ and carry out the integral over directions $\d^2 \hat{k}_+$. This gives 
\begin{align}\label{mumug2}
\left\langle a_{l m}^{\mu,*} a_{l^{\prime} m^{\prime}}^{\mu}\right\rangle_{\rm G,2}  &= (4.6)^2 \delta_{l l'} \delta_{m m'}  \int \d k_+\, k_+^2\,\, j_l(k_+ \chi_*)^2\,\, W\left(\fr{k_+}{k_s}\right)^2\int \fr{\d^3 k_1}{k_1^3 |\vec{k}_+ -\vec{k}_1|^3}\\\nn &\quad\quad\quad\quad\times \langle\cos{(c_s k_1 \tau)\cos{(c_s |\vec{k}_+ - \vec{k}_1| \tau)}}\rangle^2_p\,\, \mathcal{P}_{\mathcal{R}}(\tau_f,k_1)\,\mathcal{P}_{\mathcal{R}}(\tau_f,|\vec{k}_+ - \vec{k}_1|)\\ \nn &\quad\quad\quad\quad\times \left(\left[e^{-(\,k_1^2 + |\vec{k}_+ - \vec{k}_{1}|^{2}\,) / k^{2}_{D}(z)}\right]_{z_f}^{z_i}\right)^2.
\end{align}

\noindent Furthermore, making the replacement $\vec{k}_3 \leftrightarrow \vec{k}_4$ in the last term of \eqref{g4pt} and noticing the fact the rest of the integrand in \eqref{muac} is symmetric under $\vec{k}_3 \leftrightarrow \vec{k}_4$, we have $\left\langle a_{l m}^{\mu,*} a_{l^{\prime} m^{\prime}}^{\mu}\right\rangle_{\rm G,2} = \left\langle a_{l m}^{\mu,*} a_{l^{\prime} m^{\prime}}^{\mu}\right\rangle_{\rm G,3}$. Therefore, the non-trivial part ($l\neq 0$) part of the Gaussian contribution to $C_l^{\mu\mu}$ is given by 

\begin{align}\label{amucg}
\nn C_{l,{\rm G}}^{\mu\mu} &= 2\, (4.6)^2  \int \d k_+\, k_+^2\,\, j_l(k_+ \chi_*)^2\,\, W\left(\fr{k_+}{k_s}\right)^2\int \fr{\d^3 k_1}{k_1^3 |\vec{k}_+ -\vec{k}_1|^3} \mathcal{P}_{\mathcal{R}}(\tau_f,k_1)\,\mathcal{P}_{\mathcal{R}}(\tau_f,|\vec{k}_+ - \vec{k}_1|) \\
&\quad\quad\quad\quad\quad\quad\quad\quad\quad\times \langle\cos{(c_s k_1 \tau)\cos{(c_s |\vec{k}_+ - \vec{k}_1| \tau)}}\rangle^2_p \left(\left[e^{-(\,k_1^2 +|\vec{k}_+ - \vec{k}_{1}|^{2}\,) / k^{2}_{D}(z)}\right]_{z_f}^{z_i}\right)^2.
\end{align}

\noindent{\bf non-Gaussian $C^{\mu\mu}_l$}. To derive the non-Gaussian contribution to $\langle \mu\mu \rangle$, we insert \eqref{C4PT} in \eqref{muac}. In particular, transforming $\d^3 k_2 \to d^3 k_+$ and $\d^3 k_4 \to \d^3 k'_+$ we first take the integral over $\d^3 k'_+$ using \eqref{C4PT} ($\vec{k}_+ = \vec{k}'_+$) to obtain
\begin{align}\label{clmung}
\nn \left\langle a_{l m}^{\mu^{*}} a_{l^{\prime} m^{\prime}}^{\mu}\right\rangle_{\rm NG} &= (18.4\pi)^2 i^{l} (-i)^{l'} \int \fr{\d^3 k_+ \d^3 k_1 \d^3 k_3}{(2\pi)^{9}}\,\, Y_{lm}(\hat{k}_+)\,  Y^{*}_{l'm'}(\hat{k}_+)\,\, W\left(\frac{k_+}{k_{s}}\right)^2 j_{l}\left(k_{+} \chi_*\right) j_{l'}\left(k_{+} \chi_*\right)\\\nn
&\quad\quad\quad\quad\quad\quad\quad\quad\times \langle\cos{(c_s k_1 \tau)}\cos{(c_s  |\vec{k}_+-\vec{k}_1| \tau)}\rangle_p\,\,\langle\cos{(c_s k_3 \tau)}\cos{(c_s  |\vec{k}_+-\vec{k}_3| \tau)}\rangle_p\\
&\quad\times   \left[e^{-(k_{1}^{2}\,+\,|\vec{k}_{+}-\vec{k}_1|^2) / k^{2}_{D}(z)}\right]_{z_f}^{z_i}\, \left[e^{-(k_{3}^{2}\,+\,|\vec{k}_{+}-\vec{k}_3|^2) / k^{2}_{D}(z)}\right]_{z_f}^{z_i} T_{\mathcal{R}}\left(k_{1}, |\vec{k}_{+} - \vec{k}_1|, k_3,  |\vec{k}_{+} - \vec{k}_3|\right).
\end{align}
In \eqref{clmung}, the presence of integration over $\d^2 \hat{k}_1$ and $\d^2 \hat{k}_3$ ensures the independence of the integrand on $\hat{k}_+$ except in the arguments of $Y_{lm}$'s \cite{Chluba:2016aln}. We can therefore take the integral over $\d^2 \hat{k}_+$ using $\int \d^2 \hat{k}_+ Y_{lm}(\hat{k}_+)\,  Y^{*}_{l'm'}(\hat{k}_+) = \delta_{ll'}\delta_{mm'}$ to yield
\begin{align}\label{clmumuf}
\nn C_{l,{\rm NG}}^{\mu \mu} &\simeq \frac{(2.3)^2}{8\pi^{7}} \int \d k_{+}\,k_{+}^{2}\, j_{l}^2\left(k_{+} \chi_*\right)W\left(\frac{k_+}{k_{s}}\right)^2\int \d^3 k_{1}\d^3 k_{3}\,\,\,T_{\mathcal{R}}\left(k_{1}, |\vec{k}_{+} - \vec{k}_1|, k_3,  |\vec{k}_{+} - \vec{k}_3|\right)\\\nn
&\quad\quad\quad\quad\quad\quad\quad\quad\times \langle\cos{(c_s k_1 \tau)}\cos{(c_s |\vec{k}_+-\vec{k}_1| \tau)}\rangle_p\,\,\langle\cos{(c_s k_3 \tau)}\cos{(c_s |\vec{k}_+-\vec{k}_3| \tau)}\rangle_p \\
&\quad\quad\quad\quad\quad\quad\quad\quad\quad\quad\quad\times  \left[e^{-(k_{1}^{2}\,+\,|\vec{k}_{+}-\vec{k}_1|^2) / k^{2}_{D}(z)}\right]_{z_f}^{z_i}\, \left[e^{-(k_{3}^{2}\,+\,|\vec{k}_{+}-\vec{k}_3|^2) / k^{2}_{D}(z)}\right]_{z_f}^{z_i},
\end{align}
where we extracted $C^{\mu\mu}_l$ from \eqref{clmumuf} using the definition \eqref{defac}. 
\end{appendix}
\subsection{$C^{\mu\mu}_{l,{\rm G}}$ in the collapsed $k_+ \to 0$ limit.}\label{AppD1}
We now provide an estimate for the Gaussian $\mu\mu$ correlator \eqref{amucg} in the collapsed limit. Taking $k_+ \to 0$ in  \eqref{amucg} we have  
\beq\label{appacmug2}
C_{l,{\rm G}}^{\mu\mu} \simeq  5.9 \times 10^{-16} \int \d k_+\, k_+^2\,\, j_l(k_+ \chi_*)^2\,\int \fr{\d k_1}{k_1^4} E(k_1)^2 \left(\left[e^{-2k_1^2 / k^{2}_{D}(z)}\right]_{z_f}^{z_i}\right)^2,
\eeq
where we used $\mathcal{P}_{\mathcal{R}}(\tau_f, k_1) \simeq E(k_1) \mathcal{A}_s$ with $\mathcal{A}_s = 2.1 \times 10^{-9}$ and $E(k_1)=\{\Pi(k_1), |\alpha_{k_1}|^2\}$ for heuristic and gradient approaches respectively. Using the dimensionless variables $\tilde{k} = k_1 / k_{\rm dip}$ and $x = k_+ \chi_*$, we can rewrite \eqref{appacmug2} as
\beq\label{appacmug2f}
C_{l,{\rm G}}^{\mu\mu} \simeq  5.9 \times 10^{-16}\,\,  \mathcal{I}\left[k_{\rm dip}\right] \int \d x \, x^2\,\, j_l(x)^2 \,,
\eeq
 \begin{figure}[t!]
\begin{center}
\includegraphics[scale=0.91]{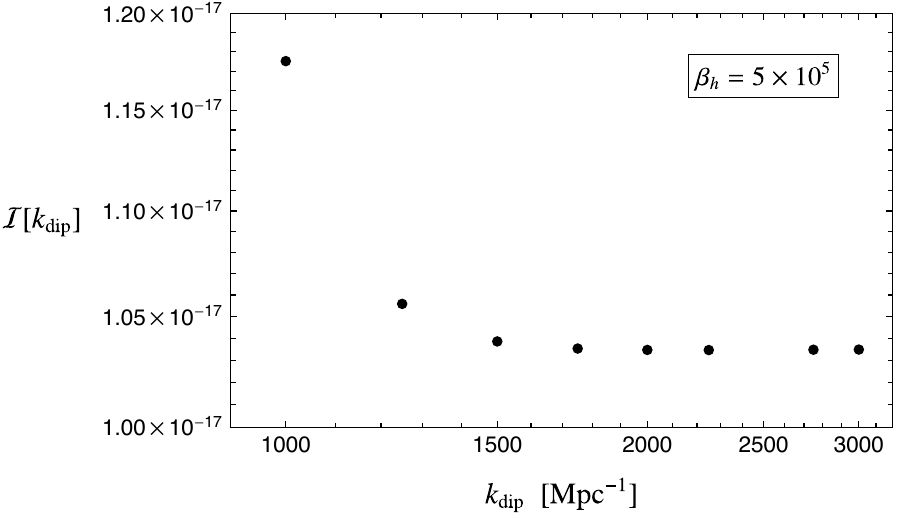}\includegraphics[scale=0.84]{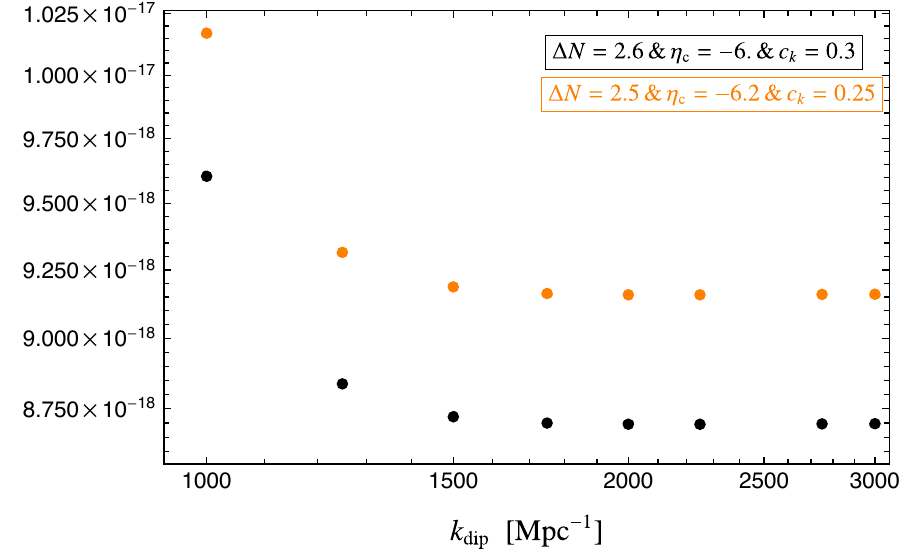}
\end{center}
\caption{\it \small The integral $\mathcal{I}$ \eqref{I} for a grid of phenomenologically interesting $k_{\rm dip}$ values using heuristic approach (Left) and the gradient expansion formalism (Right).  \label{fig:Int}}
\end{figure}\noindent
with the integral that depends on the location of $k_{\rm dip}$ (and model parameters) defined as
\beq\label{I}
\mathcal{I}\left[k_{\rm dip}\right] = \left(\fr{k_*}{k_{\rm dip}}\right)^3 \int \fr{\d \tilde{k}}{\tilde{k}^4} E(\tilde{k})^2 \left(\left[e^{-2\tilde{k}^2 / \tilde{k}^{2}_{D}(z)}\right]_{z_f}^{z_i}\right)^2,
\eeq
where we defined $k_* = \chi_*^{-1} \simeq 1/ (14\, {\rm Gpc}) \simeq 7.1\times 10^{-5}\, {\rm Mpc}^{-1}$ and  $\tilde{k}_D(z) = k_D(z)/ k_{\rm dip}$. Notice that $( k_*/k_{\rm dip})^3$  factor in \eqref{I} introduces a large suppression factor to the amplitude of $\mathcal{I}$ for phenomenologically interesting values around $k_{\rm dip} \simeq 10^3$. To illustrate this, in Figure \ref{fig:Int}, we plot $\mathcal{I}$ as a function of $k_{\rm dip}$ for representative parameter choices within both heuristic and gradient approach. On the other hand, for large enough multipoles $l$, $j_l^2(x)$ is highly peaked around $x\approx l$ and acts like a delta function in the last integral in \eqref{appacmug2f} and therefore we can approximate $\int \d x \, x^2 j_l^2(x) \approx l^2$. As a result, we anticipate that the Gaussian contribution can be approximated as for $l > 0$ as
\beq\label{mumugmax}
C^{\mu\mu}_{l,G} \approx 10^{-32}\, l^2. 
\eeq
For scales we can probe with $\mu$ distortions, we have $l_{\rm max} = 200$, this result above \eqref{mumugmax} gives $(C^{\mu\mu}_{l,\rm G})_{\rm max} \approx 10^{-28}$ and therefore we can safely conclude that in the collapsed limit, $C^{\mu\mu}_{l,\rm G} \ll C^{\mu\mu}_{l,\rm NG}$ holds for scales we would like to probe $\mu$ distortion anisotropies (See section \ref{Pmumu}).
\addcontentsline{toc}{section}{References}
\bibliographystyle{utphys}

\bibliography{paper2}

\end{document}